\documentclass[onecolumn]{aastex63}
\usepackage{graphicx,subfigure,enumerate,pstricks,latexsym,longtable,amsfonts,amsmath}
\usepackage{soul}
\usepackage{lipsum}
\usepackage{lineno}

\newcommand{\gama}{$\gamma$}

\newcommand{\beq}{\begin{equation}}
\newcommand{\eeq}{\end{equation}}

\defcitealias{Bhatta2020}{BD20}

\shorttitle{Optical variability of blazars}
\shortauthors{Bhatta G.}

\begin{document}
\title{Characterizing Longterm Optical Variability Properties of \gama-ray Bright Blazars}

\correspondingauthor{Gopal Bhatta}
\email{gopal.bhatta@ifj.edu.pl}

\author{Gopal Bhatta}
\affiliation{Institute of Nuclear Physics Polish Academy of Sciences \\
PL-31342 Krakow, Poland}

\begin{abstract}
Optical observations of a sample of 12  \gama-ray bright blazars  from four optical data archives, AAVSO, SMARTS, Catalina, and Steward Observatory, are compiled  to create  densely sampled light curves spanning more than a decade. As a part of the blazar multi-wavelength studies, several methods of analyses, e. g., flux distribution and RMS-flux relation, are performed on the observations with an aim to compare the results with the similar ones in the \gama-ray band presented in \citet{Bhatta2020}.  It is found that, similar to \gama-ray band,  blazars display significant variability in the optical band that can be characterized with  log-normal flux distribution and a power-law dependence of RMS on flux. It could be an indication of possible inherent linear RMS-flux relation, yet the  scatter in the data does not allow to rule out other possibilities. When comparing variability properties in the two bands, the blazars in the \gama-rays are found to exhibit stronger variability with steeper possible linear RMS-flux relation and the flux distribution that is more skewed towards higher fluxes. The cross-correlation study shows that except for the source 3C 273, the overall optical and the \gama-ray emission in the sources are highly correlated, suggesting a co-spatial existence of the particles responsible for both the optical and  \gama-ray emission. Moreover,  the sources  S5 0716+714, Mrk 421, Mrk 501, PKS 1424-418 and PKS 2155-304  revealed possible evidence for quasi-periodic oscillations in the optical emission with the characteristic timescales, which are comparable to those in the \gama-ray band detected in our previous work.  
\end{abstract}



\section{Introduction \label{sec:intro}}
A small class of active galactic nuclei (AGN) that possess parsec-scale radio jets pointing towards the Earth are widely known as blazars. The sources represent one of the most luminous objects (L$\sim10^{47}$ erg/s) which derive their power by accretion onto supermassive black holes at the center.  The sources shine exclusively in the Doppler boosted non-thermal emission that can be distinguished by its extreme properties e. g. high luminosity, strong polarization, and rapid flux variability.  Blazars, which illuminate the \gama-ray sky, could also be sources of cosmic neutrinos \citep[see][]{2018Sci...361.1378I,2018Sci...361..147I}. The spectral energy distribution (SED) of the broadband continuum emission from the sources is usually recognized by the two distinct hump-like features on the logarithmic Luminosity-Frequency plane. The synchrotron emission resulting from the relativistic particles abundant in the large-scale jets give rise to  the feature found in the low energy regime,  whereas inverse-Compton (IC) scattering of low energy photons by the energetic particles could result in the higher energy feature. However, various models attempt to furnish the details of the origin of the latter component. Particularly, two widely discussed leptonic models are the synchrotron self-Compton (SSC) model and the external Compton (EC) model. According to SSC (e.g., \citealt{Maraschi1992,Mastichiadis2002}) the same population of the electrons emitting synchrotron photons are responsible in up-scattering the low energy photons to produce high energy emission; whereas in the EC models the softer seed photons can come from various regions that are external to the jets, e. g., accretion disk (AD; \citealt{Dermer1993}), broad-line region (BLR; \citealt{Sikora1994}), and dusty torus (DT; \citealt{Blazejowski2000}).

Blazars are mainly grouped into two flavors:  flat-spectrum radio quasars (FSRQ), which are the more powerful sources, show emission lines over the continuum;  whereas BL Lacertae (BL Lac) sources,  the less powerful ones, show weak or no such lines. The peak of the synchrotron emission from FSRQs lies is in the low frequency regime, that is, radio to optical, and the peak of the IC component can extend up to GeV regime.  In FSRQs, we find evidence of  abundant presence of seed photons from AD, BLR and DT  \citep{Ghisellini2011} such that EC processes can result in copious amount of high energy emission. Consequently, the continuum emission from FSRQs is mostly dominated by \gama-rays.  On the other hand, in  BL Lac sources, in which the synchrotron peak lies in the UV or X-ray regions, the IC component can peak up to TeV energies making the objects an extreme class of sources.  The apparent low luminosity of this class of sources could be linked to the absence of strong circumnuclear photon fields and relatively low accretion rates. We also find another classification scheme based on the  location of  synchrotron peak frequency ($\nu_\mathrm{s}$), according to which the sources can be classified as   low synchrotron peaked blazars ( $\nu_s \leq 10^{14.0}$ Hz), intermediate synchrotron  peaked blazars ($10^{14}<\nu_s \leq10^{15.3} $ Hz),  and high synchrotron peaked blazars  ($\nu_s > 10^{15}$ Hz) (\citealt{Fan2016}, for previous similar classification see \citealt{Abdo2010}.

Blazar emission is found to be variable over a wide range of spectral and temporal frequencies  (e. g. optical;  {\citealt{Bhatta2013, Bhatta2016b} and  \citealt{Bhatta2018a}, X-ray; \citealt{Bhatta2018c}, \gama-ray; \citealt{Bhatta2020} \citep[see also recent works e. g.][]{Weaver2020}. The  studies of the multi-frequency variability  properties are utilized to explore into the physical  mechanisms occurring at the innermost regions of the source. The observed multi-wavelength (MWL) variability can be linked to a number of processes occurring either in the accretion disk and/or in the jet; e. g. the accretion disk revolving around the supermassive black hole, various magnetohydrodynamic instabilities in the disk and the jets, shocks traveling down the turbulent jets, and relativistic effects due to jet orientation \citep[e.g.][and the references therein]{Camenzind92,Wagner1995,Bhatta2013, Marscher14}. However, the exact details of the models are widely discussed and debated.  In such context, study of optical long-term variability of  blazars in relation to the similar variability in the \gama-ray,  could be important to probe the  jet dynamics, particle acceleration and energy dissipation mechanisms producing \gama-ray emission.

\begin{figure*}[ht!]
\plotone{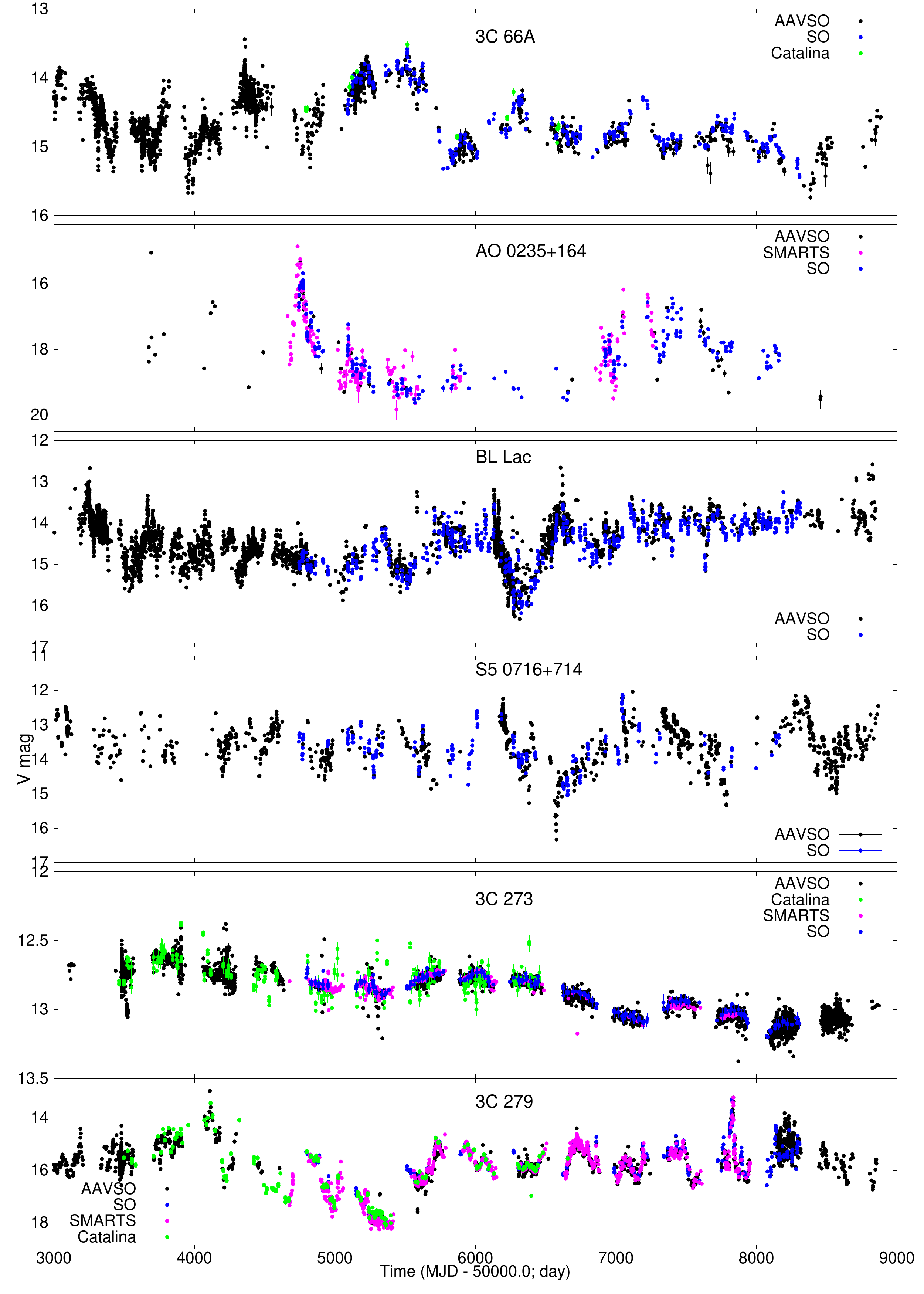}
\label{Fig:1}
\caption{Longterm optical observations of a sample of \gama-ray bright blazars as obtained from 4 optical data archives see Section \ref{sec:2}. The observation from different data archives are presented in different colors, i. e., AAVSO, Catalina  SMARTS  and Steward Observatory in black, green, magenta, and blue, respectively.
}
\end{figure*}

\renewcommand{\thefigure}{\arabic{figure} (Cont.)}
\addtocounter{figure}{-1}

\begin{figure*}[!b]
\begin{center}
\plotone{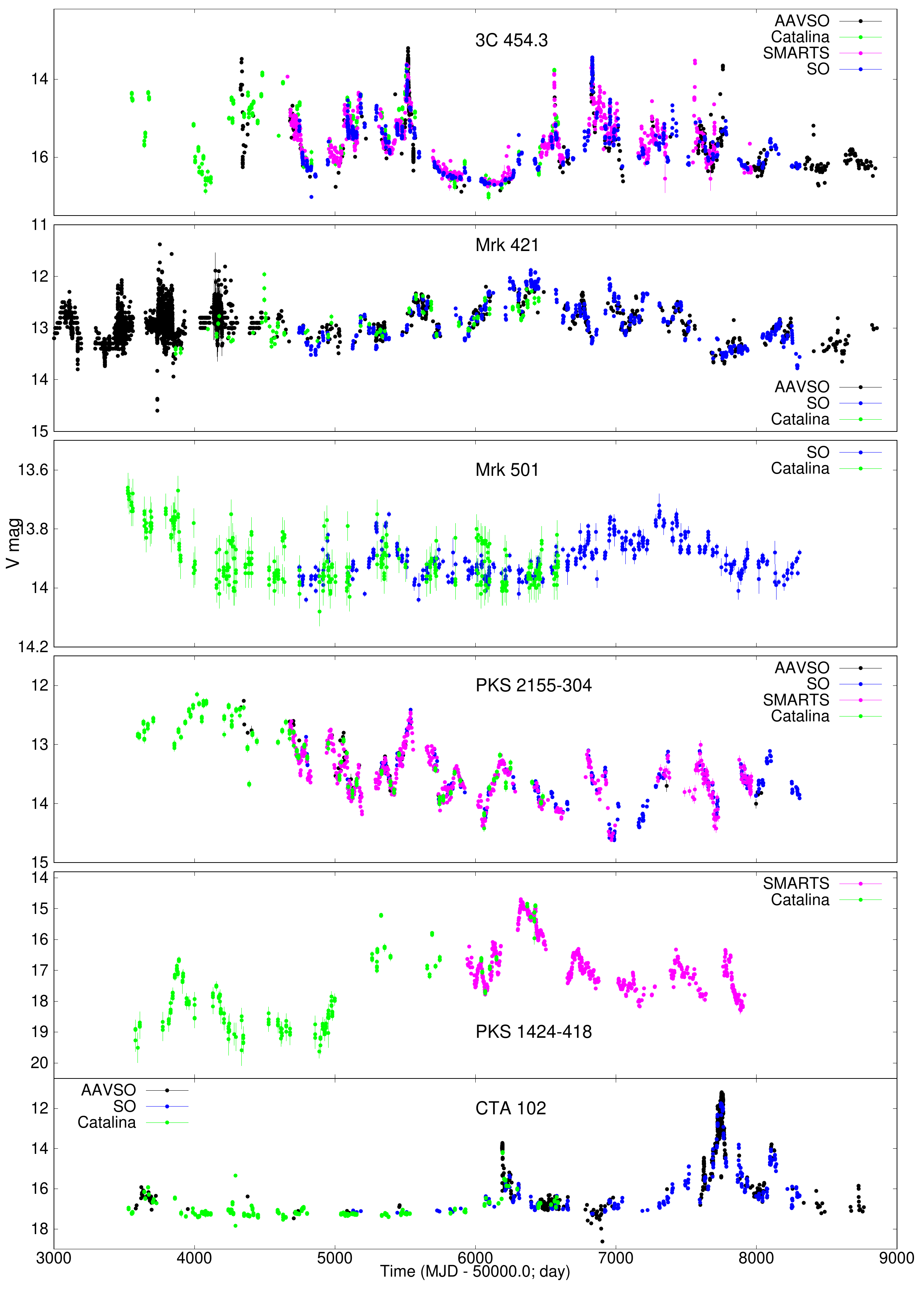}
\caption{\label{Fig:1a}}
\end{center}
\end{figure*}

Apart from the generic multi-timescale aperiodic flux variability over the wide electromagnetic frequency bands, some of the blazars are known to display periodic or quasi-periodic oscillations (QPO) in their flux  \citep[for recent review on QPOs in blazars see][]{2019arXiv190910268B,Gupta2018}. In these cases, the multi-frequency longterm light curves exhibit a characteristic timescales of a few years. In particular, in our previous work (\citealt{Bhatta2020}; henceforth \citetalias{Bhatta2020}), a detailed  study of \gama-ray emission from the blazars, we reported QPOs in the sources Mrk 421, Mrk 501, PKS 1424-418, S5 0716+714, and PKS 2155-304. In this work, the possible presence of these QPOs in the optical band is examined, and thereby attempts to establish their MWL nature.

In this work, we utilize decade long optical observations and carry out a detailed time series analysis  of a sample of 12 blazars. These blazars form a sub-sample of the sources presented in \citetalias{Bhatta2020}. Also as a part of the multi-band blazar variability studies, we compare the results with the results in the \gama-ray band from \citetalias{Bhatta2020}.  In Section \ref{sec:2}, the sample of the blazars and the data acquisition and compiling of the optical observations are outlined.  In Section \ref{sec:3}, several methods of the analysis including fractional variability, RMS-flux relation and flux distribution are discussed as well as the results of the analyses on the light curves. Then discussion on the results along with their possible implications on the nature of optical emission from the sources are presented in Section \ref{sec:4}, and  finally, summary and  conclusions of the study are outlined in Section \ref{sec:5}. 

\begin{deluxetable*}{lllllll}
\tablecaption{The source sample of the 
\gama-ray bright blazars \label{table:1}}
\tablewidth{500pt}
\tabletypesize{\scriptsize}
\tablehead{
\colhead{Source name} & \colhead{R.A. (J2000)} & 
\colhead{Dec. (J2000)} & \colhead{Red-shift} &  
\colhead{Source class} & \colhead{mean mag.$\pm$ stdv.} &\colhead{FV (\%)}\\
} 
\colnumbers
\startdata
       3C 66A 	&	 $02^h22^m41.6^s$ 	&	 $+43^d02^m35.5^s$ 	&	0.444	&	BL Lac   &	14.69	$\pm$0.38 	&	37.83	$\pm$0.23 	\\
	AO 0235+164 	&	 $02^h 38^m38.9^s$ 	&	 $+16^d 36^m 59^s$ 	&	0.94	&	BL Lac  &	18.19	$\pm$0.87  	&	115.00	$\pm$0.24 	\\
	S5 0716+714 	&	$07^h21^m53.4^s$ 	&	 $+71^d20^m36^s$ 	&	0.3	&	BL Lac  &	13.62	$\pm$0.59 	&	53.74	$\pm$0.09 	\\
	Mrk 421 	&	 $11^h04^m273^s$ 	&	 $+38^d12^m32^s$ 	&	0.03	&	BL Lac  &	12.96	$\pm$0.31  	&	29.95$\pm$0.15	\\
	3C 273 	&	 $12^h29^m06.6997^s$ 	&	 $+02^d03^m08.598^s$ 	&	0.158	&	FSRQ  &	12.80	$\pm$0.16 	&	14.44$\pm$0.28	\\
	3C 279 	&	 $12^h56^m11.1665^s$ 	&	 $-05^d47^m21.523^s$ 	&	0.536	&	FSRQ  &	15.73	$\pm$0.82 	&	80.60$\pm$	0.10	\\
	PKS 1424-418 	&	 $14^h27^m56.3^s$ 	&	 $-42^d06^m19^s$ 	&	1.522	&	 FSRQ  &	17.15	$\pm$1.01 	&	114.24$\pm$0.14	\\
	Mrk 501 	&	 $16^h53^m52.2167^s$ 	&	 $+39^d45^m36.609^s$ 	&	0.0334	&	 BL Lac  &	13.90	$\pm$0.07 	&	6.00$\pm$	1.46	\\
	PKS 2155-304 	&	 $21^h58^m52.0651^s$ 	&	 $-30^d13^m32.118^s$ 	&	0.116	&	BL Lac  &	13.49	$\pm$0.45 	&	46.01	$\pm$0.22	\\
	BL Lac 	&	$22^h02^m43.3^s$ 	&	 $+42^d16^m40^s$ 	&	0.068	&	BL Lac 	&	14.44	$\pm$0.51  &	46.22$\pm$0.07	\\
	CTA 102 	&	 $22^h32^m36.4^s$ 	&	 $+11^d43^m51^s$ 	&	1.037	&	FSRQ  &	16.34	$\pm$1.02 	&	335.27	$\pm$0.02 	\\
	3C 454.3  	&	 $22^h53^m57.7^s$ 	&	 $+16^d08^m54^s$ 	&	0.859	&	FSRQ  &	15.75	$\pm$0.63 	&$78.16\pm0.11$		\\
\enddata
\end{deluxetable*}

\renewcommand{\thefigure}{\arabic{figure}}

\section{Sample sources and data acquisition \label{sec:2}}
The sources included in the study sample are generally \gama-ray bright (mostly TeV) blazars and consist of 7 BL Lacs  and 5 FSRQs.
To carry out a systematic analysis of the optical variability in blazars, longterm optical V band observations from 4 of the optical data archives, i. e., Steward Observatory {\footnote{ \url{http://james.as.arizona.edu/~psmith/Fermi/}}\citep{Smith2009},
Catalina Surveys data archive{\footnote{\url{http://nesssi.cacr.caltech.edu/DataRelease/}} (see \citealt{Drake2009}),
the American Association of Variable Star Observers (AAVSO) {\footnote{ \url{https://www.aavso.org/data-download}} and
the Small \& Moderate Aperture Research Telescope System (SMARTS; \citealt{Bonning2012}){\footnote{\url{http://www.astro.yale.edu/smarts/glast/home.php}}, were collected and compiled to create densely sampled decade-long light curves of the sample sources.  With a goal to study multi-frequency variability properties of the blazar sources by comparing the results of the similar analysis in the \gama-ray band bands, the optical sample sources were taken from our previous work \citetalias{Bhatta2020}. For a robust comparison with the weekly \gama-ray observations,  the optical source sample was subjected to additional criteria, that the observations should have a comparable (to the \gama-ray observations) sampling in time and the least gap possible. Since not all the sources passed this criteria, out the 20 blazar presented in \citetalias{Bhatta2020}, only 12 sources were included in this study. Moreover, data are not available for every source in all 4 data archives, therefore, for some of the sources data from fewer than the 4 data archives were used. The optical  light curves that span more than a decade are presented in Figure \ref{Fig:1}, in which the observations from different data archives are shown in different colors, i. e., AAVSO, Catalina,  SMARTS  and Steward Observatory in black, green, magenta, and blue, respectively.

The source brightness in V magnitudes was converted into flux in Jansky units by using the zero points for  V band in the Cousins–Glass–Johnsons system given in Table A2 of \citet{Bessell1998}. Also, the flux were interstellar extinction corrected using the extinction  in V magnitudes for the sources as listed in the NED\footnote{\url{https://ned.ipac.caltech.edu/}}. In addition, to perform a  cross-correlation study between the optical and \gama-ray band (0.1--300 GeV), observations from the Fermi/LAT telescope were utilized. The data acquisition and the processing of the \gama-ray observations are discussed in detail in \citetalias{Bhatta2020}.

  The source names, RA, Dec, red-shift, source classification, mean magnitude with standard deviations are presented in column 1, 2, 3, 4, 5 and 6, respectively, of Table \ref{table:1}.  Of the 12 sources, the
  farthest and the nearest sources are PKS 1424-418 and Mrk 421 located at the corresponding red-shifts of 1.522 and 0.03, respectively.  Similarly, the brightest source in the sample is 3C 273 which has a mean magnitude 12.80,  and the faintest source AO 0235+164, with a mean magnitude 18.19 and standard deviation of 0.87 magnitude, is highly variable. The second faintest source, PKS 1424-418, with a mean magnitude of 17.15 is  also the most variable source as indicated by its standard deviation of magnitudes 1.01.

\section{Analysis \label{sec:3}}
With an aim to study the MWL statistical properties of blazar variability, various methods of variability analysis e. g. fractional variability, RMS-flux relation and flux distribution,  similar to those discussed in \citetalias{Bhatta2020}, were performed on the optical light curves of the sample sources. In addition, the search for QPOs was carried out  using Lomb-Scargle periodogram and comparing the possible QPO feature with the ones from the \gama-ray observations of the similar duration. The detailed discussion on the methods and the corresponding results of the analyses are presented below.

\subsection{Fractional variability}
The light curves constructed from the multi-instrument optical observation, as shown in Figure  \ref{Fig:1}, clearly show the modulations in the source flux over the period.  It is one of the primary goals of the study  to constrain the possible mechanisms playing out in the prevailing physical conditions that result in such a  dramatic variability. Fractional variability (FV), as presented in the form of Equation \ref{fv}, provides a quantified measure of the average variability in the light curve. It is expressed as,
\begin{equation}
F_{var}=\sqrt{\frac{S^{2}-\left \langle \sigma _{err}^{2} \right \rangle}{\left \langle F \right \rangle^{2}}} ,
\label{fv}
\end{equation}
and the uncertainty in the FV can be written as,
\begin{equation}
\centering
\sigma_{F_{var}}=\sqrt{ F_{var}^{2}+\sqrt{ \frac{2}{N}\frac{\left \langle \sigma _{err}^{2} \right \rangle^{2}}{\left \langle F \right \rangle^{4}}+  \frac{4}{N}\frac{\left \langle \sigma _{err}^{2} \right \rangle }{\left \langle F \right \rangle^{2}} F _{var}^{2}}} - F_{var}
\end{equation}
 where F and S stand for the flux and the variance in the flux, respectively, (\citealt{Vaughan2003}, see also \citealt{Bhatta2018a}). FV for the light curves of the sample blazars along with the associated uncertainty are listed in the 7$_{th}$ column of Table \ref{table:1} that shows that the source light curves display remarkable variability in the optical emission.  The mean FV of the sources in the sample is 79\%   with a standard deviation of  87\% --  the mean FV of BL Lacs is 47.82 \% with  standard deviation of 33.47 \% and that of the FSRQs is 124.54\% with a standard deviation of 123.20 \%.  Of the sample sources, the most variable source is FSRQ CTA 102 (z= 1.037) with FV$_{\gamma}$  $\sim$117\% and FV$_{opt}$  $\sim$335\% ; whereas the least variable source turns out to be  BL Lac  Mrk 501 with FV $\sim$ 6\%. 

To compare the variability properties in the two bands,  the linear correlation between the fractional variability in the \gama-ray and optical bands were estimated. It resulted in a strong correlation as given by the Pearson linear correlation coefficient value 0.68    with a p value of 0.01. Furthermore, comparing the FV values in Table \ref{table:1} in this work with those in BD20, it is found that sources, in general, exhibit larger variability in  the \gama-ray band with an exception of a few sources , e. g. AO 0235+164, PKS 1424-418, and CTA 102, Also, it is interesting to note that the FV of the source PKS 2155-304  being FV $\sim 46$ \%  in both of the bands, it shows very similar quantity of variability in these bands.

\subsection{RMS-flux relation}
In order to further investigate the variability properties, the variability distribution over the source optical fluxes was studied by employing excess variances as an estimator of the intrinsic source variance and thereby exploring the correlation between excess variances and the mean flux for a particular time bin in a light curve. This kind of relation between these quantities is  commonly known as \emph{RMS-flux relation} \citep[see][]{Vaughan2003}. For the purpose, the light curves were divided into N segments of equal lengths, and to ensure a robust and meaningful statistical results, each segment included at least 20 observations. For a given segment of the light curve, the excess variance is obtained by subtracting Poisson noise from the sample variance, that is, $\sigma _{XS}^{2}=S^{2}-\bar{\sigma_{err} ^{2}}$; where $S^{2}$ represents the sample variance and and the mean square of measurement error is  given by $\bar{\sigma_{err} ^{2}}=1/n\sum_{i}^{n}\sigma_{err,i} ^{2}$ \citep[see][]{Nandra1997}.  The RMS-flux plots for the sample blazar sources are shown in  Figure \ref{fig:RMS}. To characterize the trend on the  RMS-flux, the observation were fitted using a linear model both with and without an intercept as shown by magenta and blue lines, respectively, in the figure.   The resulting slope and intercept parameters of the linear fits are presented in column 2 and 5, and 3 of Table \ref{tab:table3}, and also the corresponding reduced $\chi^2$ statistics are listed in Column 4 and 6. The negative intercept in the RMS axis, equivalently a positive intercept on the flux axis, can be interpreted as a component of constant flux in the linear RMS-flux relation. On the other-hand, a positive RMS  intercept implies the presence of an excess RMS even when the RMS-flux relationship is extrapolated to zero flux, which obviously is non-physical. But it is possible that in addition to the linear RMS-flux relation, there could be another component which can dominate in the very low flux states  \citep[see][]{Gleissner2004}.

From Figure \ref{fig:RMS} and Table \ref{tab:table3}, it can be gathered that the source observations exhibit a  general a positive correlation between RMS and the mean flux, possibly a linear RMS-flux relation.  However, as a standard practice in the X-ray astronomy, in order to truly investigate the relation between the long-term mean flux and the short-term fluctuations, the segments of light curve are further divided \textbf{into} several sub-segments so that the RMS-flux relation is observed over many different frequency ranges  \citep[e. g.  see][]{Alston2019a}. This usually requires a larger number of observations, which may not be always possible in the case of the optical observations obtained from the ground based telescopes, e. g. due to the local weather and seasonal conditions. Another important caveat is that general in the radio-quiet AGN the frequency range being sampled has low intrinsic RMS (typically  $\leq$ 15 \% fractional RMS), the method is more suitable to such analysis. However in the case of blazars,  which are most variable in all frequencies, (e. g. see 7th Column of Table \ref{table:1}), there is a small chance that  a spurious linear RMS-flux relation may appear in the data  \citep[see][for more details]{Uttley2005}.  Nevertheless, the analysis presented here reveal a trend of RMS-flux relation inherent in the optical light curves of the sample blazars.

To compare  the slope parameters of the possible linear RMS-flux relation observed here and the corresponding results in \citetalias{Bhatta2020},  it is found that for FSRQs the average slope of the RMS-flux linear fit results 0.58, whereas  for BL Lacs the average value is 0.30 -- FSRQs showing a steeper slope in the RMS-flux pane. The result implies that FSRQs, in general, exhibit relatively enhanced variability during elevated flux states. A similar result was reported in the \gama-ray observations in \citetalias{Bhatta2020}.   To compare the results in the optical and \gama-ray, it is found that the slopes of the linear fit to the RMS-flux relation  is steeper in case the \gama-ray analysis, except for the FSRQs PKS 1424-418 and 3C 454.3, as reported in \citetalias{Bhatta2020}.

\begin{figure*}
\gridline{\fig{rms_Mrk501.eps}{0.35\textwidth}{}\hspace{-0.9cm}
\fig{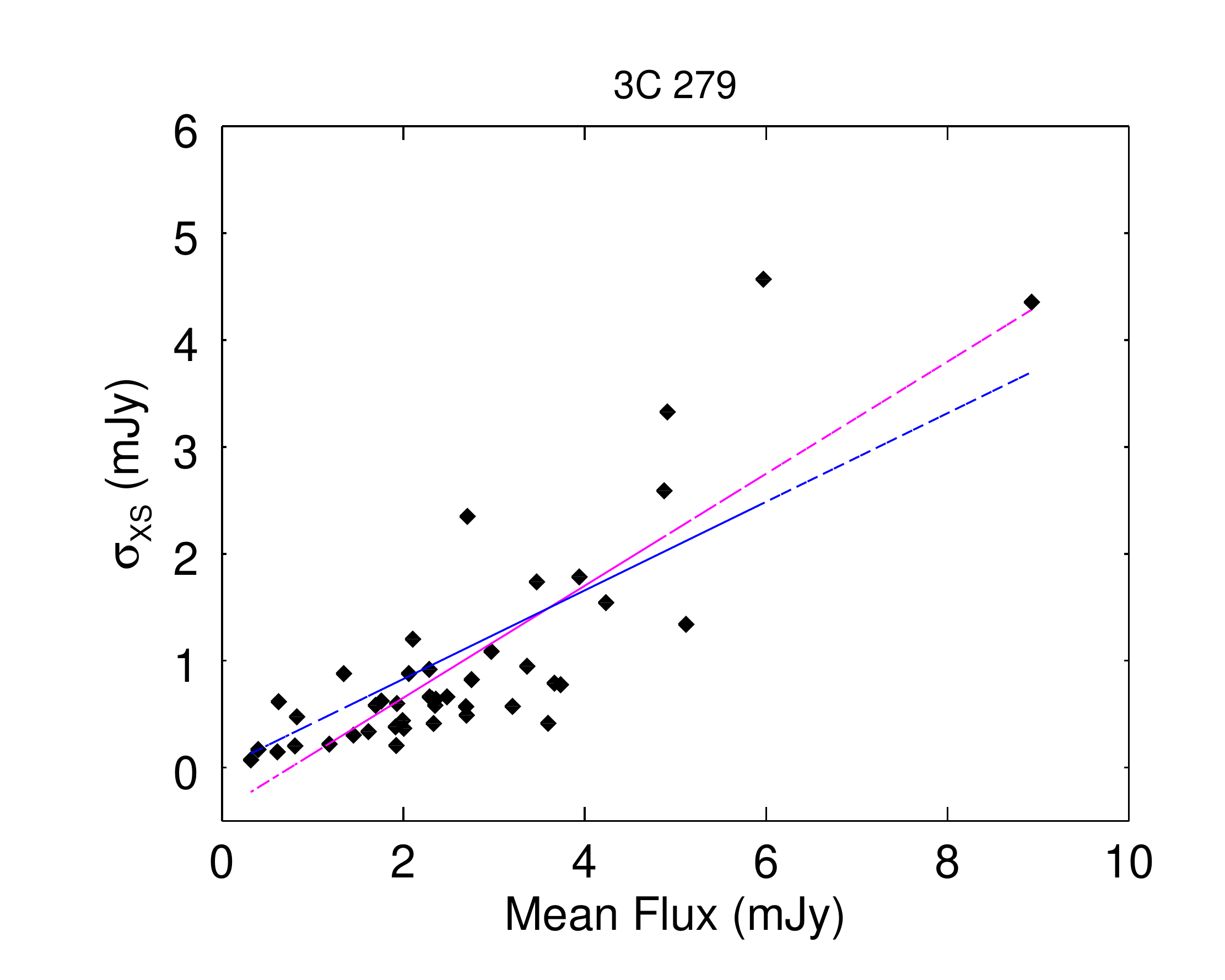}{0.35\textwidth}{}\hspace{-0.9cm}
\fig{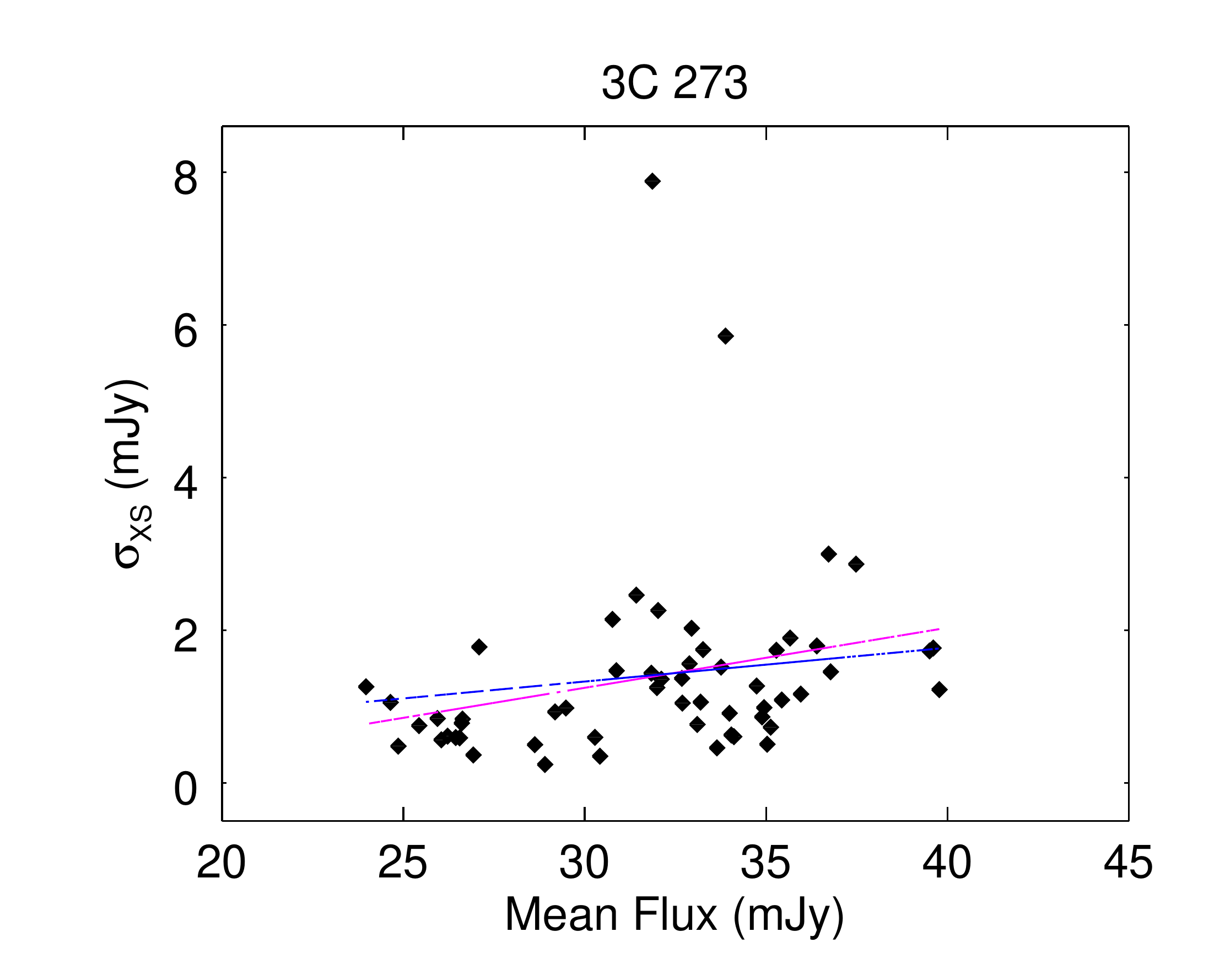}{0.35\textwidth}{}\hspace{-0.6cm}
          }
          \vspace{-1.1cm}
\gridline{\fig{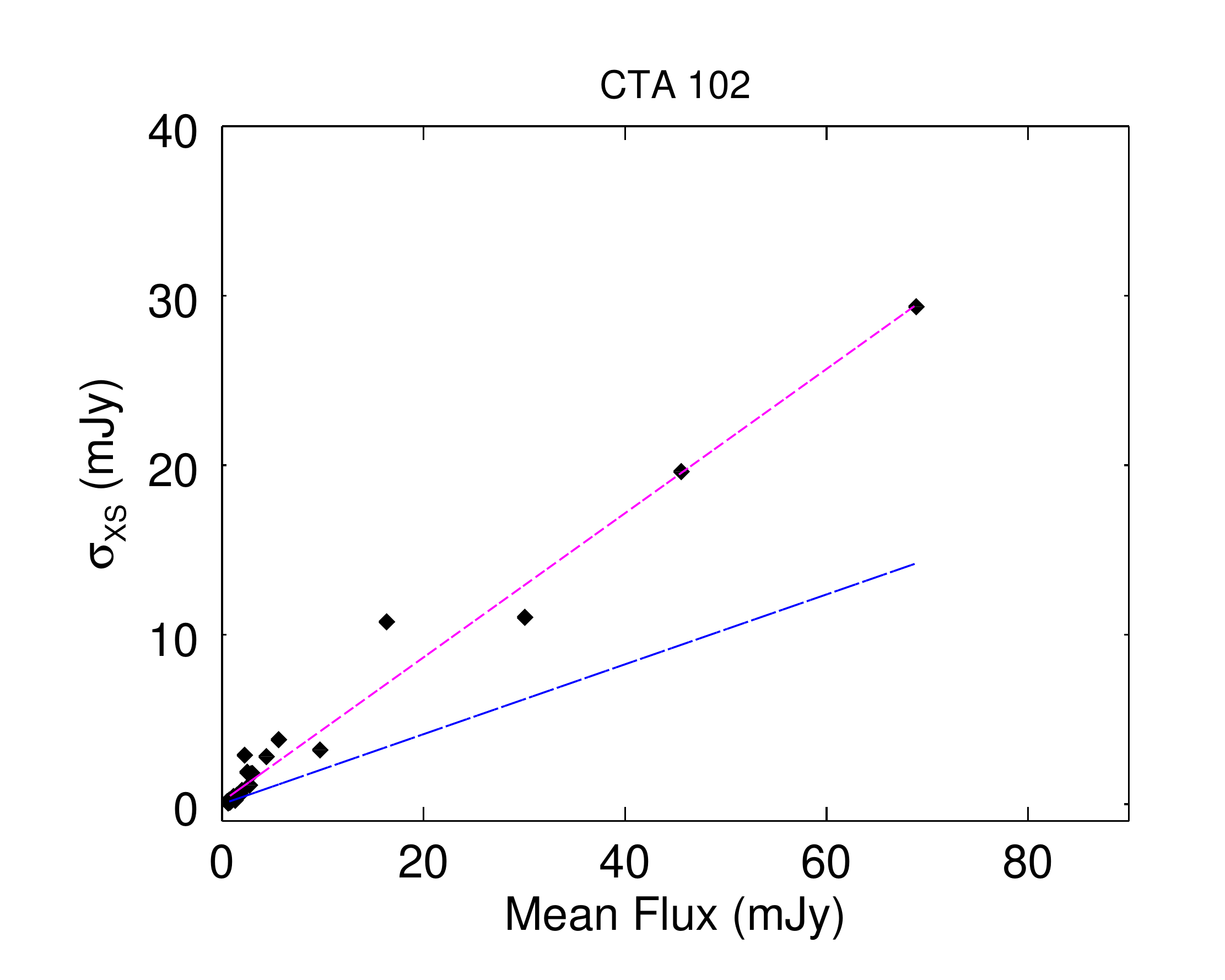}{0.35\textwidth}{}\hspace{-1.1cm}
\fig{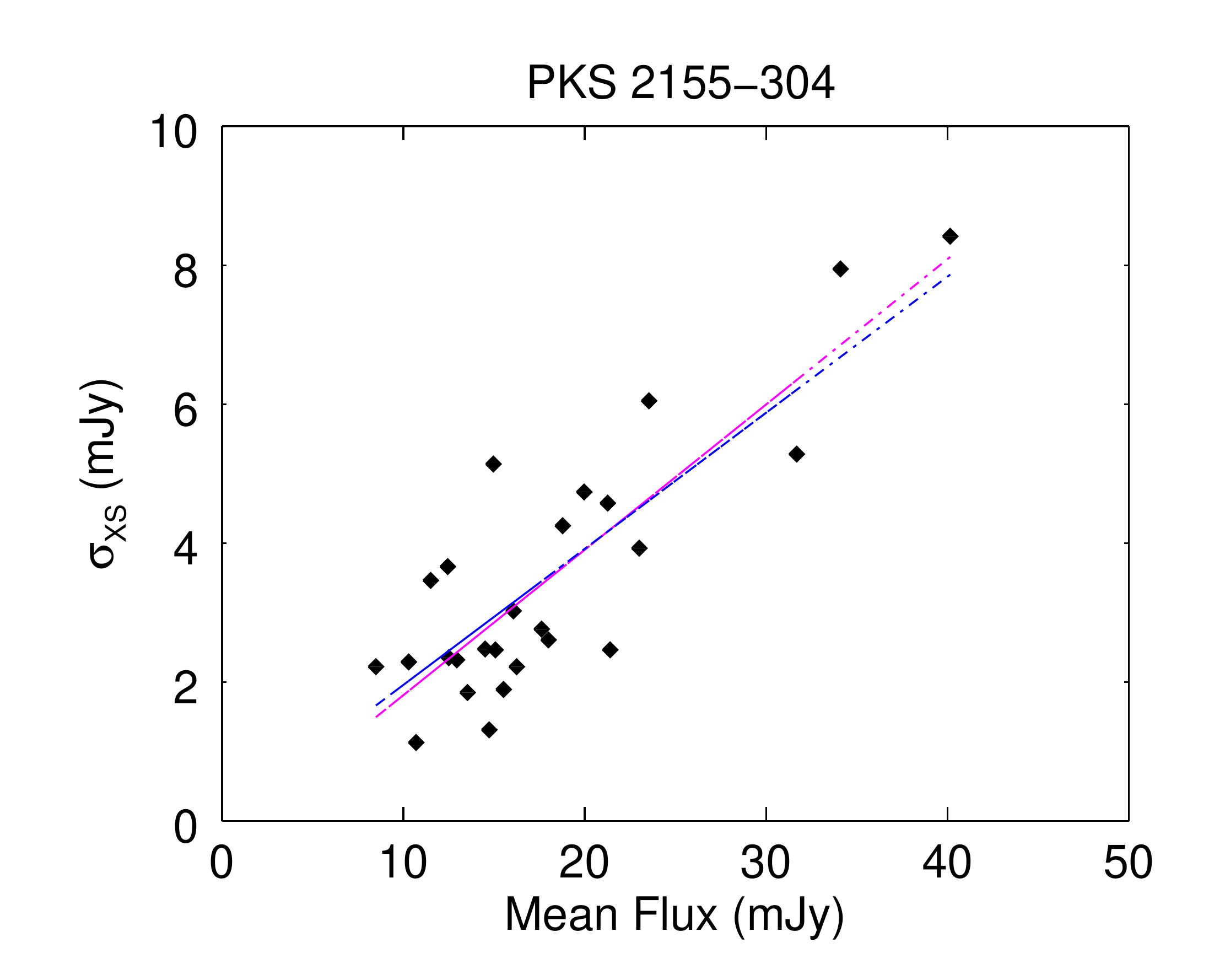}{0.35\textwidth}{}\hspace{-1.1cm}
\fig{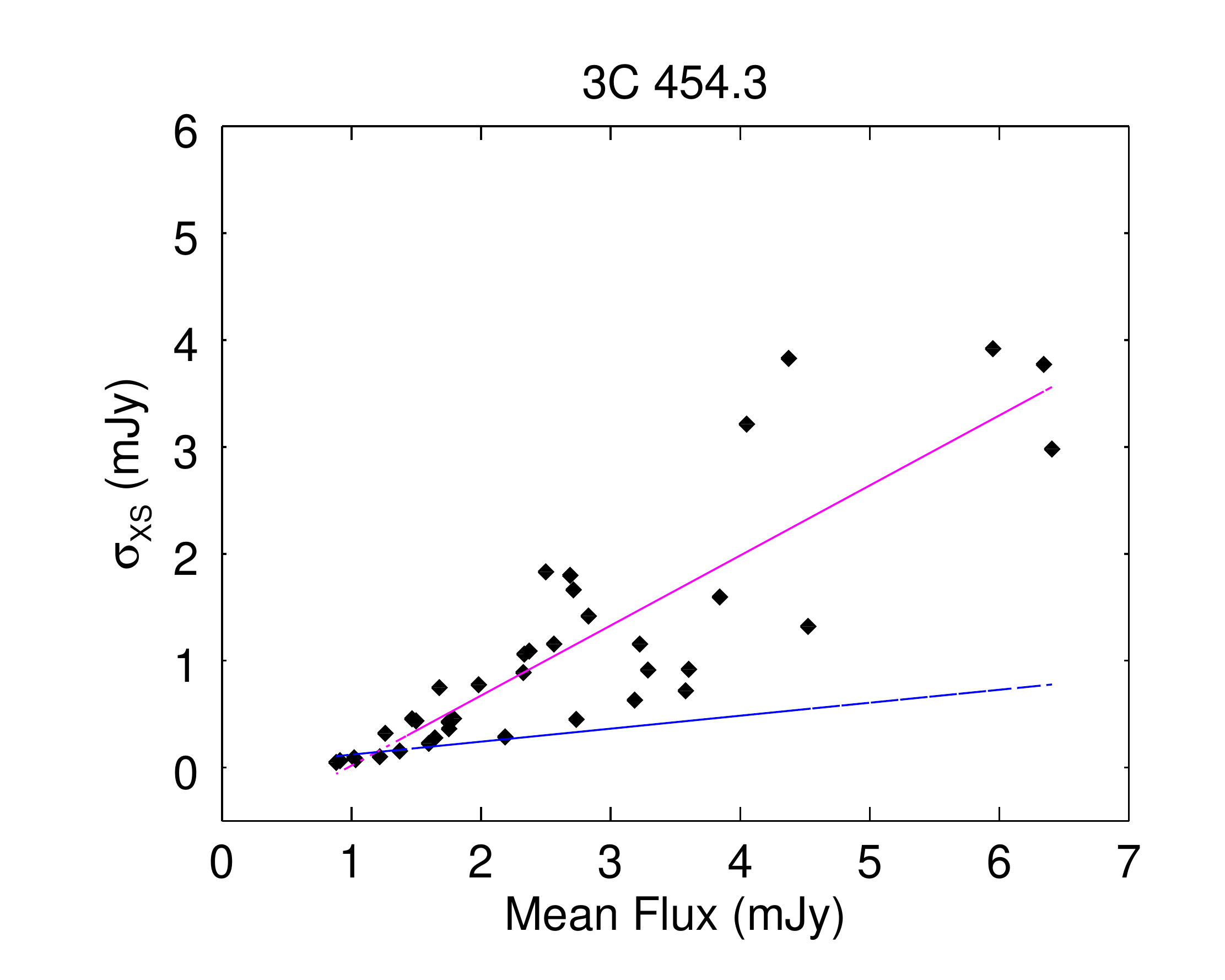}{0.35\textwidth}{}\hspace{-0.6cm}
          }
          \vspace{-0.5cm}
\gridline{\fig{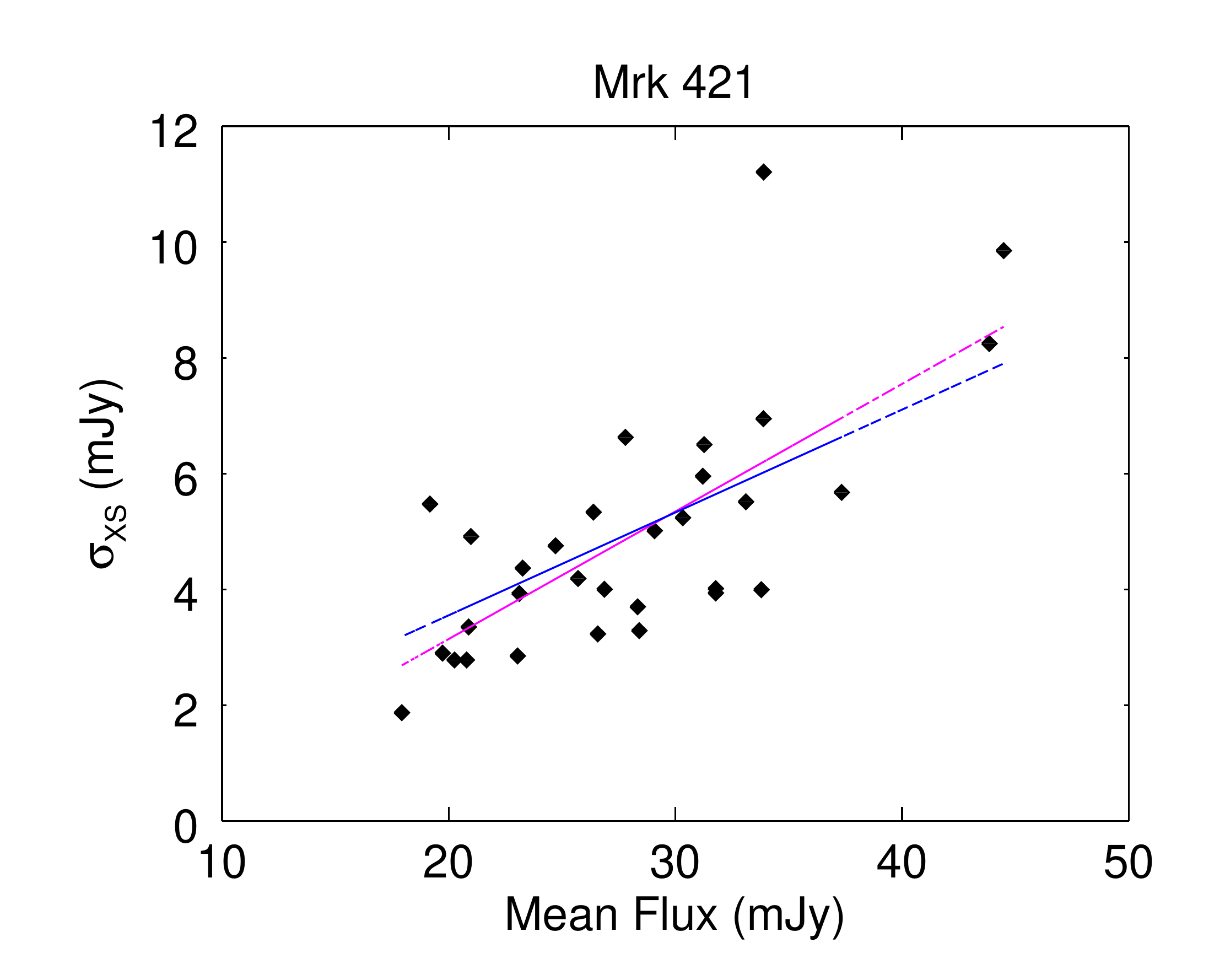}{0.35\textwidth}{}\hspace{-1.1cm}
\fig{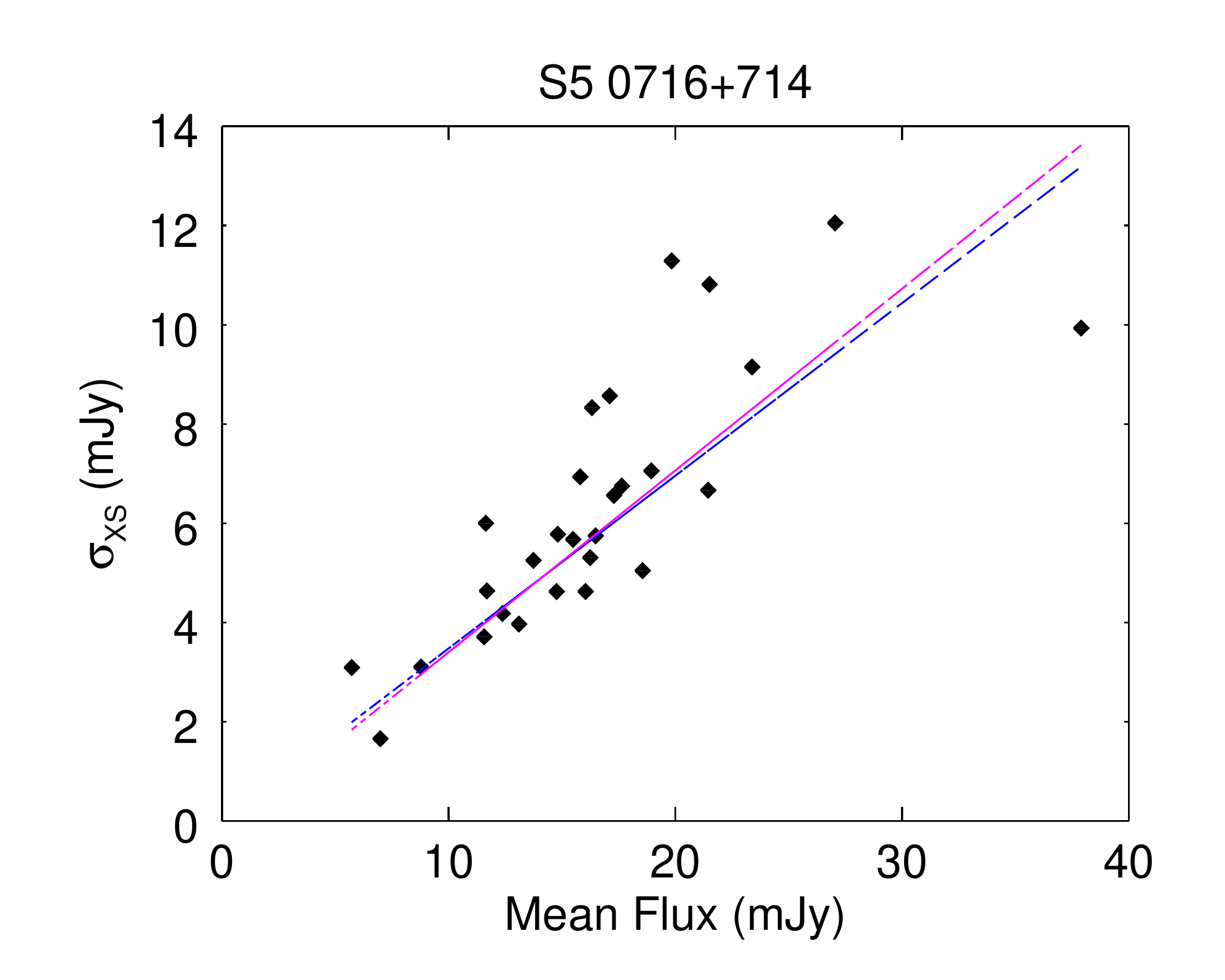}{0.35\textwidth}{}\hspace{-1.1cm}
\fig{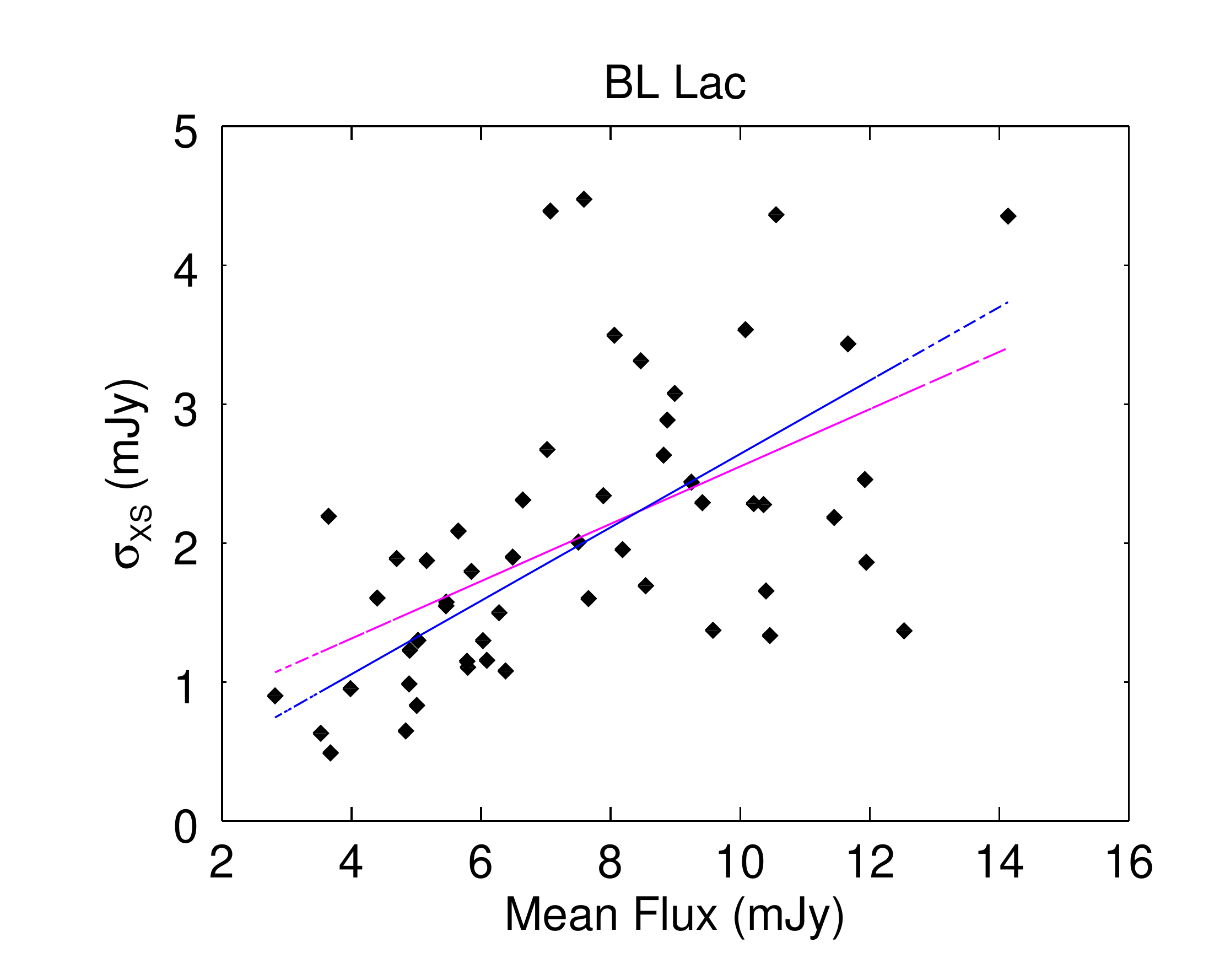}{0.35\textwidth}{}\hspace{-0.6cm}
          }
                    \vspace{-0.5cm}
\gridline{\fig{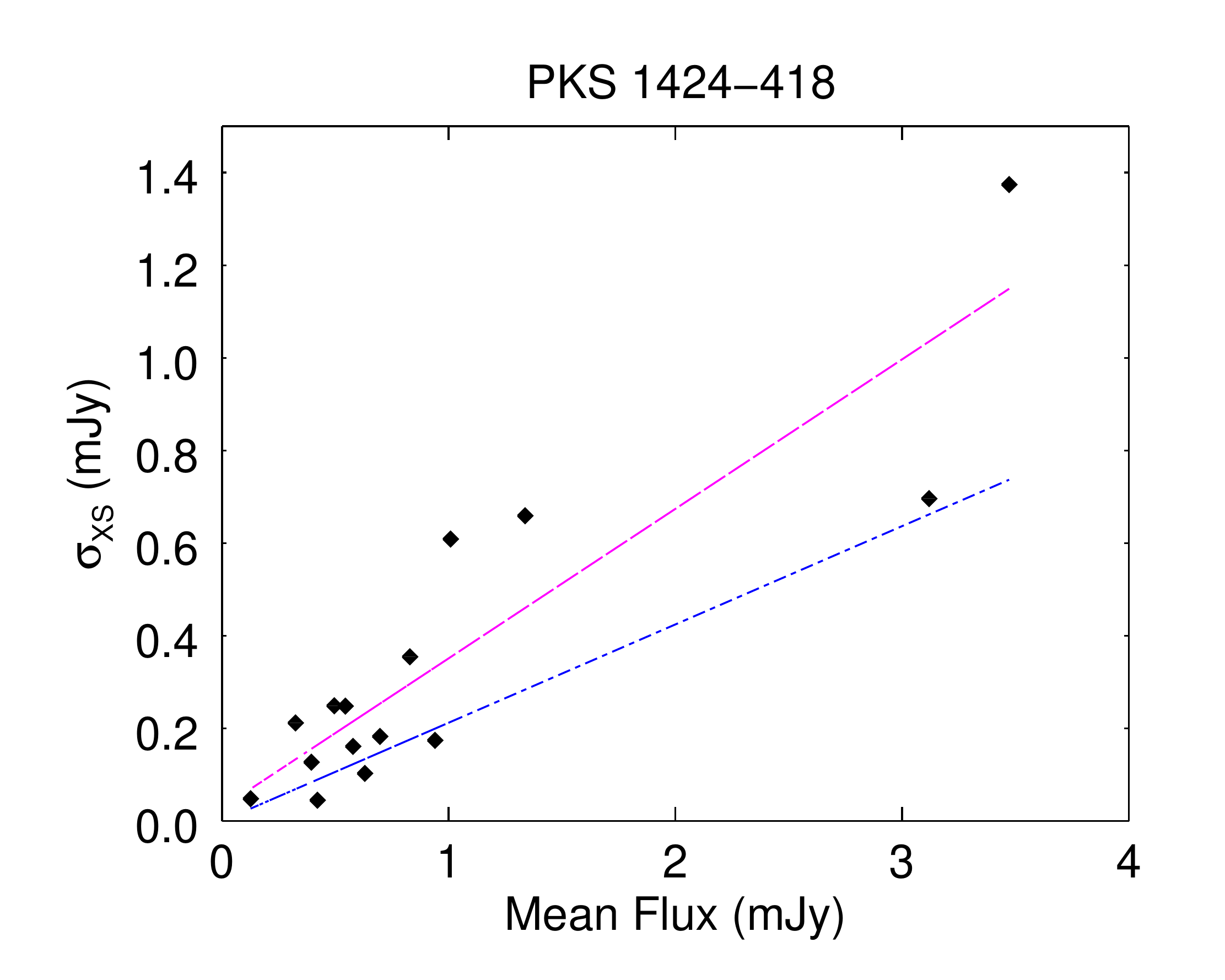}{0.35\textwidth}{}\hspace{-1.1cm}
\fig{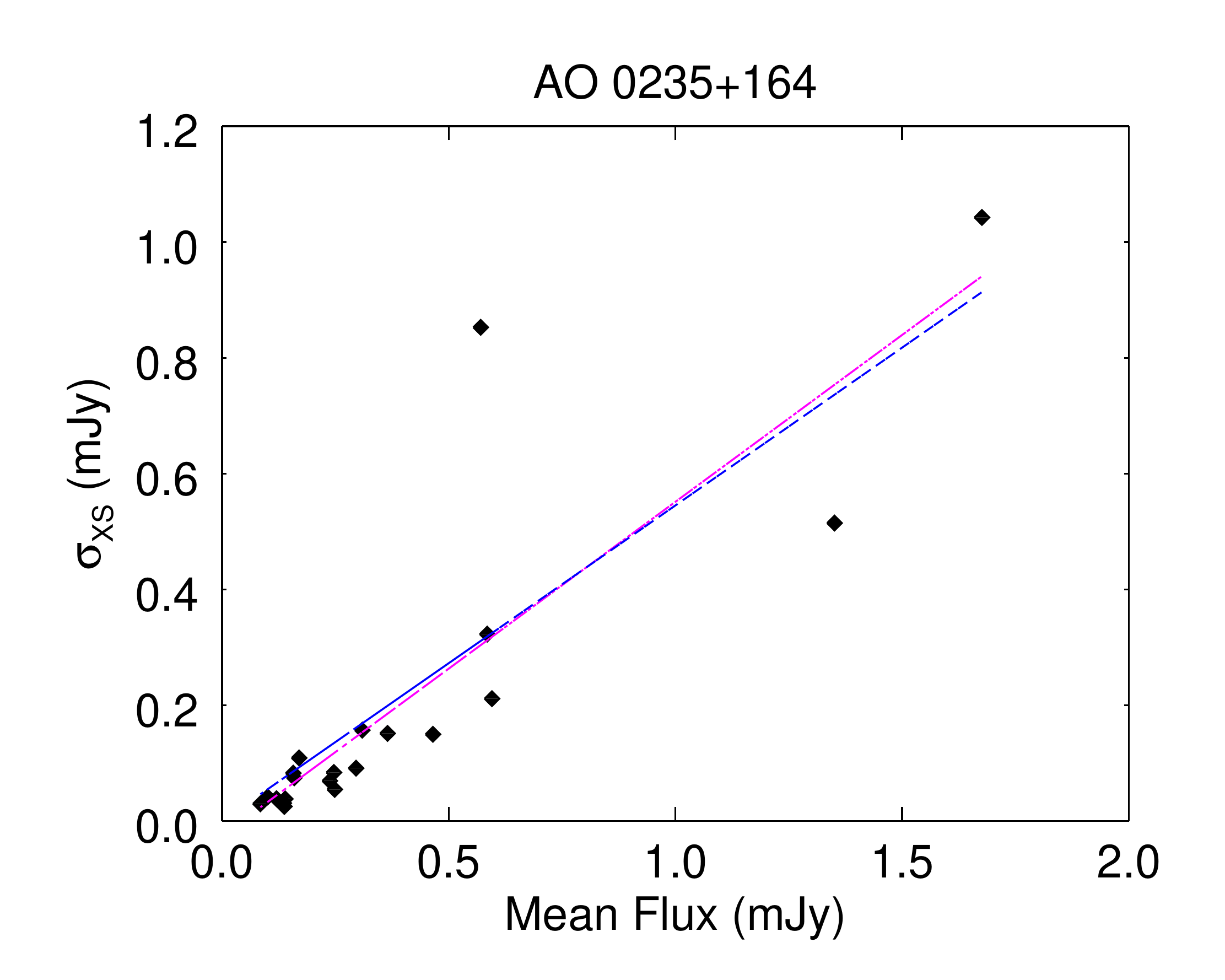}{0.35\textwidth}{}\hspace{-1.1cm}
\fig{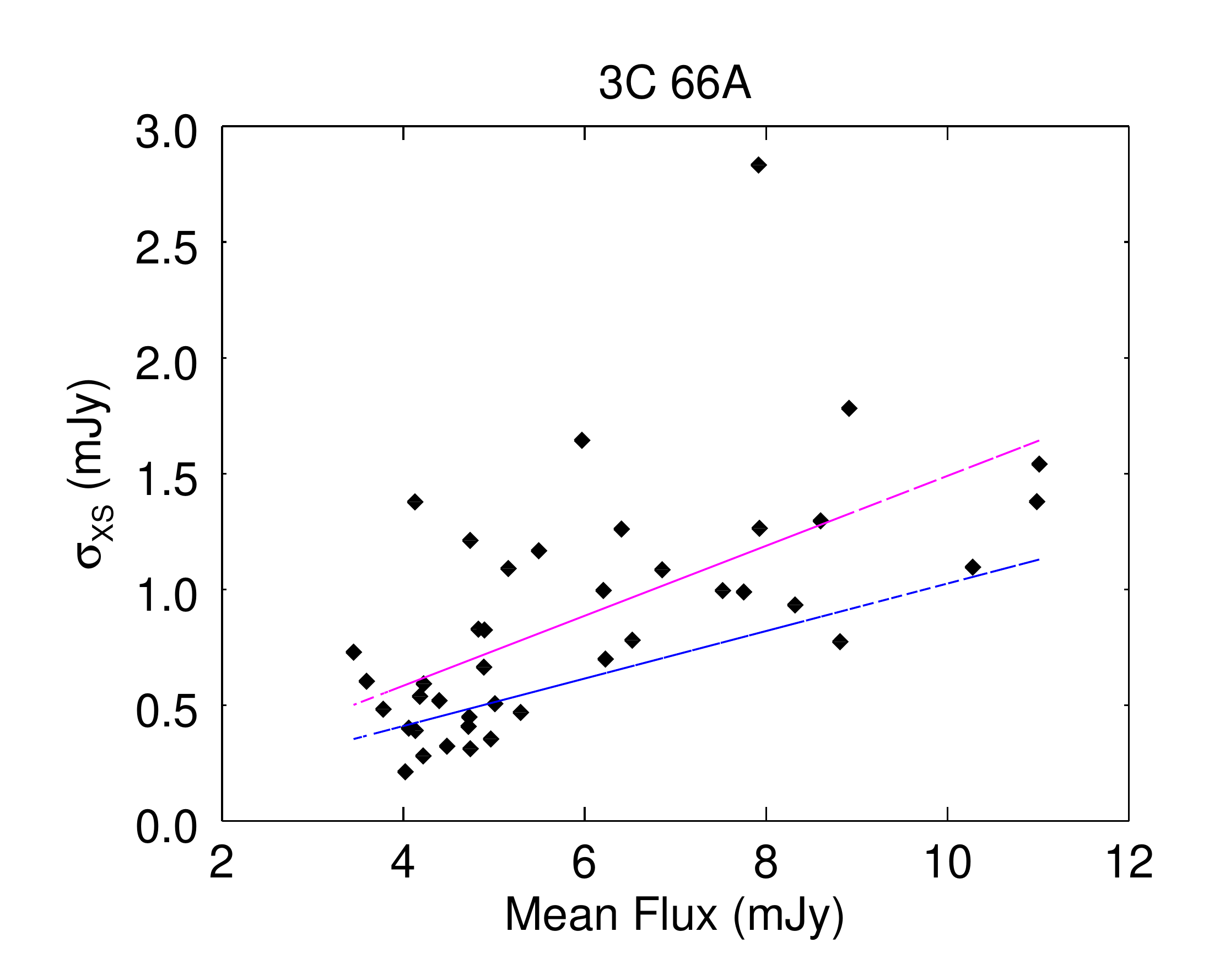}{0.35\textwidth}{}\hspace{-0.6cm}
              }
\caption{The optical RMS-flux diagram for the sample blazars.  The magenta and blue lines represent the linear fit with and without an intercept, respectively.  \label{fig:RMS}}
\end{figure*}

\begin{table}
\begin{center}
\caption{Relation between RMS and the mean flux of  optical  light curves of  blazars \label{tab:table3}}
\begin{tabular}{lcll|cl}
\hline
\hline
Source	&	&with intercept&&without intercept&		\\
\hline
	&	slope&intercept&$\chi^2$/dof&slope&$\chi^2$/dof		\\
\colnumbers\\
3C 66A 	&	$\sim$0.16&$\sim$-0.012&141.57/39=3.63&$\sim$0.15&159.20/40=3.98	\\
AO 0235+164 	&	$\sim$0.57&$\sim$-0.025&61.75/19=3.25&$\sim$0.54&69.40/20=3.47	\\
S5 0716+714 	&	0.32$\pm$0.03&-0.257$\pm$0.005&33.02/26=1.27&0.29$\pm$0.04&33.48/27=1.24	\\
Mrk 421 	&	0.22$\pm$0.03&-1.260$\pm$0.003&46.11/29=1.59&0.18$\pm$0.06&69.00/30=2.30	\\
 3C 273 	&	0.08$\pm$0.05& -1.120$\pm$0.008&40.50/54=0.746&0.05$\pm$0.01&41.25/55=0.75\\
3C 279 	&	$\sim$0.52&$\sim$-0.395&163.80/42=3.90&$\sim$0.42&213.71/43=4.97	\\
PKS 1424-418 	&	$\sim$0.32&$\sim$0.028&313.47/43=7.294&$\sim$0.21&94.78/14=6.77	\\
Mrk 501 	&	0.26$\pm$0.07&-2.616$\pm$0.004&27.82/13=2.14&$\sim$0.03&54.88/14=3.92	\\
PKS 2155-304 	&	0.32$\pm$0.02&-0.280$\pm$0.008&68.16/24=2.84&0.19$\pm$0.02&69.00/25=2.76 	\\
BL Lac 	&	$\sim$0.20&$\sim$0.488&176.46/51=3.46&$\sim$0.26&157.04/52=3.02\\
CTA 102 	&	$\sim$0.42&$\sim$0.145&87.29/29=3.01&$\sim$0.23&192.90/30=6.42\\
3C 454.3  	&	$\sim$0.65&$\sim$-0.636&172.08/36=4.78&$\sim$0.12&395.53/37=10.69\\

\hline
\end{tabular}\\
Note: The fit parameters  yielding the reduced $\chi^2$ greater than three are expressed in approximate values.
\end{center}
\end{table}

\subsection{Flux distribution: normal and log-normal }
As the flux of the variable sources undulate over the period of time, it passes through various flux states, such as high and low flux states. Therefore,  flux distribution of blazars can offer an important insight regarding the origin and nature of the variability.  To carry out a statistical analysis of optical flux distribution,  log-normal and normal probability density function (PDF)  were  fitted to the unbinned histogram of the sources fluxes. A log-normal PDF in its familiar form can be expressed as,
\begin{equation}
f_\mathrm{log-normal}(x) = \frac{1}{x s \sqrt{2\pi}} \exp({{-\frac{\left(\ln x\  -\  m\right)^{2}}{2 s^{2}}}}) \mathrm{\quad ,}
\end{equation} 
where $m$ and $s$  define the mean flux location and the scale parameters of the PDF, respectively.
Similarly,   a normal PDF  can be written as,
\begin{equation}
f_\mathrm{normal}\left(x\right) = \exp(-\frac{(x - \mu)^{2}}{2\sigma^{2}}) \mathrm{\quad ,}
\end{equation}
where $\mu$ and $\sigma$ are the mean and the standard deviation of the PDF, respectively.

To  implement the statistical analysis, the Maximum Likelihood Estimation (MLE) method was applied on the source flux distribution using the PDF fitting the software package \emph{fitdistrplus}\footnote{\url{https://cran.r-project.org/web/packages/fitdistrplus/index.html}} \citep{Delignette2015} publicly available in the MASS library of R. The package attempts to fit the PDFs to the unbinned  flux distribution following MLE method. The source histogram  fitted with normal and log-normal PDFs are presented in Figure \ref{fig:PDF}. The residuals corresponding to the model fit are shown in the lower panels of the plots. Similarly, the mean of the flux, scale  for the log-normal fitting are listed in column 2, 3, and 4, respectively, of Table \ref{table:2};  and the corresponding fitting statistics, that is, the log-likelihood (LL), Bayesian information criterion (BIC) and Akaike information criterion (AIC)  are listed in column 4, 5, and 6, respectively. Similar quantities for the normal PDF are listed in the columns 7, 8, 9, 10 and 11.   It can be seen that log-normal PDF fitting result in the smaller AIC and BIC  compared to the normal PDF fitting. This indicates that the asymmetric  log-normal PDF, with a heavy tail extending towards higher flux, better represents the blazar flux distribution in the optical.  A similar conclusion in the results of the \gama-ray  flux distribution of blazars was reported in \citetalias{Bhatta2020}. Moreover, it is seen that compared to optical flux distribution,  the  \gama-ray flux distribution shows a large spread towards higher flux as indicated by the larger scale parameter values, with the exception of the sources AO 0235+165 and PKS 1424-418.

\begin{deluxetable*}{llllll|lllll}
\tablecaption{ Lognormal and normal distribution fit statistics for the optical flux distribution of the sample sources using maximum likelihood estimation method.  \label{table:2}}
\tablewidth{500pt}
\tabletypesize{\scriptsize}
\tablehead{
\colhead{} & 
\multicolumn{3}{c}{Lognormal fit} & \multicolumn{3}{c}{Normal fit} \\
\hline
\colhead{Source name} &
\colhead{$m$} &\colhead{$s$} &\colhead{$LL$} & 
\colhead{AIC}  &  \colhead{$BIC$} &
\colhead{$\mu$} & \colhead{$\sigma$} &\colhead{$LL$} &
\colhead{AIC}  &  \colhead{$BIC$} 
} 
\colnumbers
\startdata
3C 66A &	1.89$\pm$0.01&0.36$\pm$0.01&-2085& 4174& 4183&	7.08$\pm$0.09&2.70$\pm$0.06&-2208& 4421& 4430\\
AO 0235+164 &	-1.26$\pm$0.05&0.85$\pm$0.03&    2&    1&    8&	0.44$\pm$0.03&0.59$\pm$0.02& -282&  568&  575\\
S5 0716+714 &	2.84$\pm$0.02&0.55$\pm$0.02&-2282& 4568& 4576&	19.77$\pm$0.43&10.85$\pm$0.31&-2365& 4735& 4744\\
Mrk 421 &	3.30$\pm$0.01&0.29$\pm$0.01&-4097& 8198& 8208&	28.21$\pm$0.24&8.37$\pm$0.17&-4189& 8382& 8392\\
3C 273 &	3.46$\pm$0.00&0.14$\pm$0.00&-3295& 6595& 6605&	32.29$\pm$0.14&4.61$\pm$0.10&-3301& 6607& 6617\\
3C 279 &	0.97$\pm$0.02&0.77$\pm$0.02&-2372& 4747& 4757&	3.43$\pm$0.08&2.73$\pm$0.06&-2703& 5409& 5419\\
PKS 1424-418 &	-0.07$\pm$0.04&0.90$\pm$0.03& -568& 1140& 1148&	1.41$\pm$0.07&1.48$\pm$0.05& -832& 1668& 1676\\
Mrk 501 &	2.43$\pm$0.00&0.06$\pm$0.00& -427&  858&  866&	11.37$\pm$0.03&0.69$\pm$0.02& -438&  880&  888\\
PKS 2155-304 &	2.97$\pm$0.02&0.41$\pm$0.01&-2082& 4168& 4177&	21.36$\pm$0.40&9.83$\pm$0.29&-2197& 4398& 4406\\
BL Lac &	2.09$\pm$0.01&0.47$\pm$0.01&-3259& 6521& 6531&	9.01$\pm$0.12&4.24$\pm$0.09&-3376& 6757& 6767\\
CTA 102 &	0.66$\pm$0.05&1.20$\pm$0.03&-1434& 2871& 2880&	6.21$\pm$0.65&16.46$\pm$0.46&-2667& 5338& 5346\\
3C 454.3 &	0.99$\pm$0.02&0.59$\pm$0.02&-1455& 2914& 2923&	3.28$\pm$0.09&2.61$\pm$0.07&-1839& 3681& 3691\\
\enddata
\tablecomments{For the normal fit $\mu$ and $\sigma$ are presented in the unit of  flux in mJy, whereas for the lognormal fit $m$ is in the unit of natural log of flux.}
\end{deluxetable*}

\begin{figure*}
\gridline{\fig{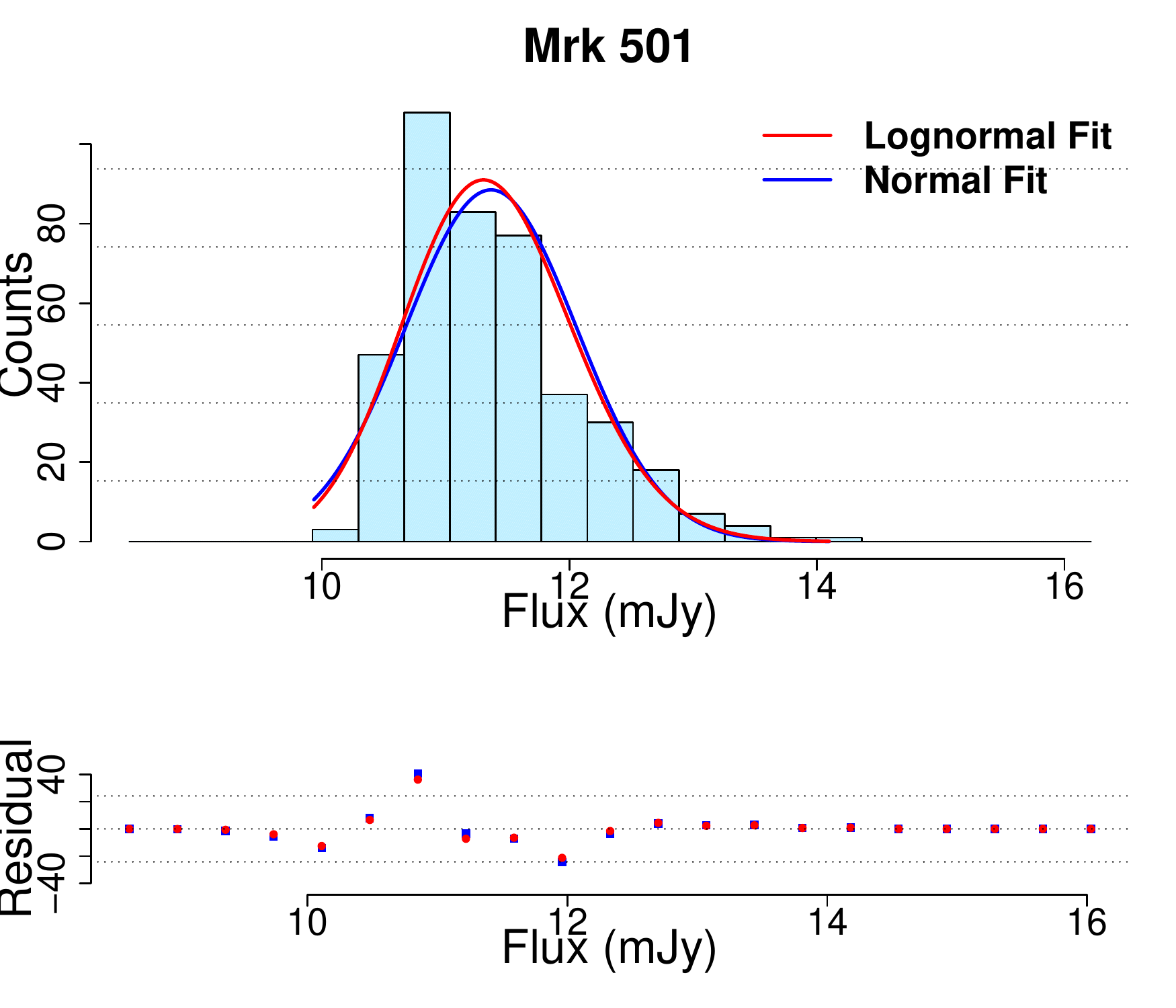}{0.35\textwidth}{}\hspace{-0.8cm}
\fig{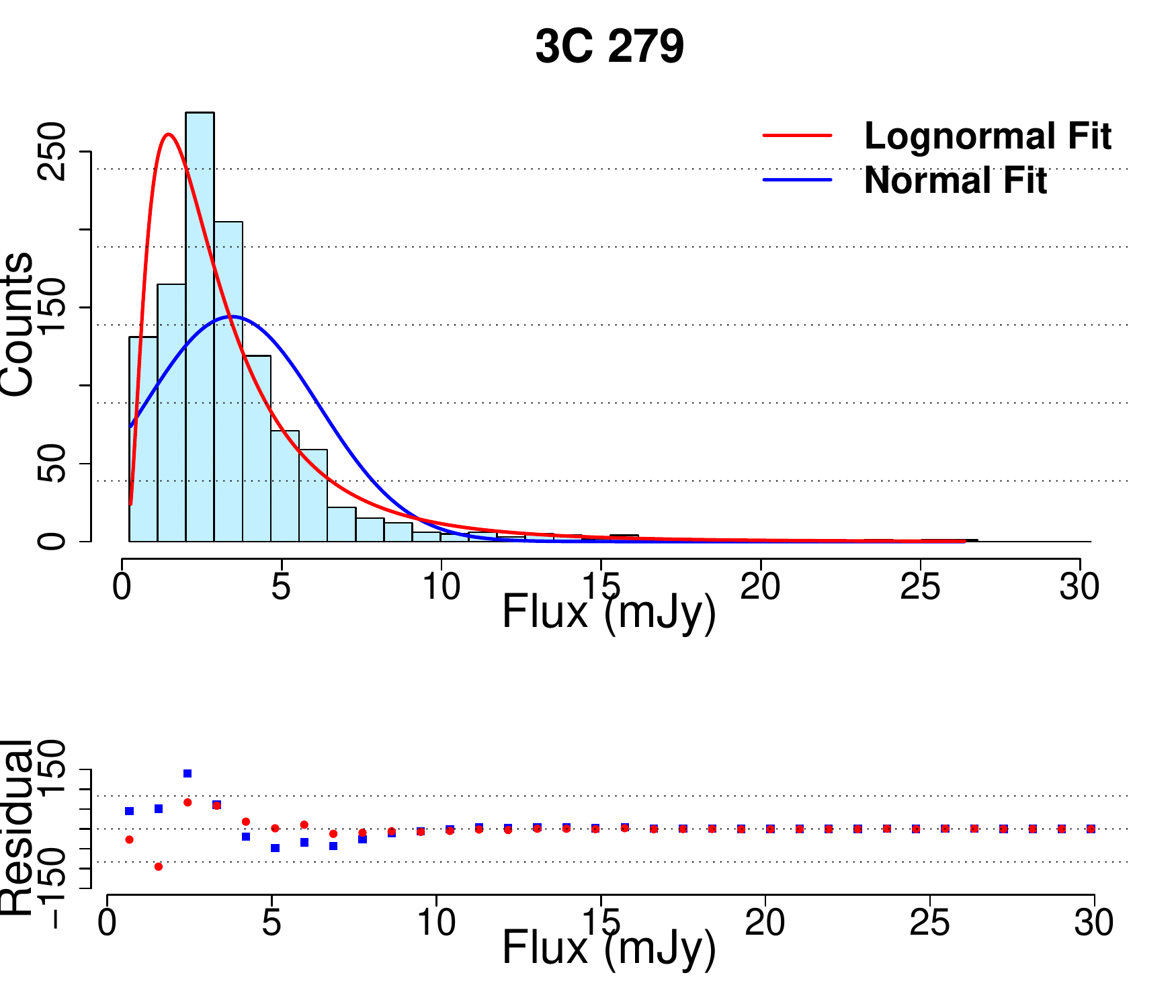}{0.35\textwidth}{}\hspace{-0.8cm}
\fig{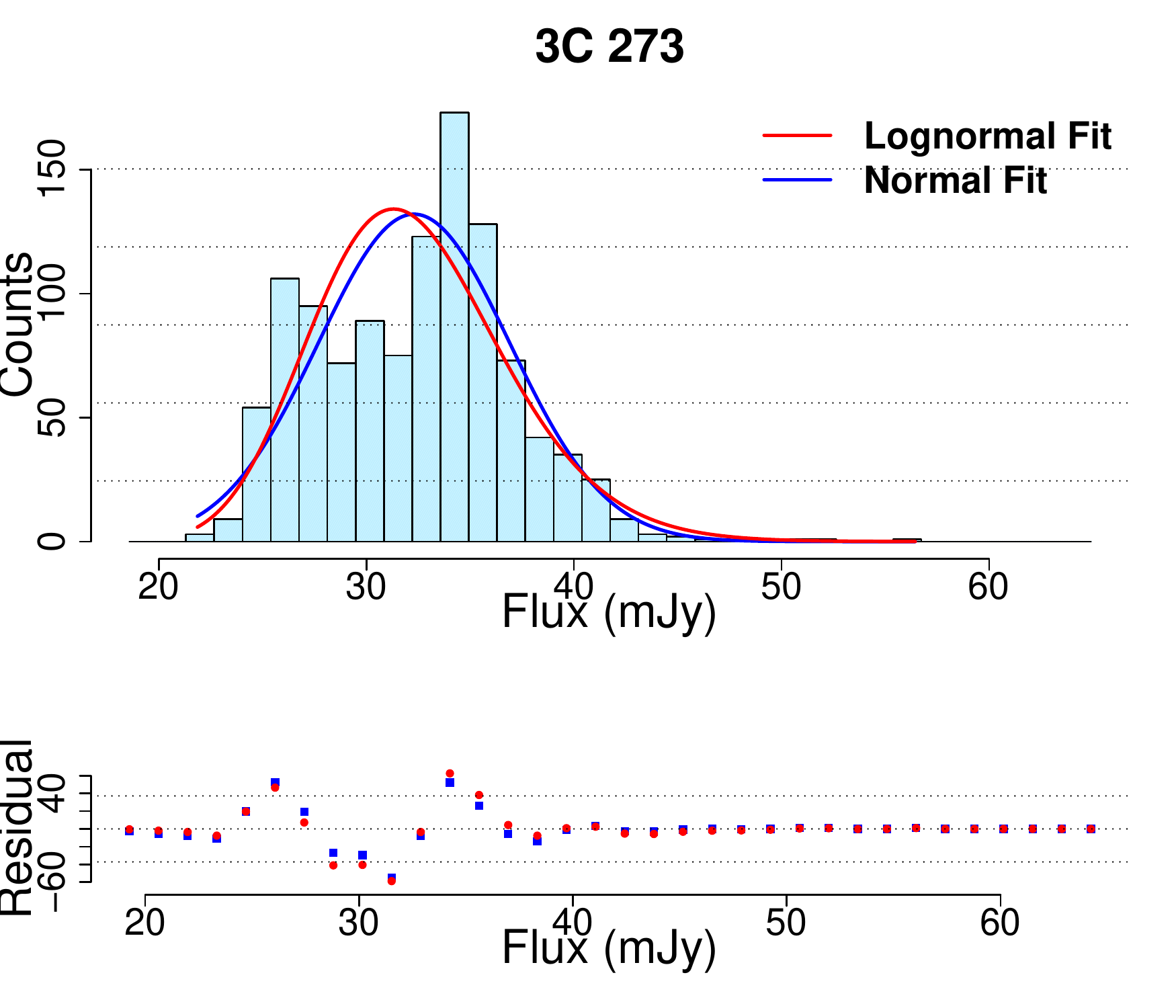}{0.35\textwidth}{}\hspace{-1.1cm}
          }
          \vspace{-1.1cm}
\gridline{\fig{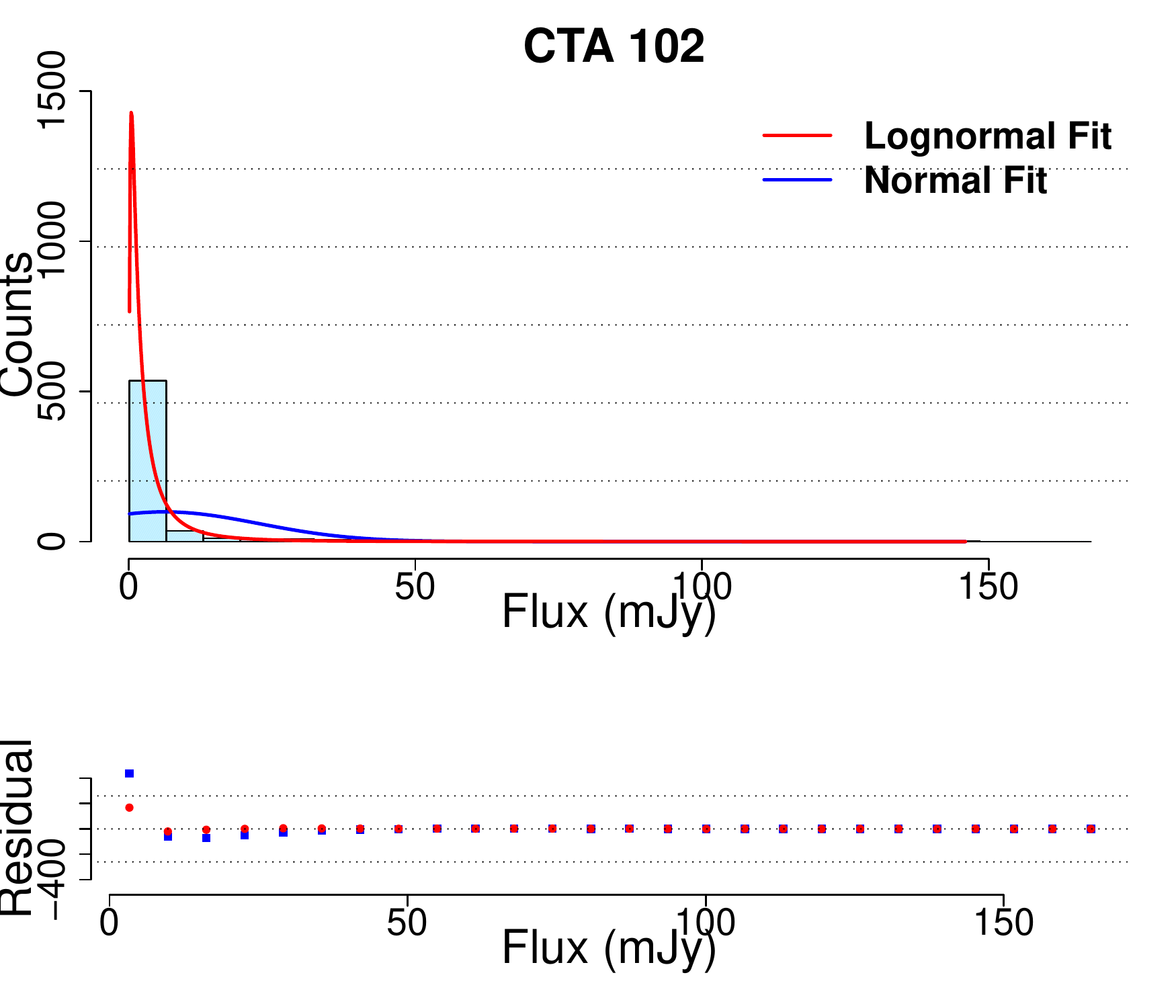}{0.35\textwidth}{}\hspace{-0.8cm}
\fig{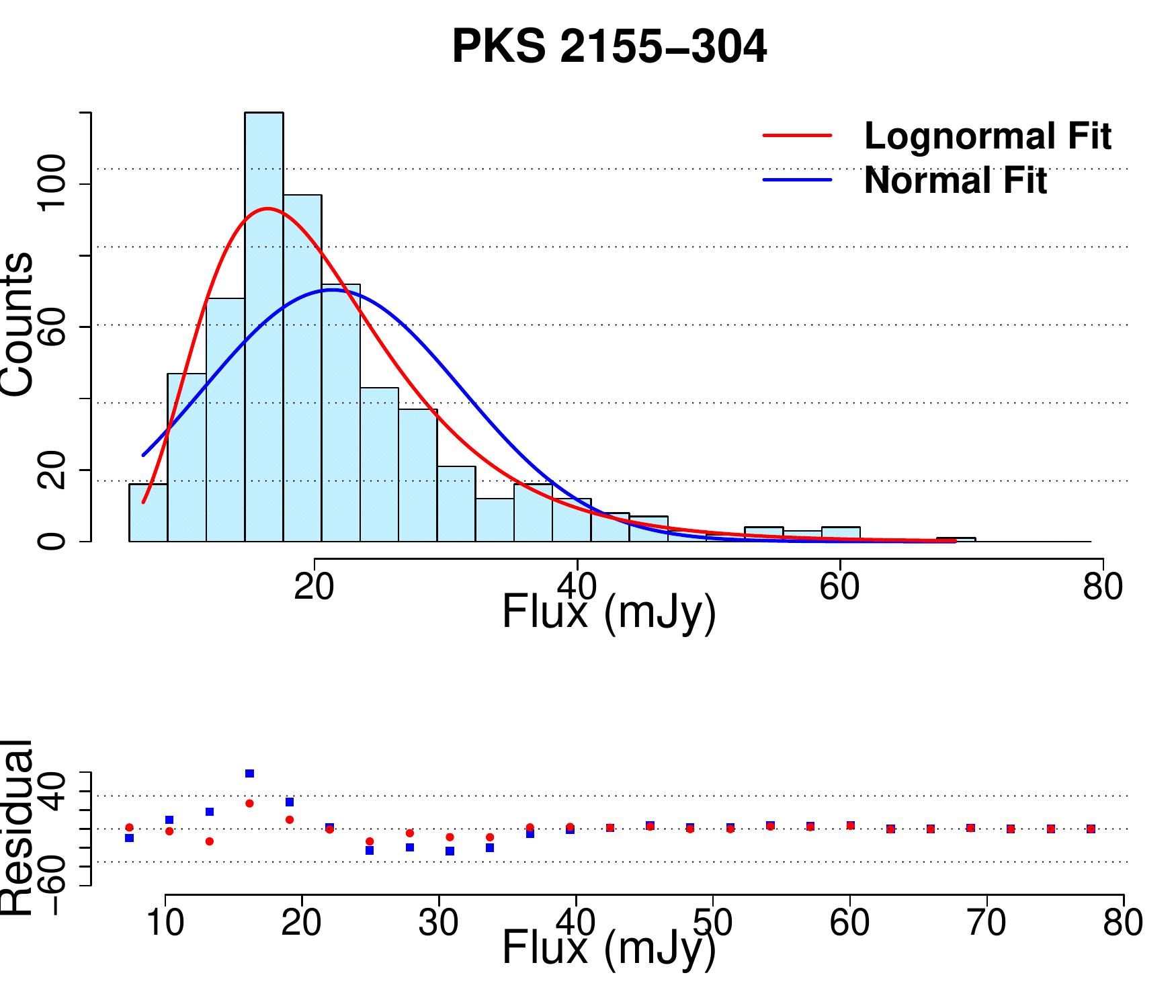}{0.35\textwidth}{}\hspace{-0.8cm}
\fig{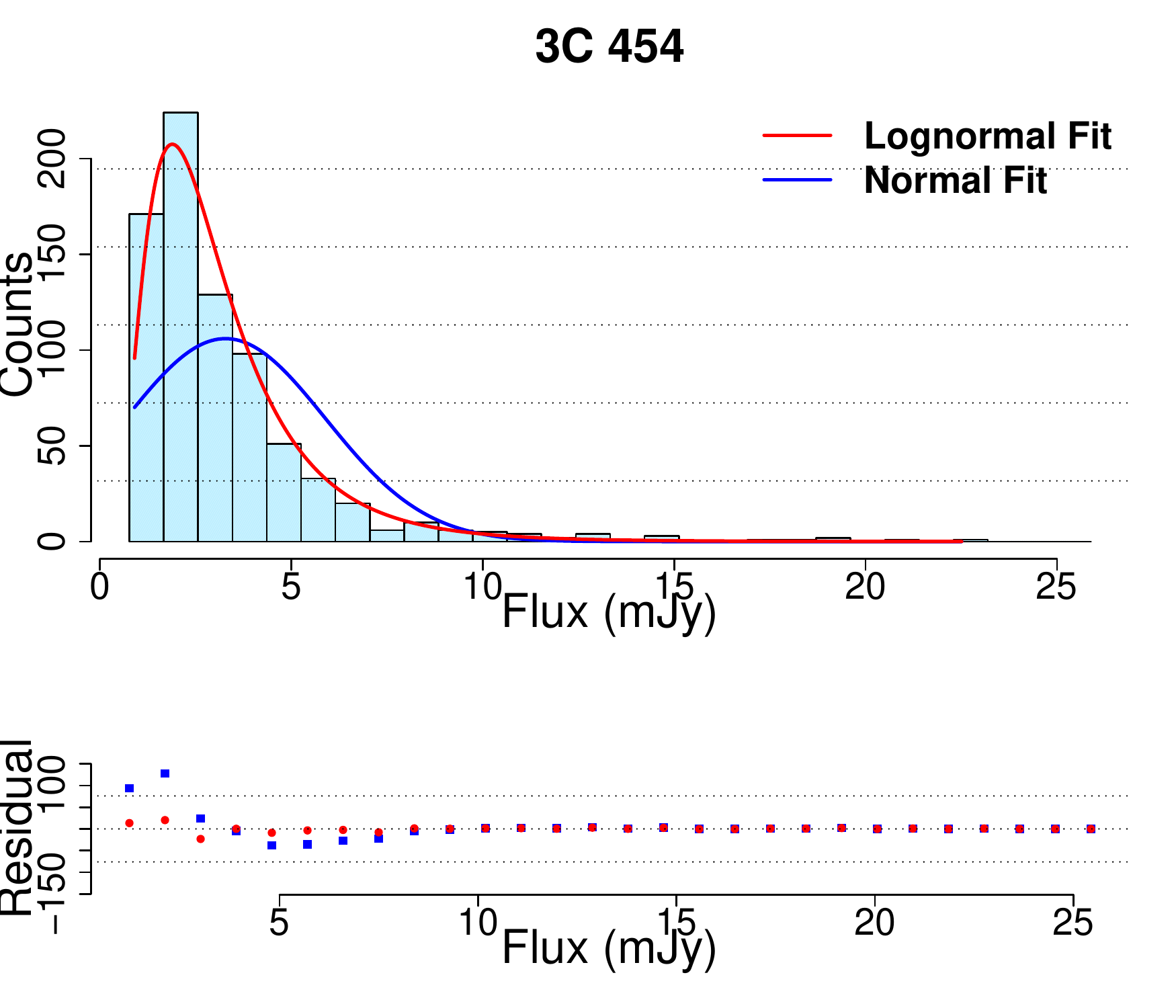}{0.35\textwidth}{}\hspace{-1.1cm}
          }
          \vspace{-0.5cm}
\gridline{\fig{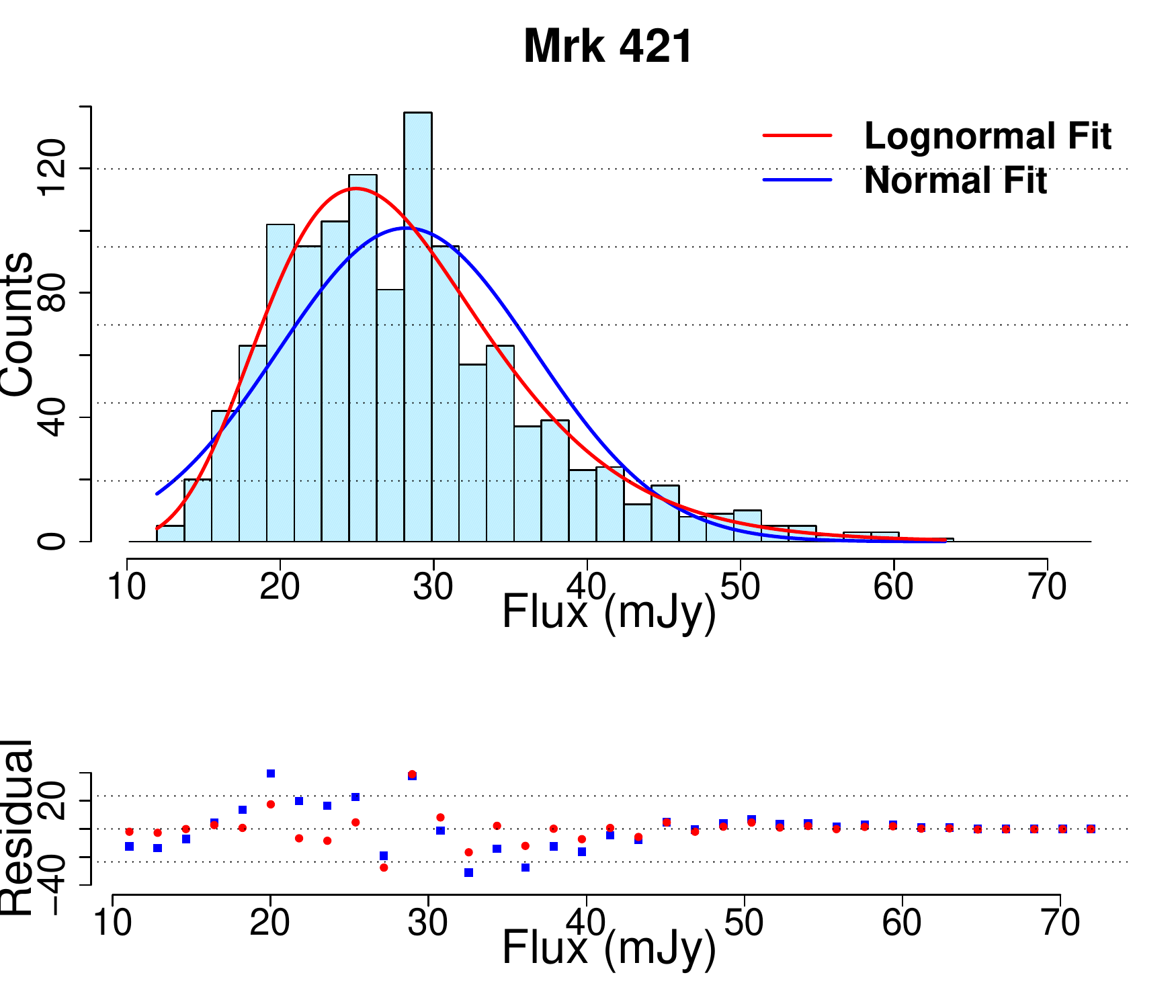}{0.35\textwidth}{}\hspace{-0.8cm}
\fig{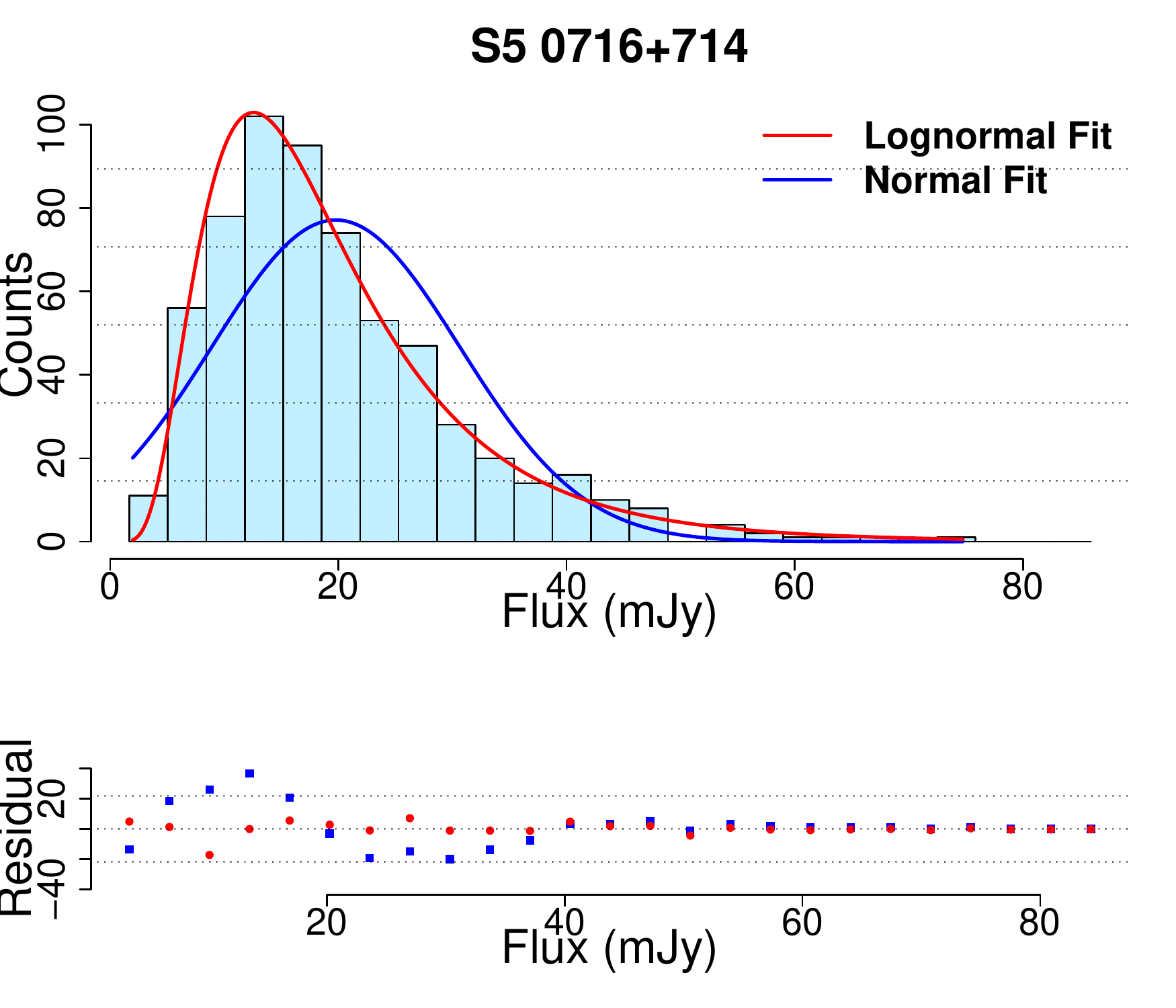}{0.35\textwidth}{}\hspace{-0.8cm}
\fig{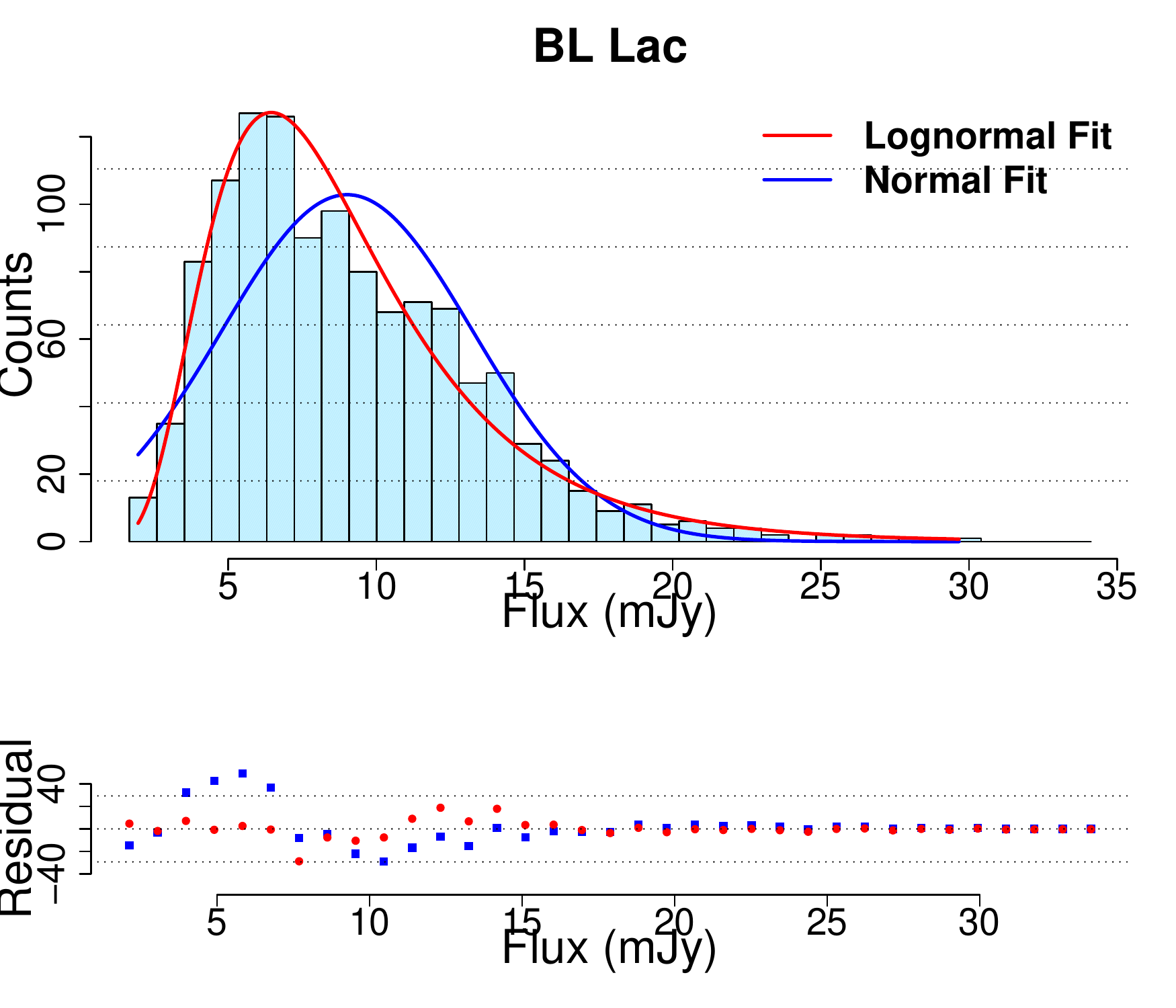}{0.35\textwidth}{}\hspace{-1.1cm}
          }
                    \vspace{-0.5cm}
\gridline{\fig{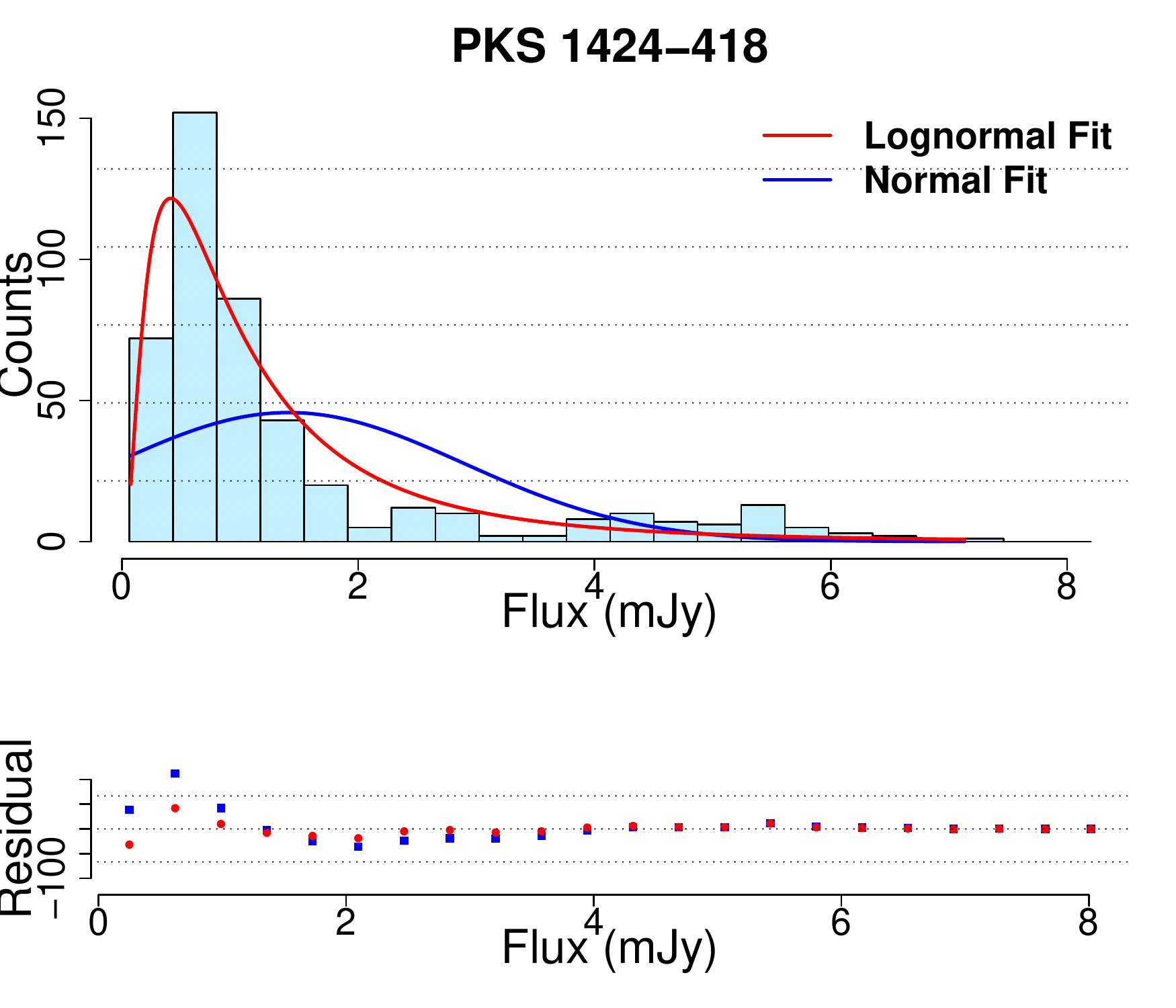}{0.35\textwidth}{}\hspace{-0.8cm}
\fig{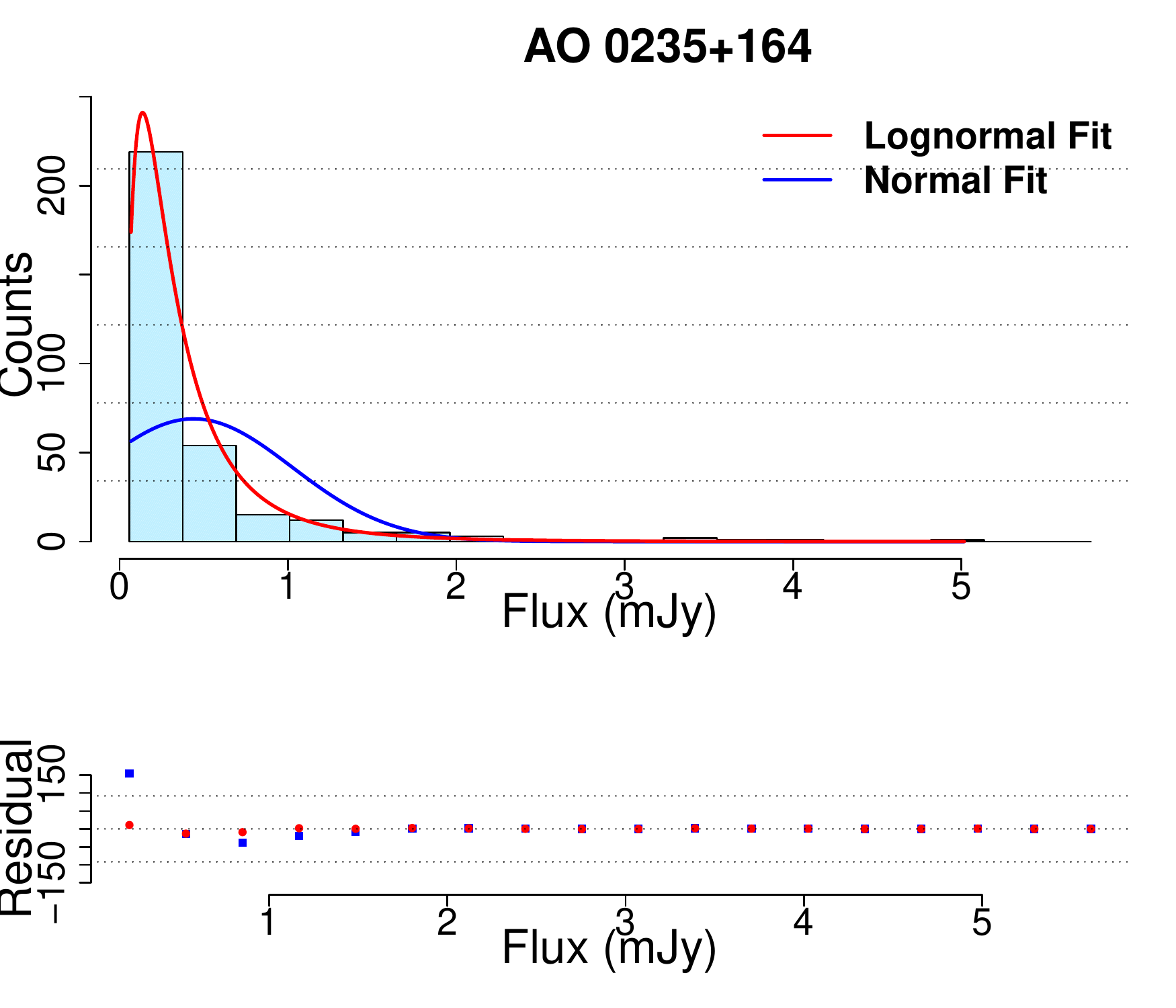}{0.35\textwidth}{}\hspace{-0.8cm}
\fig{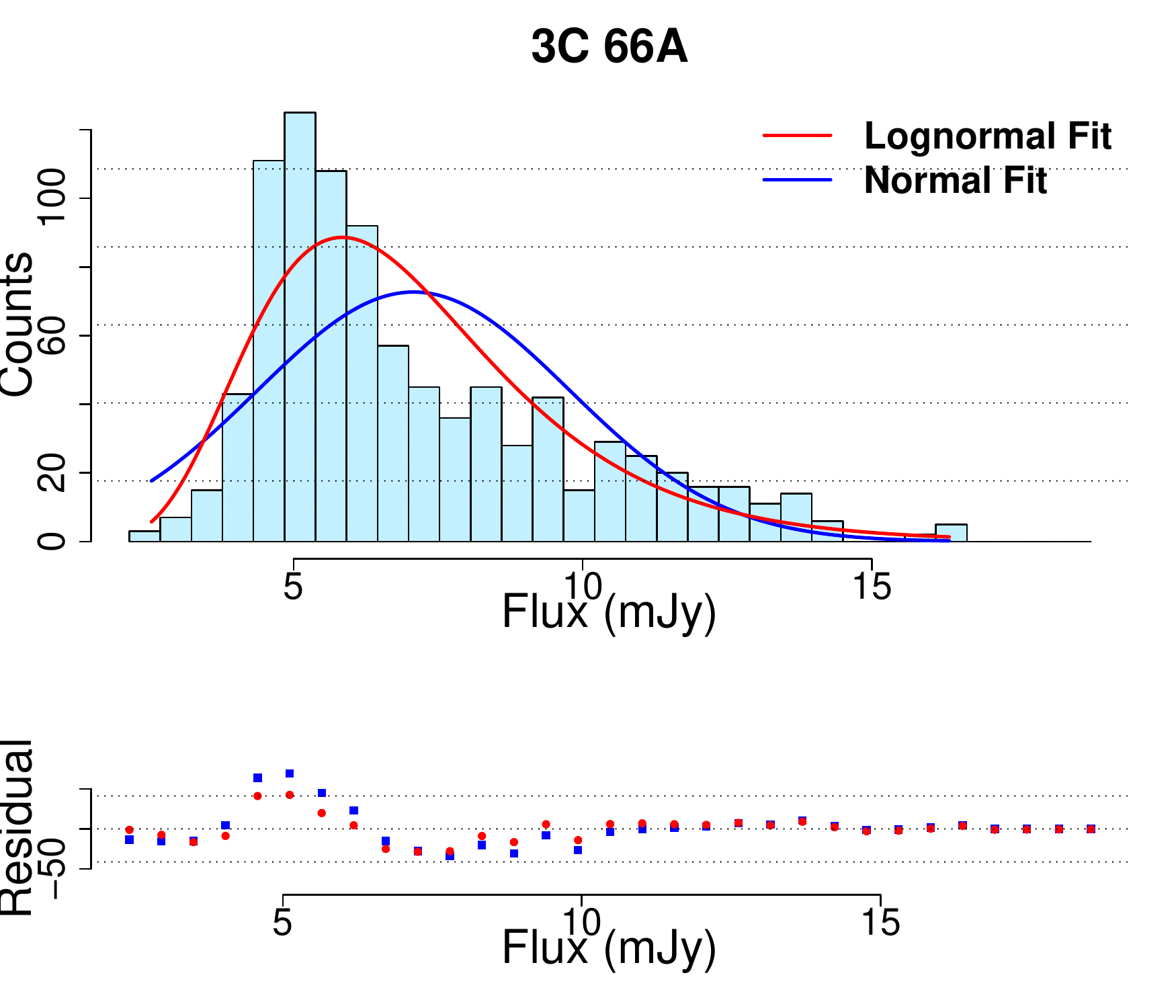}{0.35\textwidth}{}\hspace{-1.1cm}
              }
\caption{Histogram of the optical flux of the sample  blazars. The blue and the red curves represent the normal and log-normal PDF fit to the observations, respectively. The corresponding residual plots are shown in the lower panels.\label{fig:PDF}}
\end{figure*}

\subsection{Cross-Correlation between the optical and the \gama-ray band \label{sec:dcf}}
A study of the correlation among emissions in different energy bands  can shed light on the structure of the blazar emission regions, and the dominant radiation mechanisms participated by various distribution of the particles.
  In this work, a cross-correlation study between the long-term optical light curves and the \gama-ray light curve from the Fermi/LAT telescope is conducted using the  method based on the discrete correlation function (DCF; \citealt{Edelson1988}). The method is particularly known to deal with the unevenly spaced time series. To follow the method, the unbinned DCFs are computed as,
\begin{equation}
UDCF_{ij}=\frac{\left ( x_{i}-\bar{x} \right )\left ( y_{j}-\bar{y} \right )}{\sqrt{\left ( \sigma _{x}^{2} -e_{x}^{2}\right )\left ( \sigma _{y}^{2} -e_{y}^{2}\right )}}
\label{UDCF}
\end{equation}
where $\bar{x}$ and $\bar{y}$ represent mean fluxes of the two light curves with their corresponding  variances, $\sigma^{2}$s, and uncertainties in the flux $e^{2}$s. These discrete pairs are subsequently binned in equal-width time bins  that are comparable to the sampling of the light curves. Then the average DCF of the bin containing  the M pairs of UDCFs is given as, 
\begin{equation}
DCF(\tau )=\frac{1}{M}UDCF_{ij}
\label{DCF}
\end{equation}

 However, as the sampling distributions of DCF highly are skewed, using sample variance as an estimator of the  DCF uncertainty can be  inaccurate. To address the issue,  the DCFs, representing the cross-correlation coefficients, r, are z-transformed  (ZDCFs) according to the following Fisher transformations,
  \begin{equation}
 z=\frac{1}{2}ln\left ( \frac{1+r}{1-r} \right ), \quad \rho=tanhz, \quad \varsigma =\frac{1}{2}ln\left ( \frac{1+\rho}{1-\rho} \right ), 
  \end{equation}
which now approximately follow  normal distribution, and therefore the variance of the distribution provides a more robust  estimation of DCF uncertainties. Once the mean and the variance of the z-transformed distributions are computed,  they are inverse-transformed to the original distribution (for details see \cite{Alexander2013}). Furthermore, in contrast to the original DCF method in which regularly spaced time bins are used, in ZDCF the unbinned DCFs are averaged from the bins that contains a fixed number of populations, at least 11 pairs of observations.

The cross-correlation  between the \gama-ray and V  band optical observations  of the sample sources was implemented by using publicly available software\footnote{\url{http://www.weizmann.ac.il/particle/tal/research-activities/software}}. The  duration of the optical observations was chosen to be similar to that of the  \gama-ray observations, and moreover,  the light curves were binned weekly to match the sampling of the weekly-binned \gama-ray observations as presented in  \citetalias{Bhatta2020}. The diagram for lag against ZDCF value of the sample blazars are shown in Figure \ref{fig:zdcf}. The lag between the emissions was estimated by computing the centroid of the ZDCF peak, i. e., $\rm{\tau_c = \sum_i \tau_i ZDCF_ i /\sum_i ZDCF_i }$, where $\tau_c$  is calculated using the ZDCFs around the DCF peak that are greater than 0.8 times the ZDCF peak value \citep[see e. g.][]{Lobban2020,Buisson2017}. A positive lag implies \gama-ray variability features preceding those in the optical observations. The results of the cross-correlation study between the optical and the \gama-ray emission from the sample blazars are summarized in Table \ref{tab:table4}. The lead/lag of the most significant ZDCF peak and its ZDCF value in a source are listed in columns 2 and 3, respectively. To estimate the significance of the observed peaks against possible peaks which might have originated owing to the underlying red-noise variability in the two emission bands, 1000 artificial light curves were generated using single power-law model \citet{TK95}. As the optical emission and \gama-ray emission have different nature of variability as indicated by two different spectral indices as listed in the 5th and 6th column, respectively, of the Table \ref{tab:table4}, and these  spectral parameters were taken from \citet{Nilsson2018} and \citet{Bhatta2020}, respectively.  The distribution of ZDCFs between a random set of the simulate light curves and the real optical light curve were utilized to compute the 90\% and 99\% significance curves that are presented in the magenta and the red color, respectively, in  Figure \ref{fig:zdcf}.

\begin{figure*}
\gridline{\fig{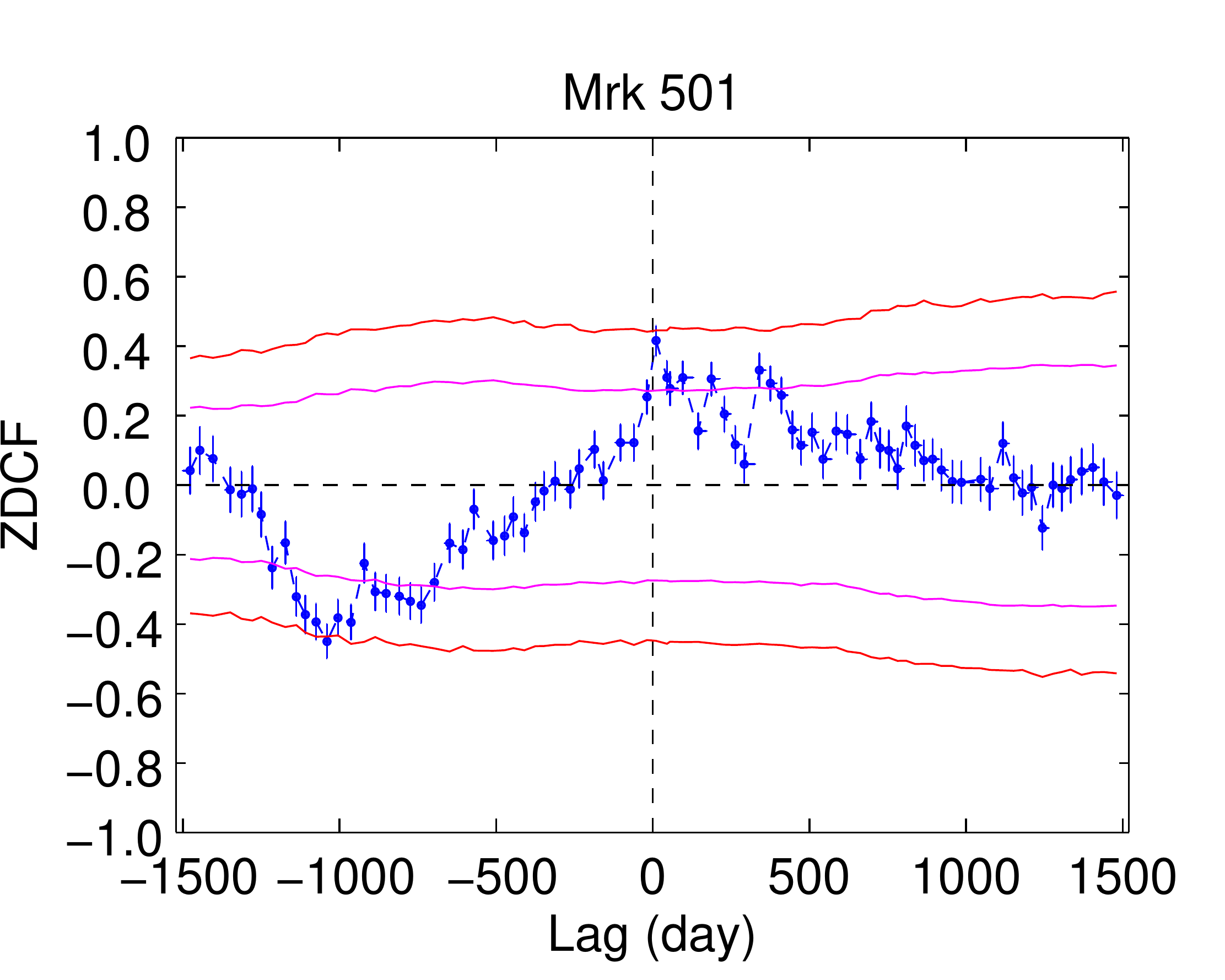}{0.33\textwidth}{}\hspace{-0.8cm}
\fig{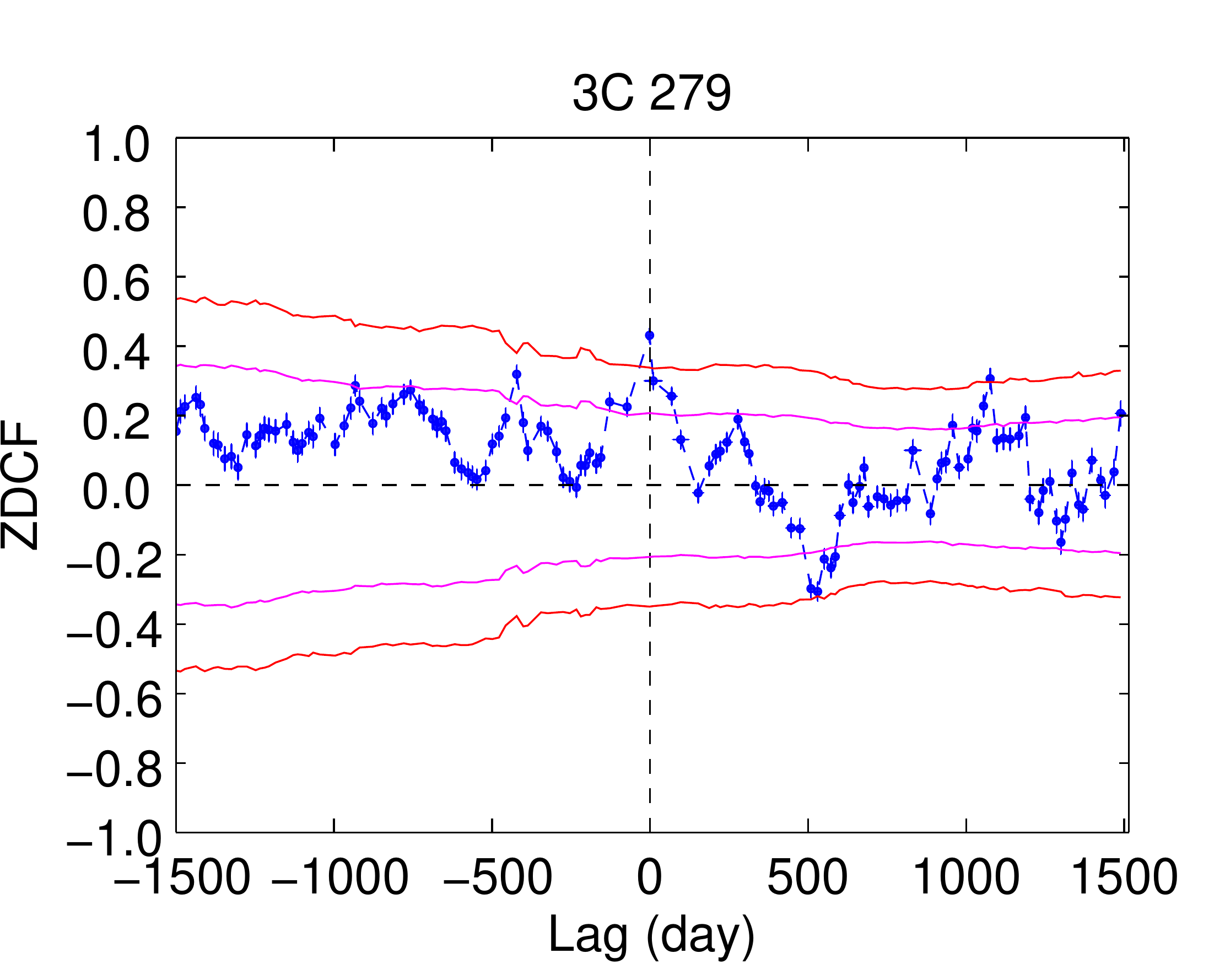}{0.33\textwidth}{}\hspace{-0.8cm}
\fig{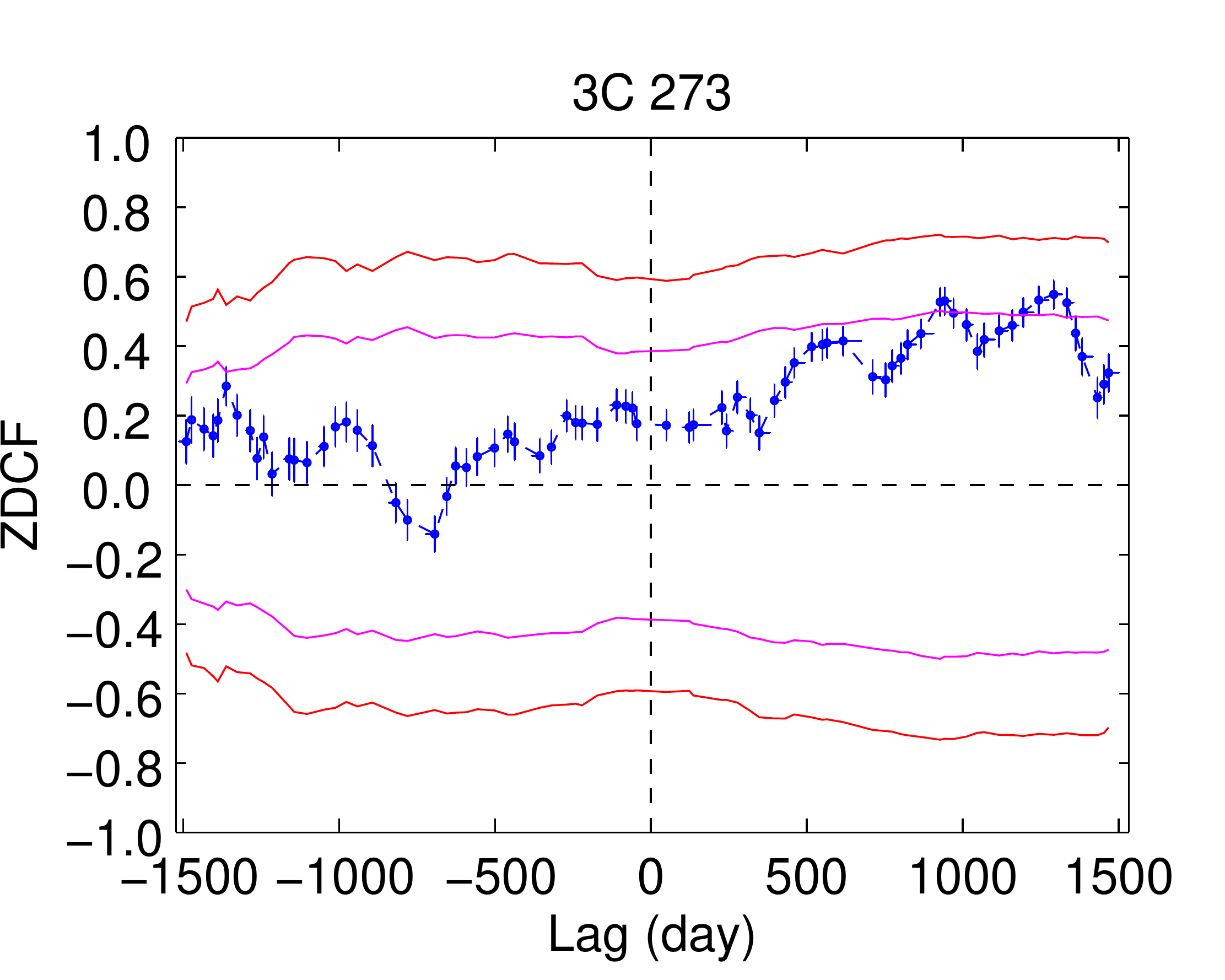}{0.33\textwidth}{}\hspace{-1.1cm}
          }
          \vspace{-0.7cm}
\gridline{\fig{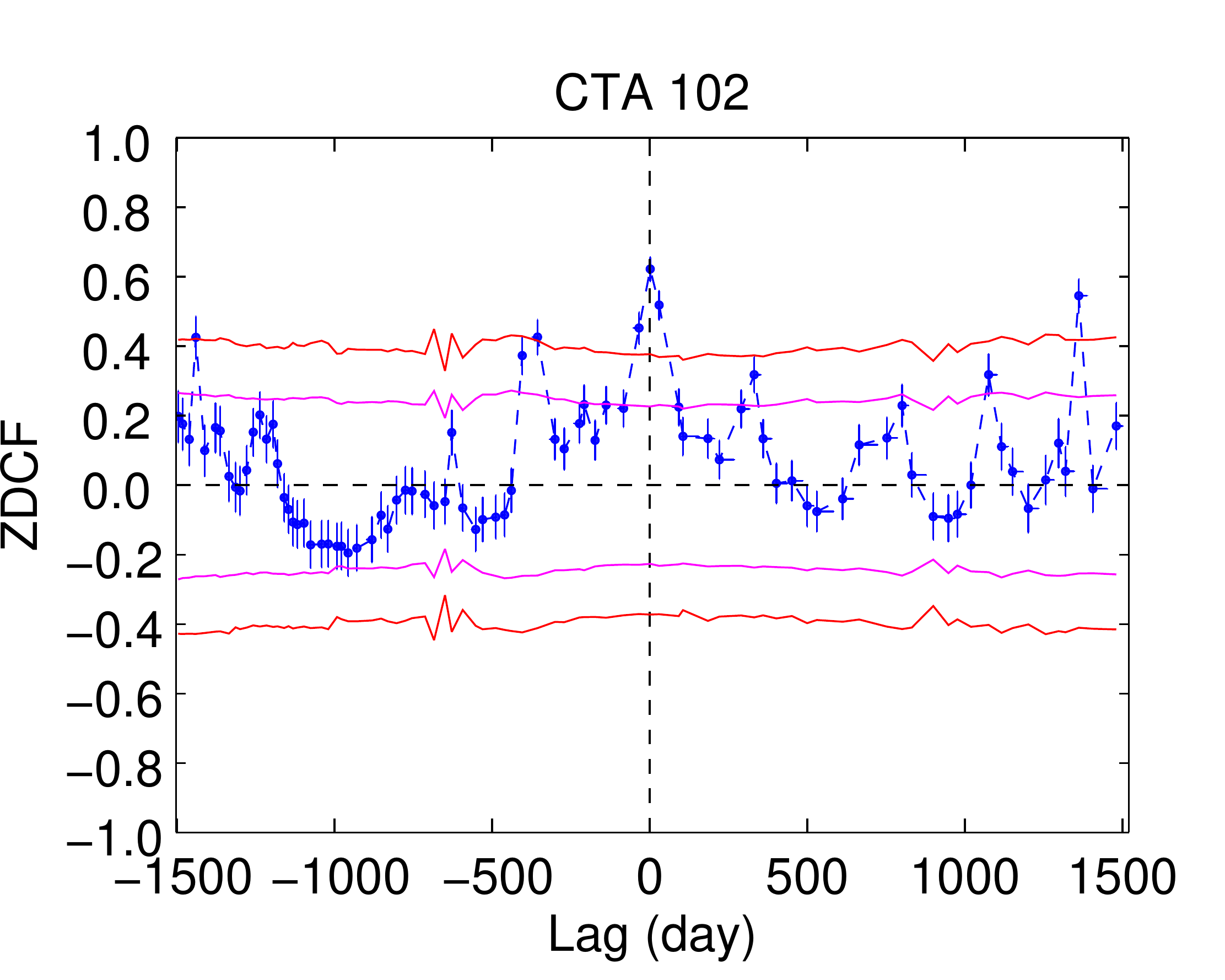}{0.35\textwidth}{}\hspace{-0.8cm}
\fig{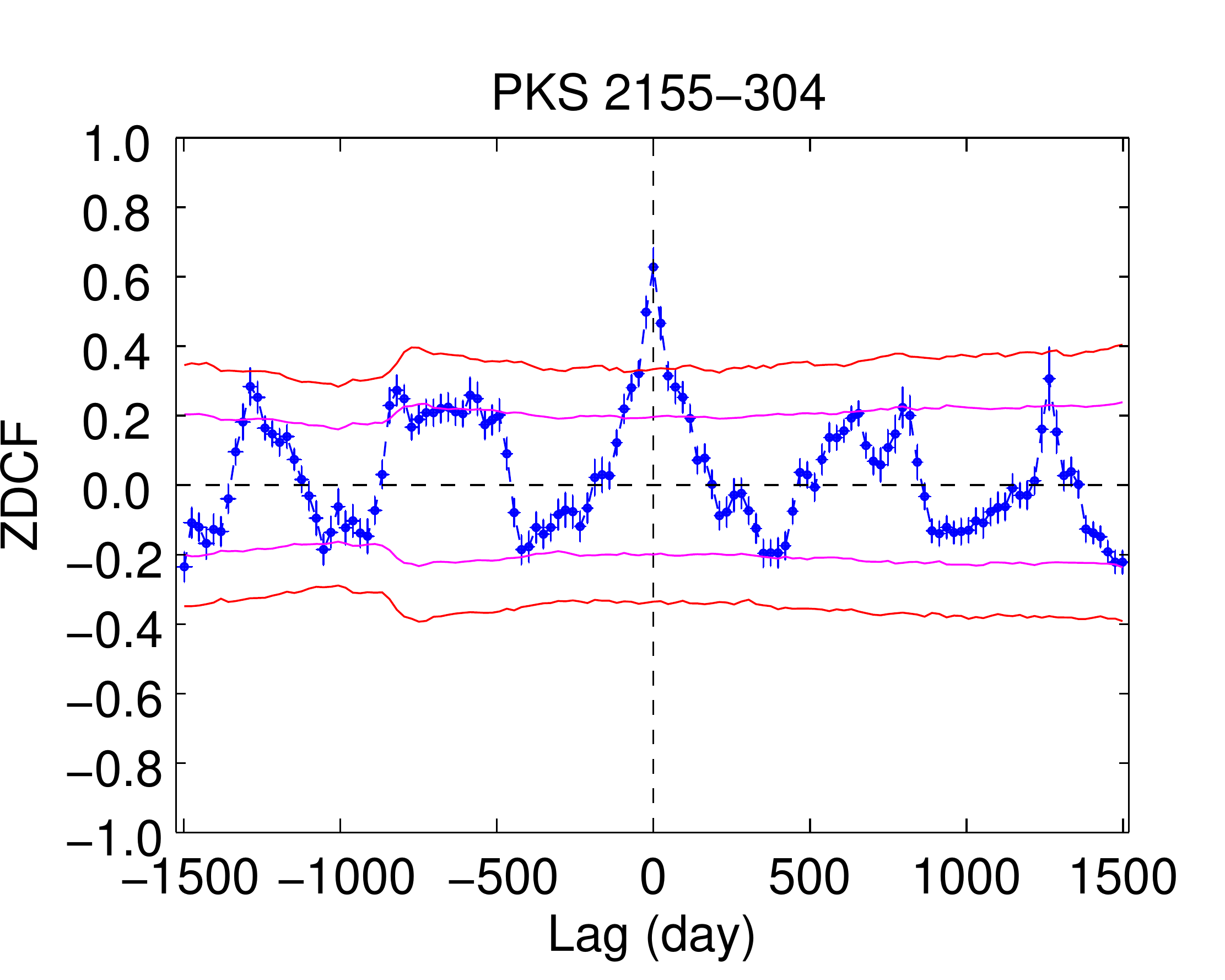}{0.35\textwidth}{}\hspace{-0.8cm}
\fig{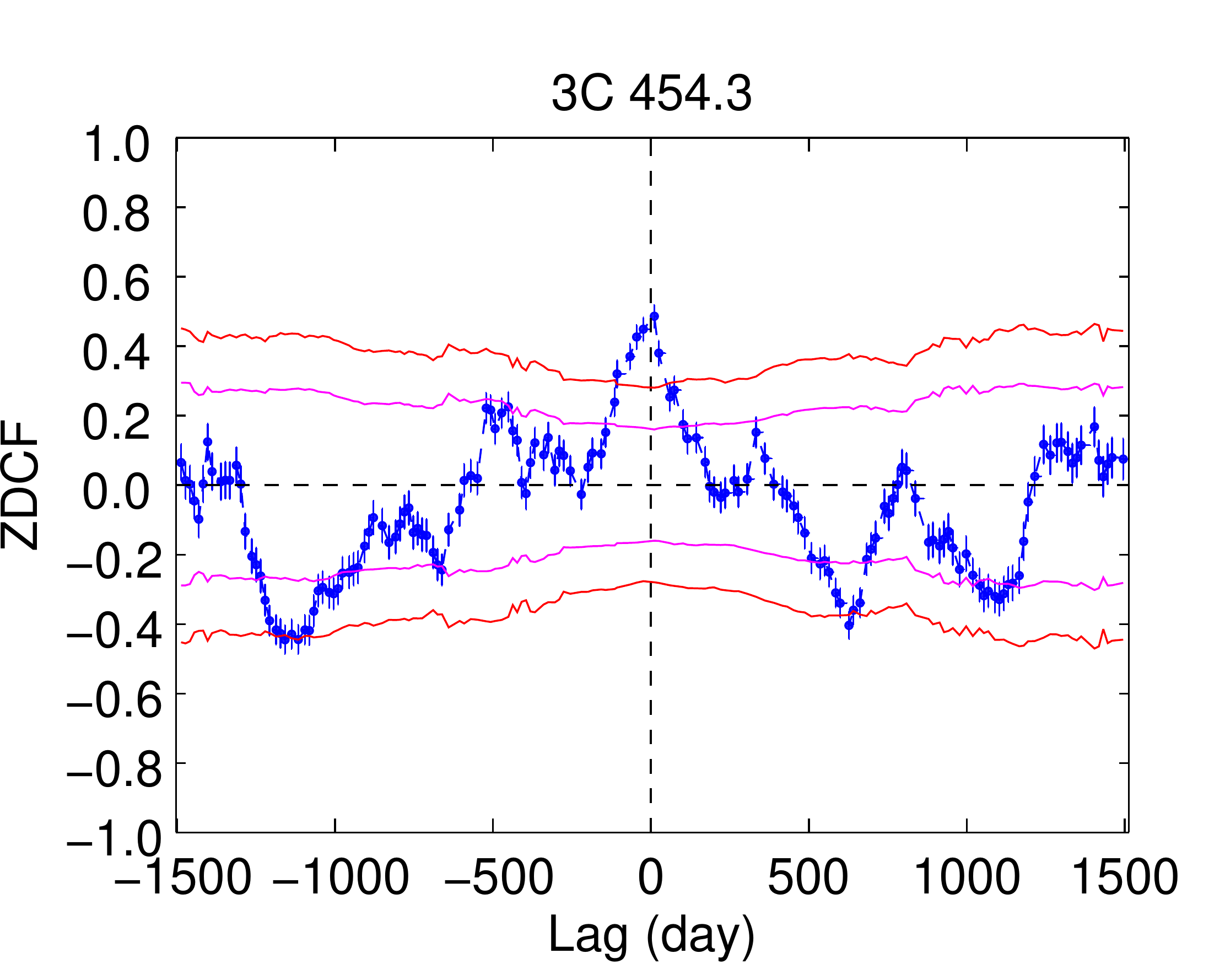}{0.35\textwidth}{}\hspace{-1.1cm}
          }
          \vspace{-0.5cm}
\gridline{\fig{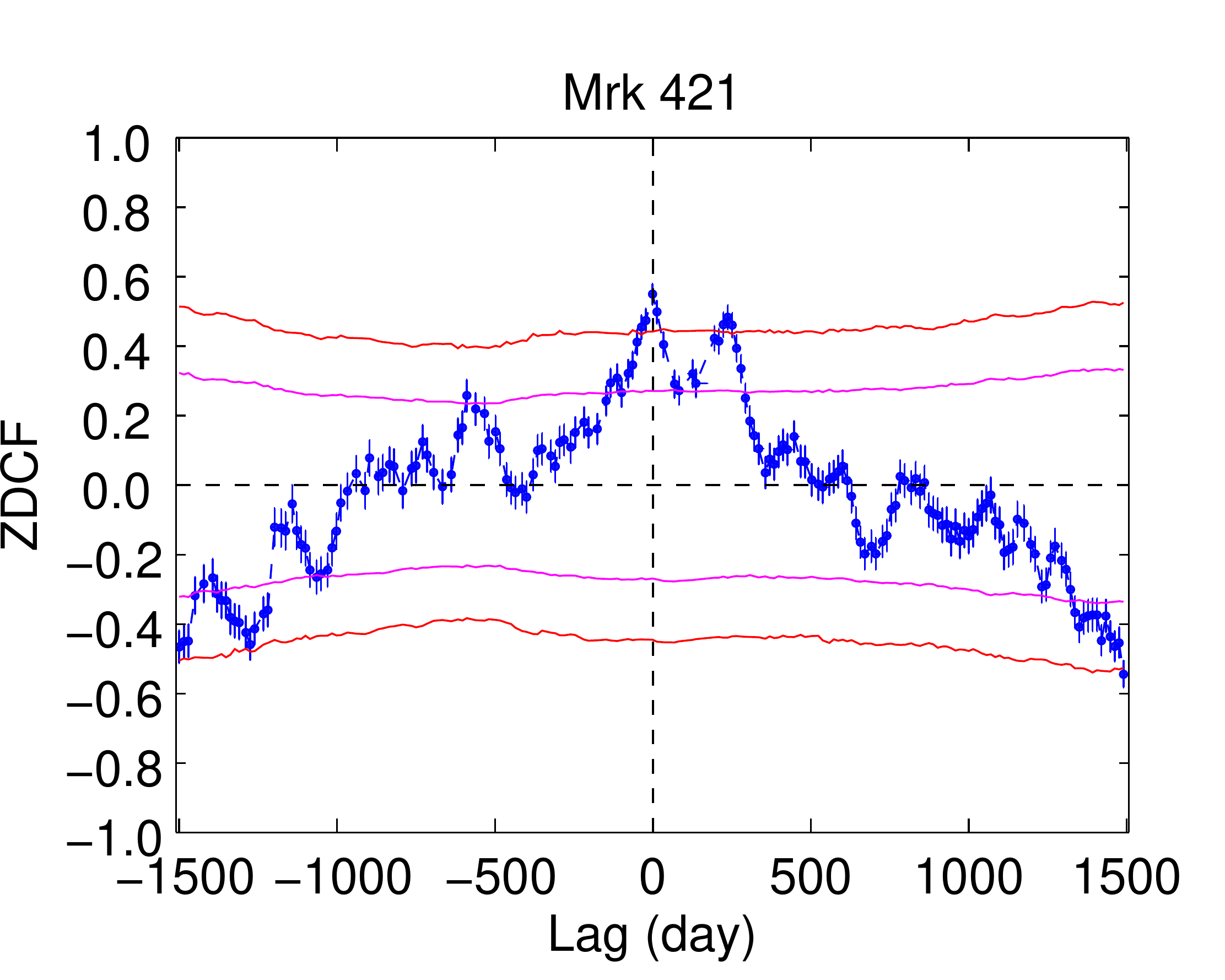}{0.35\textwidth}{}\hspace{-0.8cm}
\fig{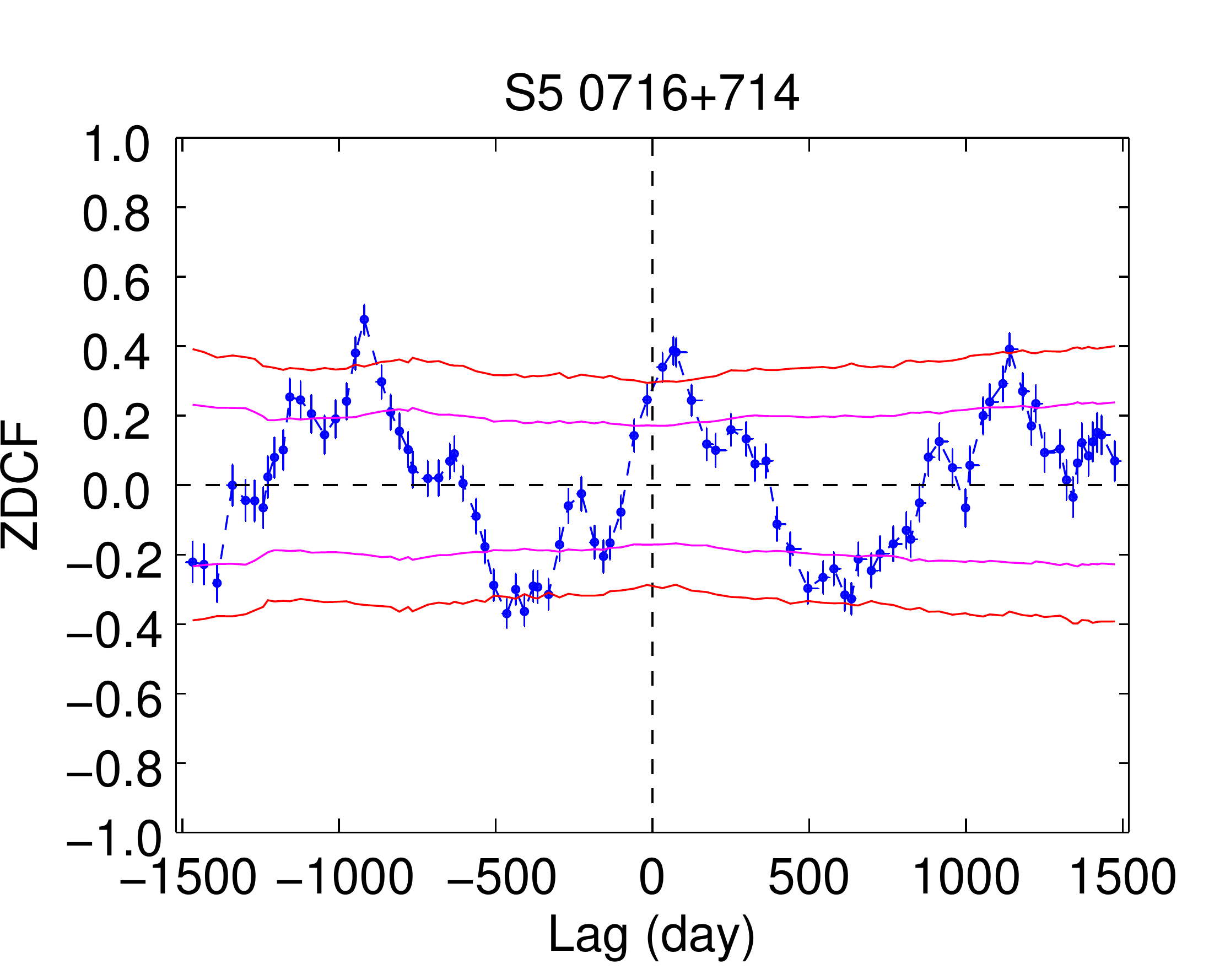}{0.35\textwidth}{}\hspace{-0.8cm}
\fig{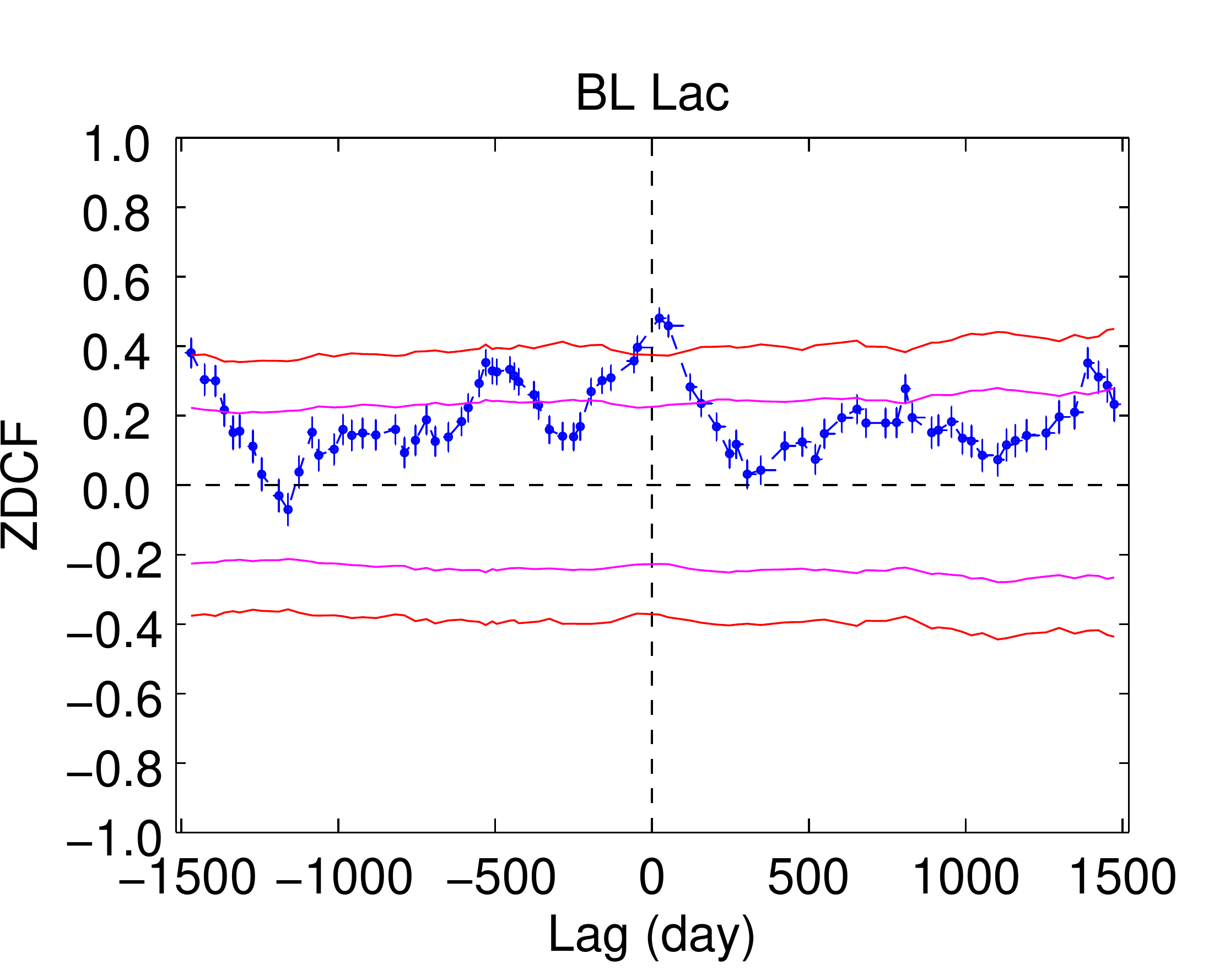}{0.35\textwidth}{}\hspace{-1.1cm}
          }
                    \vspace{-0.5cm}
\gridline{\fig{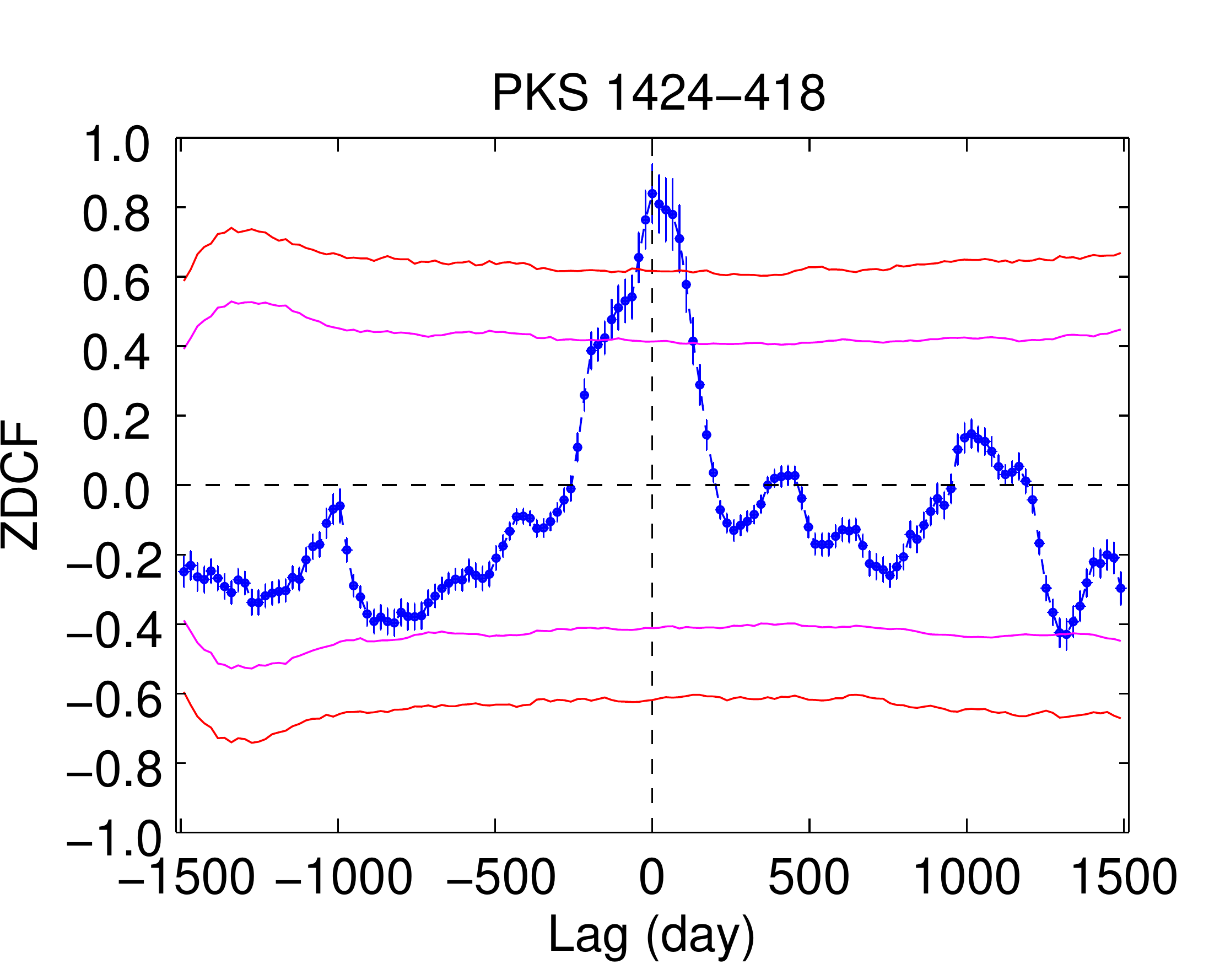}{0.35\textwidth}{}\hspace{-0.8cm}
\fig{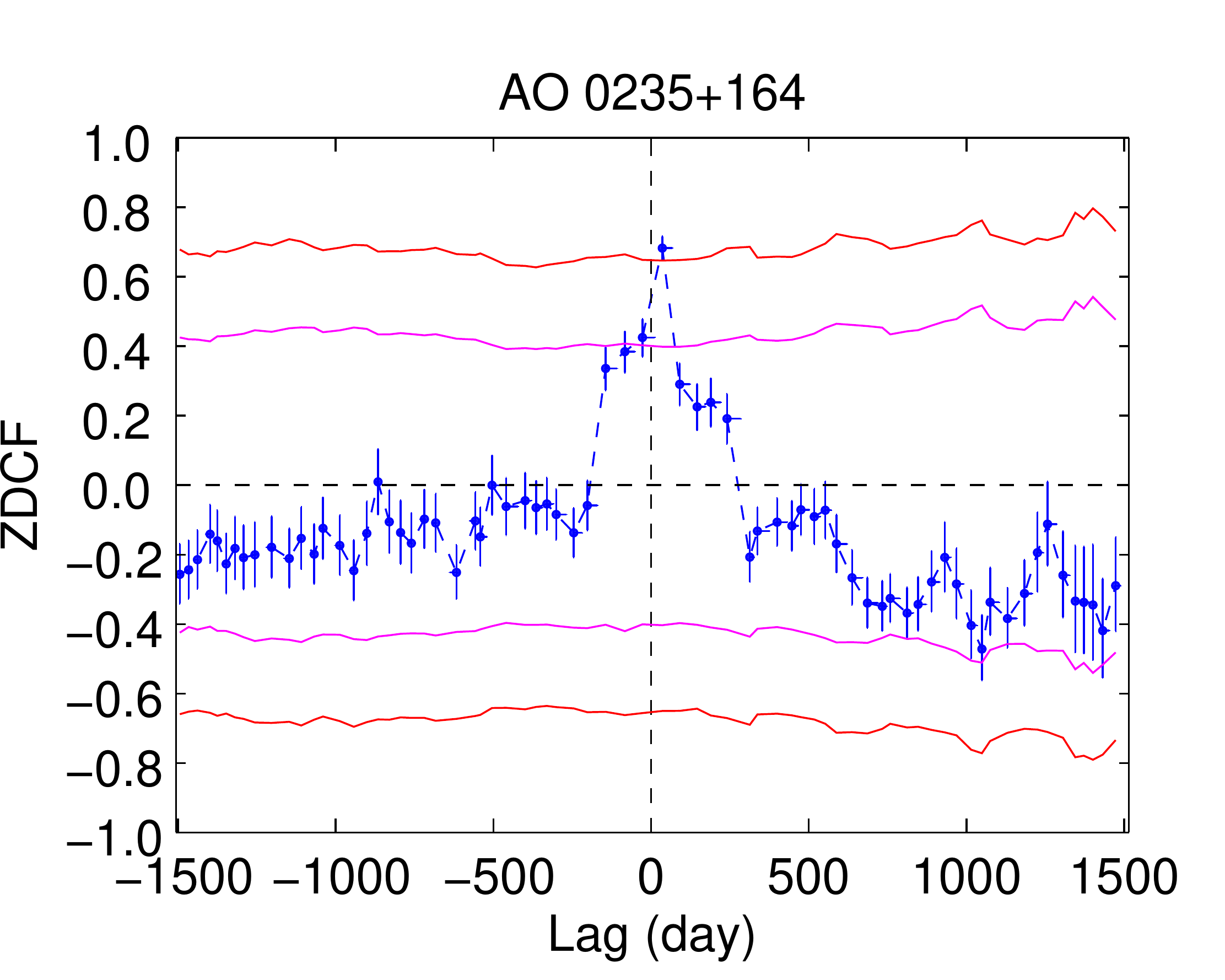}{0.35\textwidth}{}\hspace{-0.8cm}
\fig{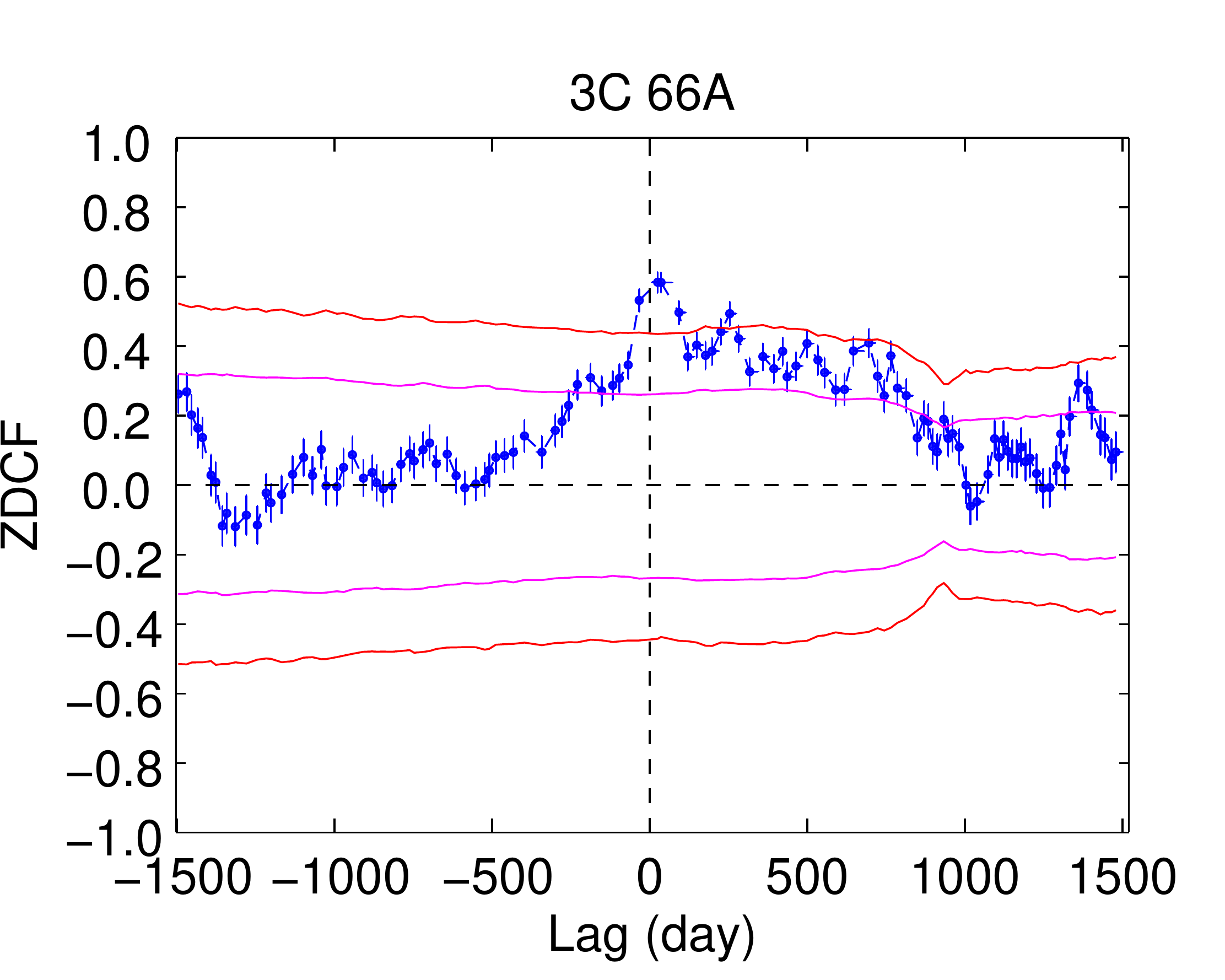}{0.35\textwidth}{}\hspace{-1.1cm}
              }
\caption{Z-transformed discrete cross-correlation between the longterm  \gama-ray and optical observations of the sample sources. A positive lag indicates variability features in optical emission lagging in time behind the similar features in \gama-ray emission of the sources. The magenta and red curves mark the 90\% and 99\% significant contours, respectively, estimated using MC simulations. The most significant lag and the associated significance, estimated taking into account of the correlated red-noise is presented in Table \ref{tab:table5} \label{fig:zdcf}}
\end{figure*}

In order to estimate the uncertainty in the observed time lag,  flux redistribution/random subset selection (FR/RSS)  as described in \citet{Peterson1998,Peterson2004} was followed. From 2000 Monte Carlo simulations of FR/RSS realizations, a distribution of the centroid lags was created,  and the 1$\sigma$ of the distribution was taken as a measure of the uncertainty in the time lag. As an illustration, the distribution of lag centriod for the blazar S5 0716+714 is presented in Figure \ref{Fig:FR_RSS}.  In the figure, the vertical line shown in magenta color marks the location of the peak centriod corresponding to the observed time lag.

As the plots in Figure \ref{fig:zdcf} and the corresponding Table \ref{tab:table5} show, in most of the cases, the optical and \gama-ray emission seem to be largely correlated. Moreover, the light curves in both  optical and \gama-ray are presented in the normalized flux units in Appendix, which show that in most of the instances the optical variability track the \gama-ray variability very closely, or \emph{vice versa}. The result is mostly consistent with the results from the earlier works \cite[e.g. see][]{Liodakis2018,Ramakrishnan2016,Cohen2014, Chatterjee2012}.  In the sources 3C 279, Mrk 421, CTA 102, Mrk 501,  PKS 0454-234, PKS 1424-418 and PKS 2155-304 the most significant ZDCFs have the lag within the error bars, or the lead/lag less than the sampling time bin, i. e. 7 days, meaning no significant average lead/lag between the emissions can be claimed. The sources BL Lac, 3C 66A and 3C 454.4 display positive lags of a few days. The most distinct significant lag is observed in the source  S5 0716+714, which is ~66 d. Although the correlation coefficient is $\sim$ 0.4, it shows higher significance (98.5\%) against correlated noise. Also, it is noted that sources Mrk 421 and 3C 66A reveal a complex pattern of cross-correlation, in the sense that  in addition to the prominent zero lag, the sources also reveals another significant peak around 300 days.  Of all the sources, FSRQ PKS 1424-418 shows the highest value (0.86) of correlation coefficient with a large significance of 99.90\%, followed by BL Lac AO 0235+164 0.68 with the significance of 99.40 \% and FSRQ CTA 102 0.62 with 99.70\% significance. Also it is interesting to note that, of all the sample sources, blazar 3C 273 clearly shows no significant correlation between the optical and \gama-ray emission within a few hundred days.

\begin{figure}[]
\includegraphics[width=0.45\textwidth]{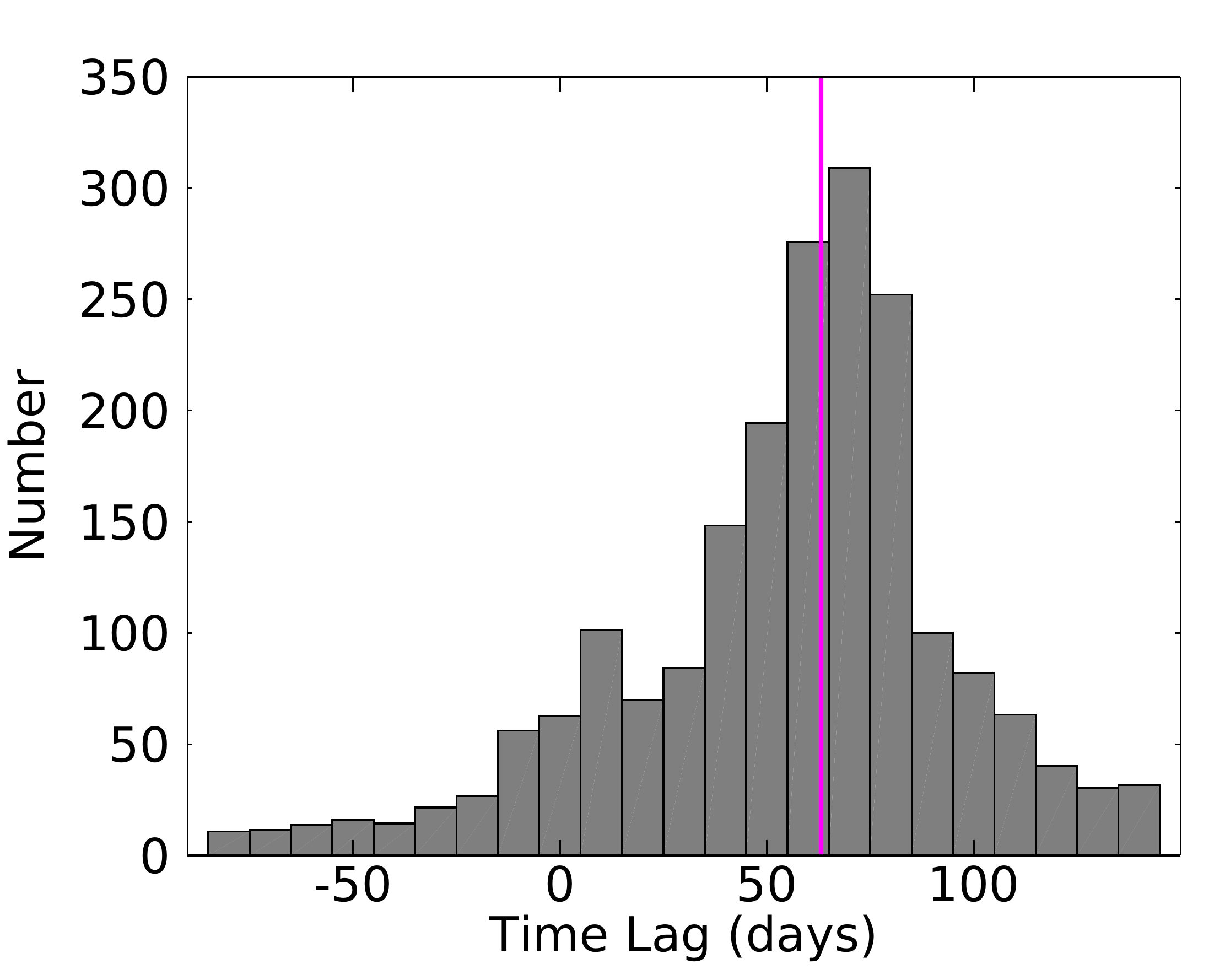}
\caption{Histogram of the lags corresponding to the centroids of the ZDCF peaks resulting from the optical \gama-ray  emission correlation in the blazar S5 0716+714 allows us to estimate the uncertainty in the observed lag. The light curve data were randomly resampled and the ZDCFs were computed for 2000 realizations following the method presented in \citet{Peterson1998,Peterson2004}. The vertical line in magenta marks the lag corresponding to centroid of the observed ZDCF peak.}
\label{Fig:FR_RSS}
\end{figure}

\begin{table*}[!t]
\begin{center}
\caption{Cross-correlation between \gama-ray and optical light curves. Here positive lag implies variability features in optical emission lagging in time behind the similar features in \gama-ray emission from the sources.\label{tab:table4}}
\begin{tabular}{lccccc}
\hline
\hline
Source & Lag (d)  & ZDCF& significance (\%)& $\alpha_{\gamma}$& $\alpha_{opt}$\\
\colnumbers\\
\\
3C 279 & -1.61 $\pm$7 & 0.43 $_{-0.04}^{+0.04}$ & 99.91& 1.10& 1.52\\
Mrk 421 & -1.22 $\pm$6 & 0.55 $_{-0.04}^{+0.04}$ & 99.96& 1.00& 1.38\\
S5 0716+714 & 66.75 $\pm$11 & 0.39 $_{-0.05}^{+0.05}$ & 99.92& 1.00& 1.18\\
CTA 102 & 1.35 $\pm$5 & 0.62 $_{-0.05}^{+0.04}$ & 99.91& 1.20& 1.60\\
BL Lac & 23.42 $\pm$15 & 0.48 $_{-0.04}^{+0.04}$ & 99.96& 1.00& 1.27\\
3C 66A & 25.15 $\pm$13 & 0.58 $_{-0.04}^{+0.04}$ & 99.97& 0.90& 1.40\\
3C 273 & -80.35 $\pm$19 & 0.23 $_{-0.06}^{+0.06}$ & 76.71& 1.20& 1.50\\
Mrk 501 & 10.41 $\pm$8 & 0.42 $_{-0.06}^{+0.06}$ & 98.03& 1.10& 1.65\\
AO 0235+164 & 35.66 $\pm$9 & 0.68 $_{-0.05}^{+0.04}$ & 99.42& 1.40& 1.55\\
3C 454.3 & 10.90 $\pm$5 & 0.49 $_{-0.04}^{+0.04}$ & 99.93& 1.30& 1.50\\
PKS 1424-418 & -4.90 $\pm$10 & 0.86 $_{-0.02}^{+0.02}$ & 99.99& 1.50& 1.75\\
PKS 2155-304 & 2.80 $\pm$7 & 0.62 $_{-0.02}^{+0.02}$ & 99.98& 0.90& 1.55\\
\hline
\end{tabular}
\end{center}
\end{table*}

\subsection{ Quasi-Periodic Oscillations }
The power spectral density (PSD) analysis performed on long-term optical light curves of a large number of blazar sources suggests that the blazar periodograms  can be largely represented by a single power-law PSD in the logarithmic frequency-power plane \citep[][]{Nilsson2018}. A similar conclusion was presented during the PSD analysis of 20 blazar \gama-ray light curves in \citetalias{Bhatta2020}. Nonetheless, evidence of presence of QPOs can be derived from  the periodogram which peaks at some characteristic frequencies corresponding to timescales of a few years, although occasional low-frequency peak can also arise owing to the red-noise like behavior of blazar variability.   In the literature, several AGN, both radio-loud and radio-quiet, are known to show QPOs in their light curves in different energy bands on wide range of timescales \citep[see][and the reference therein]{Bhatta2019,Gupta2018,  bhatta16c}. Probably the first case of MWL QPO, a characteristic timescale of  ($\sim$ 2-year) was reported in the blazar PG 1153+113 using the Fermi/LAT, X-ray, optical, and GHz radio observations \citep[see][]{Ackermann15}. In the optical, blazar OJ 287 is famous for its characteristic double-peak feature in its historical optical light curve that keeps repeating approximately after every $\sim$12 years \citep[e. g.][]{Sillanpaa88,Kidger1992,Valtonen2006}.  Search for QPOs in the optical emission of a sample of blazars has been carried out in several works \citep[e.g., see][]{Sandrinelli2017,Sandrinelli2016}.  In \citetalias{Bhatta2020} a systematic search of QPOs in a sample of \gama-ray bright blazars carried out and presence of significant QPOs in the sources Mrk 421, Mrk 501, PKS 2155-304 and PKS 1424-418 was reported. One of the chief goals of the this paper is to examine the MWL nature of the QPOs in these sources, by conducting similar analysis  in the optical observations of the similar duration.

To search for possible periodic flux modulations, the decade-long optical observations of the blazars were analyzed using the Lomb-Scargle periodogram (LSP; \citealt{ Lomb76,Scargle82}). The widely used method is expressed as
 \begin{equation}
P=\frac{1}{2} \left\{ \frac{\left[ \sum_{i}x_{i} \cos\omega \left( t_{i}-\tau \right) \right]^{2}}{\sum_{i} \cos^{2}\omega \left (t_{i}-\tau \right) } + \frac{\left[ \sum_{i}x_{i} \sin\omega \left( t_{i}-\tau \right) \right]^{2}}{\sum_{i} \sin^{2}\omega \left( t_{i}-\tau \right)} \right\} \, ,
\label{modified}
\end{equation}
where $\tau$ is defined by $\tan\left( 2\omega \tau \right )=\sum_{i} \sin2\omega t_{i}/\sum_{i} \cos2\omega t_{i} $\,.  The periodogram is evaluated for $N_{\nu}$ number of regularly spaced  frequencies between the minimum and the maximum frequencies given as, $\nu_{min} = 1/T$ and $ \nu_{max}=1/(2 \Delta t)$, respectively.  

 In particular, the LSP of the optical and the \gama-ray observations were compared to see if the \gama-ray QPO found in the earlier paper are also present in the optical observations. For the purpose, the observation period of the optical light curves were made nearly equal to the \gama-ray light curves and the optical light curves were weekly binned to match the binning of the \gama-ray light curves.  The LSPs of \gama-ray (from \citetalias{Bhatta2020})  and the optical (from this work) are presented in the black and blue color, respectively,  in Figure \ref{fig:6}. The uncertainty  in the peak timescale was estimated using Equation 52 from \cite{VanderPlas2018}, which accounts for the dependence of the uncertainty on the number of observations and their average signal-to-noise ratio. A moderate value of signal-to-noise ratio, i . e. 10, was used during the estimation. From the figure it can be seen that, for some cases, optical LSP peaks can be observed to coincide with the \gama-ray LSP peaks that were shown to be QPOs in \citetalias{Bhatta2020}.  The timescales corresponding to large peaks present at the similar temporal frequencies in both of the bands, and therefore considered significant, are listed in the 2nd (\gama-ray) and 3rd (optical) column of Table \ref{tab:table5}. Moreover, to present a measure of the significance of the observed LSP features against the variability owing to inherent power-law noise,  10000 MC simulated light curves were employed. The spectral power-law indexes required for the simulations were taken from \citet{Nilsson2018}, which uses the similar length of optical light curves of the sources. The simulated light curves were re-sampled to match the sampling and gaps of the real observation. In addition, to account for the observed linear RMS-flux relation with the slopes listed in Table \ref{tab:table3},  the simulated light curves were exponentiated \citep[][]{Alston2019,Uttley2005}. The 90 and 95\% significance contours resulting from the distribution of the simulated LSP are presented in magenta and red curves, respectively.   The possible QPOs present in each of the sources are discussed below.

\begin{figure*}
\gridline{\fig{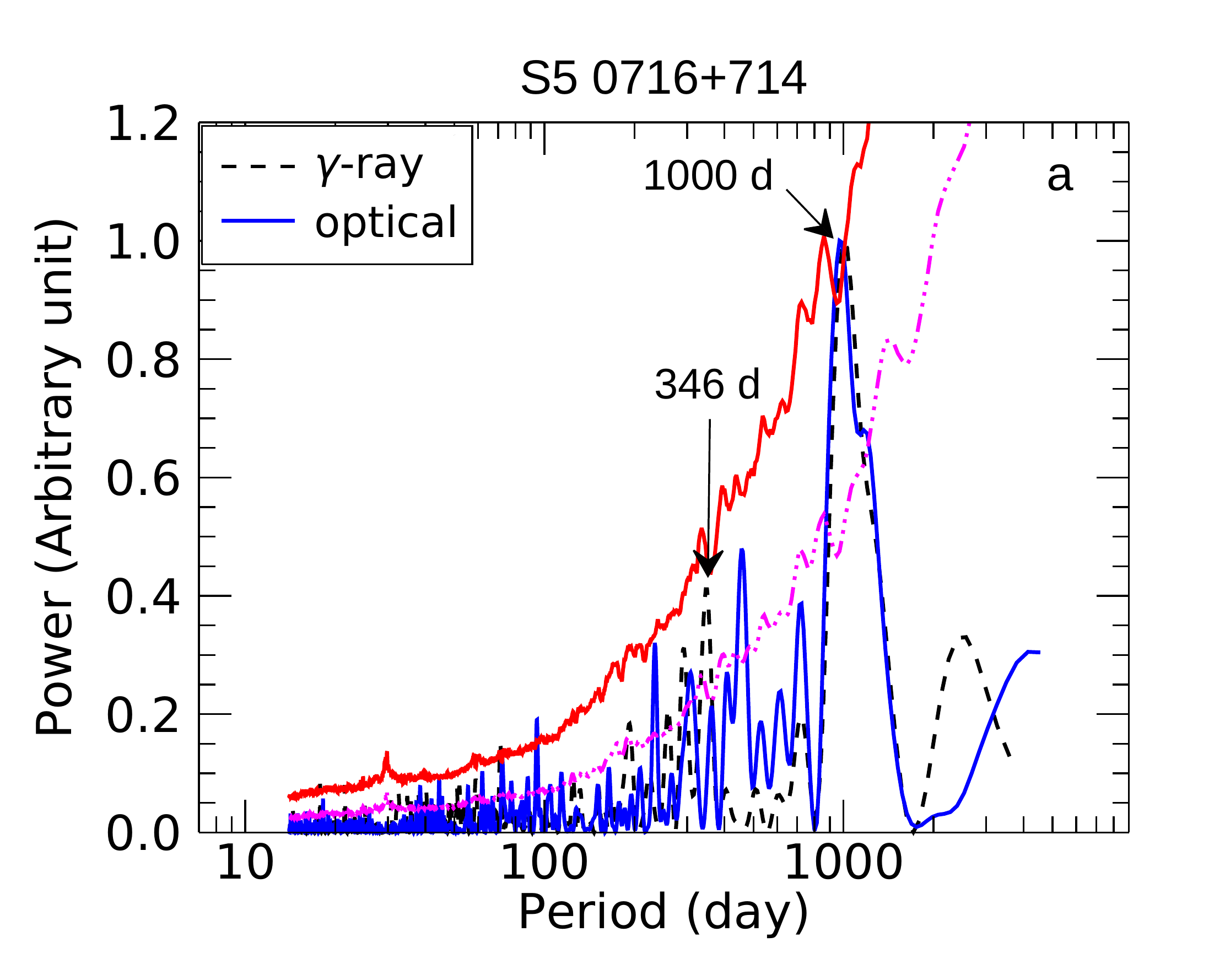}{0.5\textwidth}{}\hspace{-1cm}
          \fig{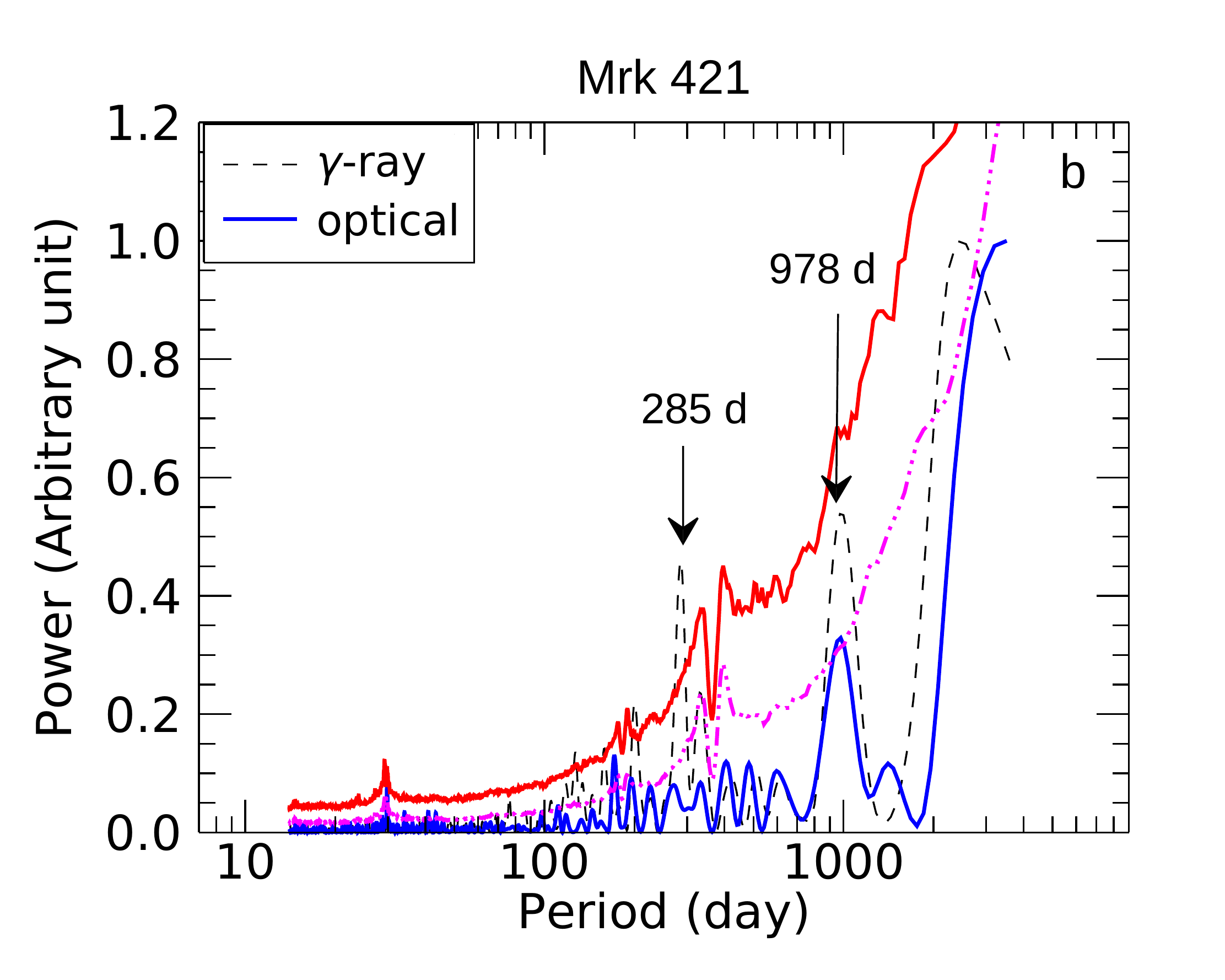}{0.5\textwidth}{}\hspace{-0.5cm}
		}
\vspace{-1.0cm}
          \gridline{\fig{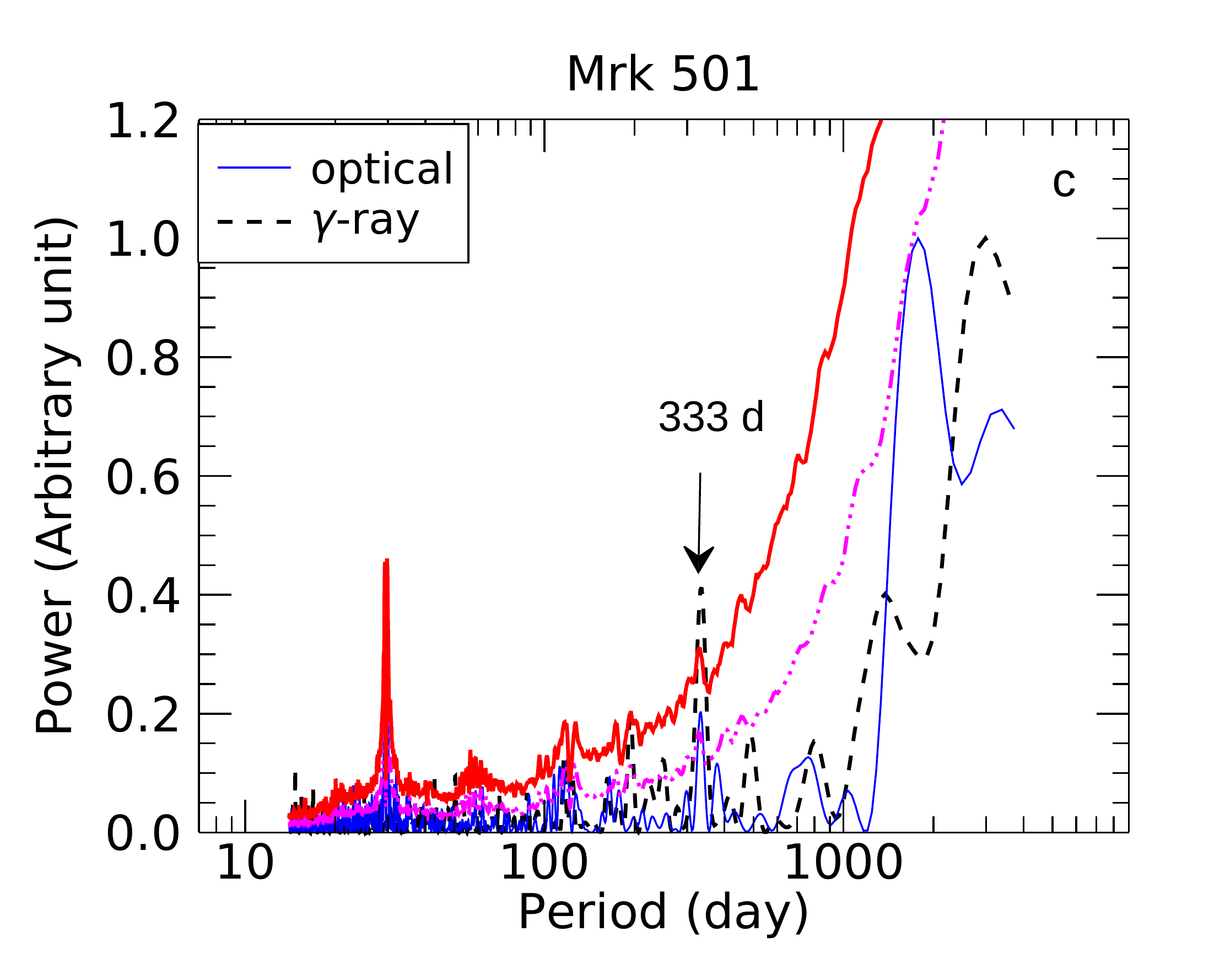}{0.5\textwidth}{}\hspace{-1cm}
			\fig{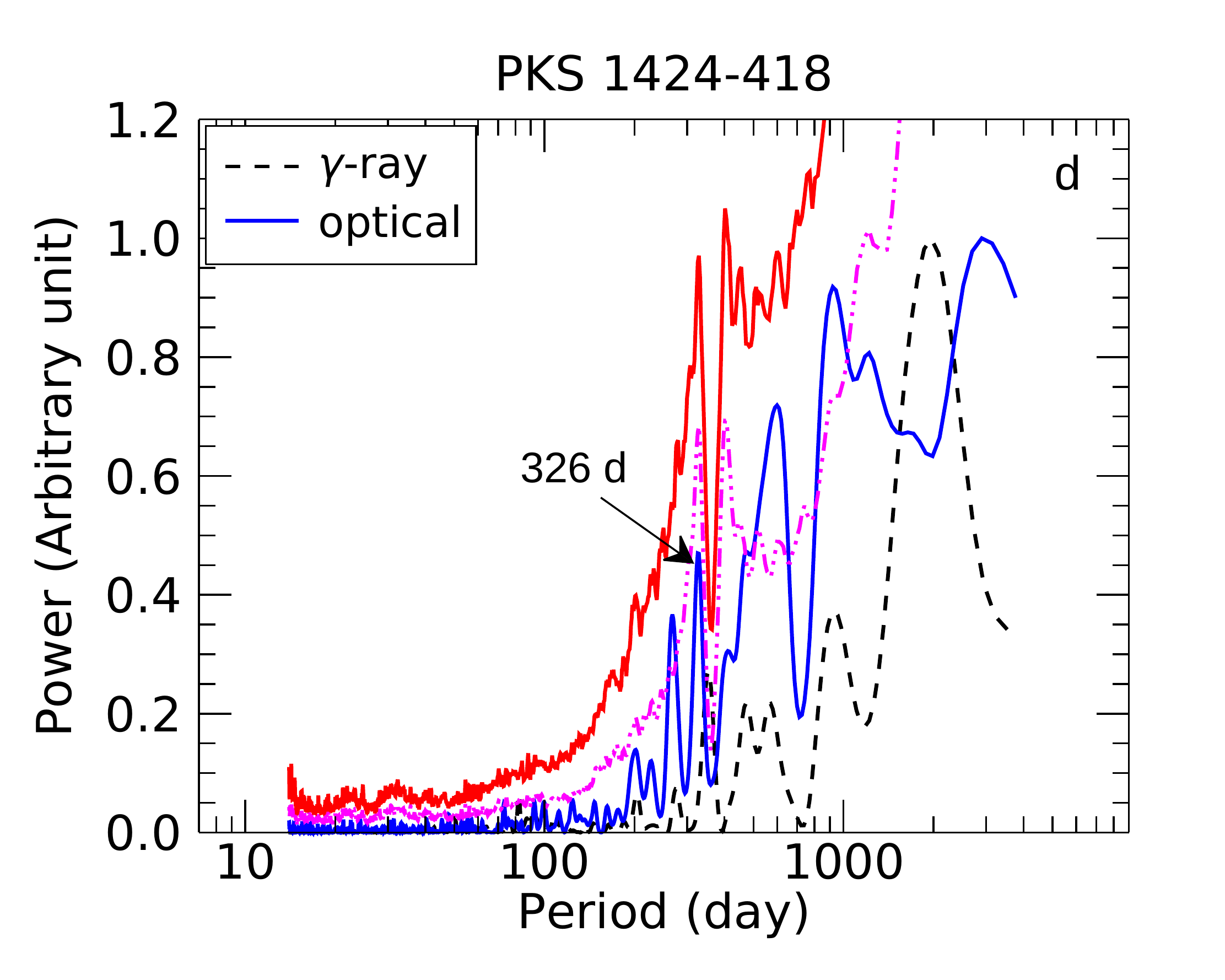}{0.5\textwidth}{}\hspace{-0.5cm}
                     }

                    \vspace{-1.0cm}
\gridline{	\fig{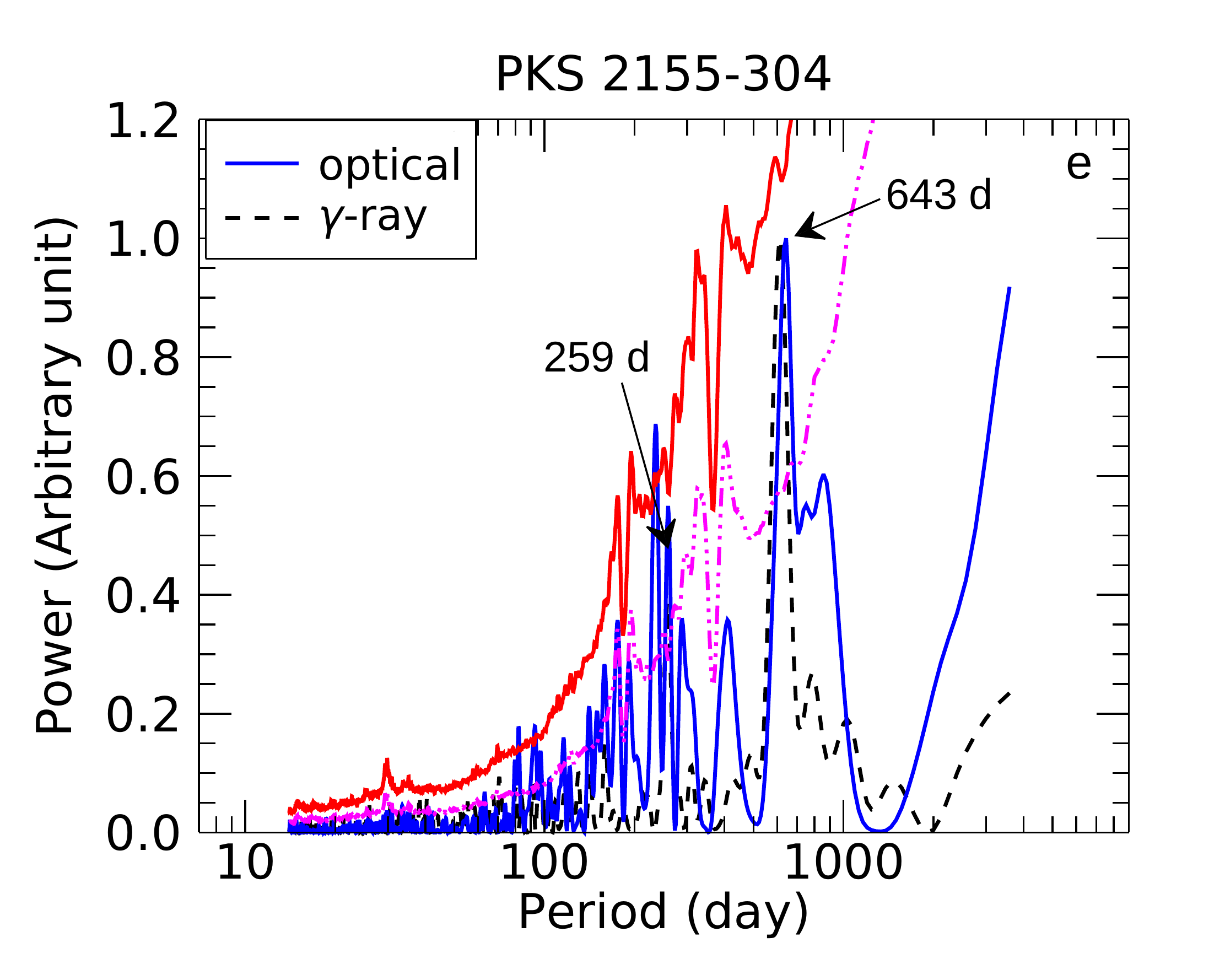}{0.5\textwidth}{}\hspace{-1cm}
                 \fig{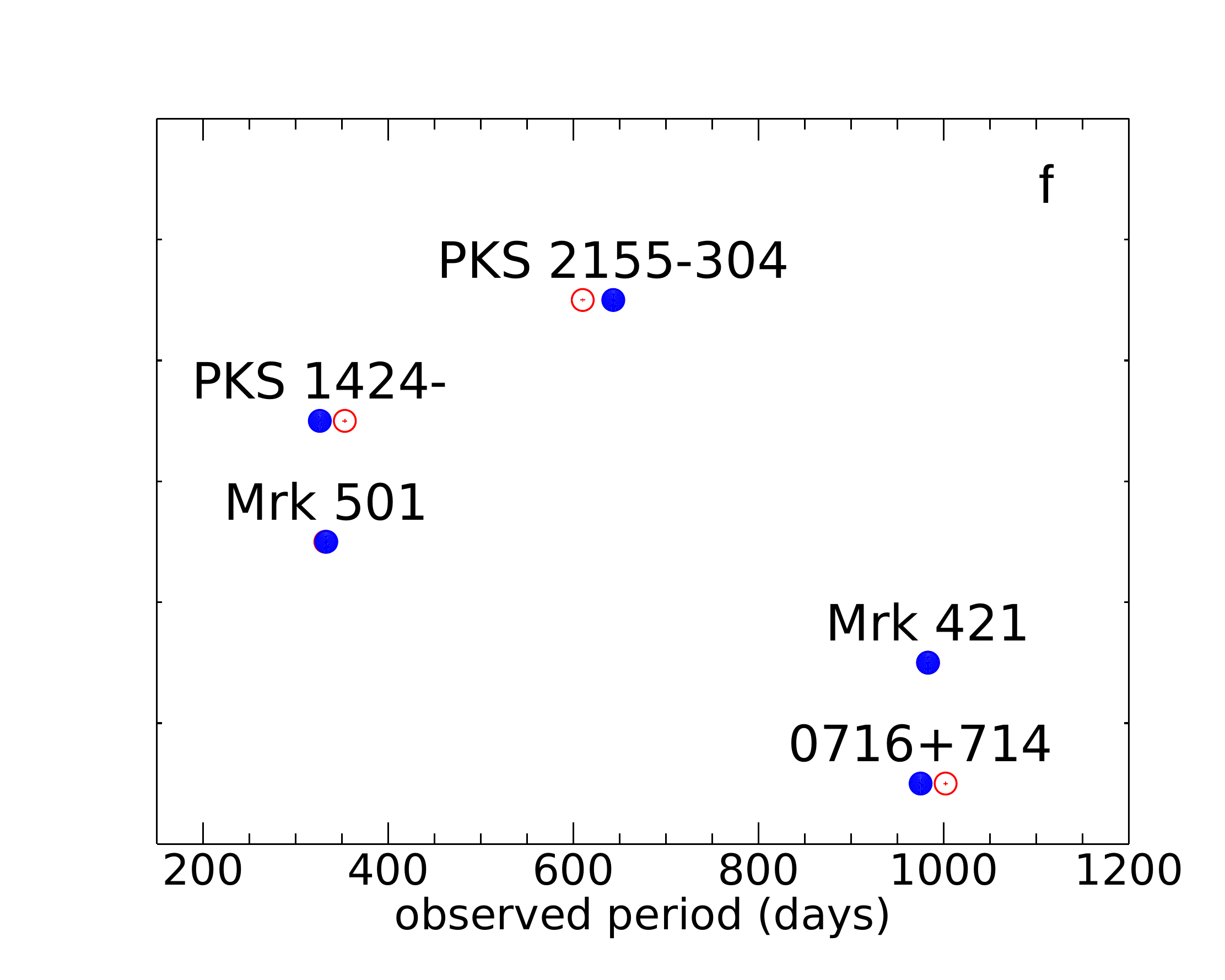}{0.5\textwidth}{}\hspace{-0.5cm}
		          }
\caption{Comparison between the blazar LSPs  in optical (blue solid line) and \gama-ray band (black dashed line). The significance of the optical LSP features against red-noise like variability  are shown by the 90 (dotted-dashed magenta line) and 99 \% (red solid line) contours. The figure in Panel f presents the temporal  coincidence of the possible optical QPOs with the  \gama-ray QPOs as reported in \citetalias{Bhatta2020}.    \label{fig:6} } 
\end{figure*}

\begin{table}[!b]
\begin{center}
\caption{List of the blazars in the sample that show significant QPO in the \gama-ray light curves\label{tab:table5}}
\begin{tabular}{lccc}
\hline
\hline
source &\gama-ray period (d)  & optical period ( d)& $\alpha_{\rm{opt}}$ \\
\colnumbers\\
\hline
S5 716+714	&	1002	$\pm$ 	0.71	&	1000	$\pm$ 	0.50&1.18 	\\
Mrk 421	&	981$\pm$ 	0.46	&	978$\pm$ 0.52& 1.38		\\
Mrk 501	&	332$\pm$ 	0.53	&	333$\pm$ 	0.24&1.65 	\\
PKS 1424-418	&	353	$\pm$ 	0.68	&	326	$\pm$ 	0.14&1.75 \\
PKS 2155-304	&	610	$\pm$ 	0.64	&	643	$\pm$ 	0.38&1.55 \\
\hline
\end{tabular}
\end{center}
\end{table}

\begin{itemize}
\item S5 0716+714:  The QPO in the optical observations corresponding to the  \gama-ray QPO at 346 d,  as reported in \citetalias{Bhatta2020} and marked in the panel a) of Figure \ref{fig:6},  was not observed.  But a QPO with a period of $\sim$1000 d QPO with more than 99\% significance is clearly visible with an appreciable overlapping in timescales in both the optical and \gama-ray bands. Such overlapping provides strong evidence for the existence of real and physical QPO, as opposed to mere artifact owing to inherent red-noise.    A similar optical QPO of 1223 d in the 20-d binned R-band light curve spanning longer duration was detected previously  \citet{Raiteri2003}. In addition to these, another possible QPO around 100 d period was found to be significant above 99\%.

\item  Mrk 421:    An optical  LSP peak around $\sim$ 978 d period was detected very close to the 983 d \gama-ray QPO reported in \citetalias{Bhatta2020}.  It is interesting to note that  both the peaks were found be to  90\% significant against underlying power-law noise in the corresponding bands. LSP in the \gama-ray and optical bands are shown in black and blue color, respectively, in panel b) of Figure \ref{fig:6}. The more significant \gama-ray  QPO  at 285 d in the source (see \citetalias{Bhatta2020}), however, was not detected in the V-band light curves. Similarly, 310 d optical QPO seen after removing flares in the optical observations, including AAVSO data partly, reported by \citet[][]{Benitez2015} and  the QPO with a characteristic timescale of $\sim$ 480 day in the rest frame, equivalent to observed $\sim$ 490 days, as reproted by \cite{Nilsson2018}  were not observed in the LSP analysis

\item  Mrk 501:    In the blazar  Mrk 501, the optical LSP peak with 95\% significance centered at 333 d was found to clearly coincide with the  332 d \gama-ray QPO as reported in \cite{Bhatta2019}, as shown in panel c) of Figure \ref{fig:6}. The significant temporal coincidence further strengthens the evidence for the multi-frequency nature of the observed QPO. 

\item PKS 1424-418:  A \gama-ray QPO with 353 d characteristic timescale was reported for the first time in \citetalias{Bhatta2020}. The LSP analysis of the optical observations of the source also revealed a peak at  326 d, which is is coincident with the \gama-ray QPO within the range of the associated uncertainty. However, the significance of the possible QPO turns out to be below 90\% against the power law noise of the spectral index 1.75. The optical and the \gama-ray LSPs of the blazar PKS 1424-418  are shown in panel d) of Figure \ref{fig:6} in blue and black color, respectively. 

\item  PKS 2155-304: The LSP of the optical light curves implied the presence of a 643 d QPO in the blazar  PKS 2155-304 as shown in the panel e) of Figure \ref{fig:6}.  This QPO, 97\% significant against red-noise in the optical light curve, is closer within the error to 610 d \gama-ray periodic flux oscillations reported in \cite{Bhatta2020} and 620 d QPO reported in \citet{Sandrinelli2018}. The 317 d timescale reported by  \citet[][]{Zhang2014RAA}  in the R-band light curves spanning 35 yrs, however, was not detected in the V-band light curve presented here.  But on the other hand,  a 96\% significant LSP peak around 259 d, which is very close to the 257 d \gama-ray QPO reported in \citetalias{Bhatta2020}, was also observed.

\end{itemize}

\section{Discussion \label{sec:4}}
In this section,  possible implications of the results obtained from the analyses on the decade-long optical observation of blazars and the comparison between the variability properties in the optical and \gama-ray band are discussed. The results are interpreted within the context of standard model of blazars.
\begin{itemize}
\item Optical variability in blazars:\\
The optical light curves of the blazar sources, as presented in Figure \ref{Fig:1}, exhibit brightness modulations of various degrees. To conduct variability analysis, the observed variability was quantified  by employing fractional variability, which serves as an estimator of the average flux modulations in the light curves over the period. The numbers listed in the 7th column of Table \ref{table:1} suggest that blazar sources are distinctly characterized by their remarkable activity in the optical band. Furthermore, it is seen that, on average, FSRQs are more variable than the BL Lacs. This is consistent with the variability in the $\gamma$-ray band as reported in \citetalias{Bhatta2020}. The origin of the optical emission from blazars can be of both thermal and non-thermal nature. In particular, in FSRQs the signatures of thermal emission in optical/UV related to the accretion disk and emission lines from the broad-line regions can be observed.  In the context of multi-colored thin disk approximation, for a black hole with a mass M$_8 $, expressed in the units of $1\times10^8$ solar masses,  accreting at a rate $\dot{m}$, expressed in the units of the Eddington accretion rate $\dot{m}_{Edd}$, the temperature profile  of the accretion disk as a function of distance from the central black hole can be written as
\begin{equation}
T\left ( r \right )\sim 1\times10^6 \left ( \frac{\dot{m}}{\dot{m}_{Edd}} \right )^{1/4}\left ( \frac{1}{M_8} \right )^{1/4}\left ( \frac{r}{r_g} \right )^{-3/4},
\end{equation}
 \citep[see][]{Peterson1997} where the gravitational radius is given by
$r_g = GM/c^2$. With the assumption of black body radiation, the temperature can be translated into the wavelength of the optical emission. For a typical blazar mass of $1\times 10^9 M_\odot$,  the region that emits V band optical emission, with an effective wavelength of 0.545 $\mu$m, lies at the distance of $\sim1000 \ r_g$, a distance that falls within the inner region of the disk. In such a scenario, the observed long-term variability can result due to the changes at the disk, e. g., perturbation in the viscosity, accretion rate, turbulence etc.  in the corresponding characteristic timescales \citep[see][]{Czerny2006}.  However, in blazars the flux variability originating at the disk could easily be modified by the Doppler boosted emission from jets in course of their propagation along the jets. Nevertheless, it could be possible that the signatures of the modulations should propagate along the jet through disk-jet coupling mechanisms \citep[e. g. see][]{Bhatta2018b,Blandford1982}. 

  In the scenario of non-thermal emission from the jets, shock waves propagating the jet can result in a power-law distribution of energetic particles $N(\gamma)\propto \gamma ^{-p}$. These particles in the ambient magnetic field emit synchrotron emission (optical emission is the present case) with a power-law spectral profile.   In the context of  leptonic models, if we make a simple and crude approximation that each electron radiates all of its power at the single frequency  given by $\nu \approx \gamma^2 \nu_G$, where $\nu_G =qB/2\pi m$ represents gyration frequency of a particle with mass m and charge $q$ and moving into a magnetic field $B$, and  further assume a typical magnetic field of 1 Gauss, the Lorentz factors of these synchrotron electrons responsible for the non-thermal optical emission can estimated to be as high as $\sim10^4$. In hadronic models, it is also possible for the protons to contribute to the synchrotron emission. But in such scenario, proton being nearly 1836 times more massive than electron, Lorentz factor and/or ambient magnetic field must be larger in the similar proportion.  
 In the jet models, the observed optical variability could be driven by a combination of factors, e. g.,  modulations  in the distribution of particles and ambient magnetic field at the emission region. Moreover, it is also possible that a part of the observed  variability  could also be merely projection effects, e. g.,  arising owing to twists and bends along the jets, such that changes in either the velocity and/or the project angle of the emission region can lead to an amplified flux modulation\citep[see e.g.][]{Bhatta2018d,Raiteri2017} 
 
 In order to be able to utilize the light curves to extract information about the dynamical states of the central engine and thereby achieve a complete characterization of the variability phenomenon, it would require an approach invoking non-linear differential equations in the framework of relativistic gravito-magneto-hydrodynamics, an extremely daunting task. In the absence of such treatment, it is often convenient to model the observed variability in blazars as being driven by linear and non-linear stochastic processes. Nonetheless, more recent studies reveal  deterministic content in the light curves with an indication that the dynamical evolution at the center can be tracked in the long-term light curves \citep[see][and the references therein]{Bhatta2020b} .

\item Flux distribution and RMS-flux relation:\\
In order to study the distribution of of fluxes from the modulating optical light curves of the sample blazars,  normal and lognormal PDFs --two widely employed PDFs-- were fitted to the unbinned flux histogram. Resulting AIC and BIC values suggests that the observed flux histogram of blazars could be realization of the underlying  log-normal processes yielding a heavy tail skewed towards higher fluxes. Recently, log-normal PDF has been applied to characterize the blazar MWL fluxes distribution in a number of sources, e. g., \citealt[][]{Bhatta2020,Shah2018} (\gama-ray), \citealt{Chakraborty2020,MAGIC2020} (Mrk 421 multi-frequency).  Similarly, the RMS-flux relation is widely observed in astrophysical systems, especially ubiquitous among black hole X-ray binaries \cite[see][and the references therein]{Heil2012}.  A linear RMS-flux relation as well as log-normal flux distribution has been reported in a number of blazar sources in several frequency bands e.g. \citealt{Giebels2009} (X-ray), \citealt{Edelson2013} (optical), \citealt{Bhatta2020,Kushwaha2017} (\gama-ray).  Recently, \citet{Bhattacharyya2020} reported presence of the linear RMS-flux relation in X-ray light curves of the blazars Mrk 421, PKS 2155-304, and 3C 273,  the sources among the ones studied in \citetalias{Bhatta2020} and this work.  The linear RMS-flux relation suggest that the variability properties of AGN are largely correlated during multiple flux states and thereby could be an indication of the underlying non-linear processes driving the observed variability resulting  in the flux distribution that is skewed towards higher flux, such as log-normal PDF \citep[see][]{Uttley2005}. Moreover, the observed log-normal distribution of the blazar flux might imply the presence of multiplicative processes in operation, e. g., multiplicative coupling of the perturbations occurring at either the disk and/or the jet, as opposed to additive coupling as in shot-noise-like perturbation  \citep{Arevalo2006,Lehto1989}. However, such observations could as well be the manifestation of linear processes  \citep[for a recent critical review on non-linearity, the RMS-flux, and log-normal flux distributions readers are directed to][]{Scargle2020}. Similarly, log-normal distribution of jet powers, resulting from the multiplicative process in the jets operating in long timescales, could explain the observed ultra-high energy cosmic rays \cite[see][]{Matthews2021}.

 Furthermore, in the accretion disk models,  log-normal flux distribution and the linear RMS-flux relation could be associated to the uncorrelated fluctuations in the $\alpha$-parameter, governed by viscosity fluctuations,  that take place at different radii and gradually propagate outwards modulating the mass accretion rates at larger scales \citep[see][]{Lyubarskii1997}.   Such disk modulations can propagate through the jet via strong disk-jet connection in radio-loud AGN. In case of blazars, the modulations subsequently are amplified owing to relativistic beaming effects, and thereby could be affected by  projection effects.   In the relativistic jets models, highly skewed flux distribution could be produced in the jets-in-jets scenario in which the jets dominated by Poynting flux can give rise to the condition for the production of  isotropically distributed mini-jets \citep[see][]{Giannios2009}. The  distribution of the emission produced in such a case has been found to hold the RMS-flux relation \citep[see][]{Biteau2012}.  In hadronic model, such scenario can be efficient in obtaining a large bulk Lorentz factor required for the synchrotron emission from protons.

  \item Optical \gama-ray correlation: \\
  Cross-correlation between the optical and the \gama-ray emission from the sample blazar sources was studied applying the ZDCF. The method provides an estimator for the average correlation between the variable emission features in the two spectral bands. As we observe from Table \ref{tab:table5} and Figure \ref{fig:zdcf}, although highly significant, the ZDCF values seem to be relatively moderate. This is most likely due to difference in the overall nature of the statistical variability properties  in two different bands as characterized by the different PSD slope indexes.
The observed strong long-term correlation between the variability features in the optical and the  \gama-ray emission finds a natural explanation within the framework of leptonic blazar models.  In such models, both synchrotron and IC processes, that result in the low energy photons (optical emission in this case), and  the \gama-ray emission, respectively, take place within the jet. As a result, a positive correlation between the low and high energy emission can be expected. Moreover, in the SSC flavor of the leptonic models, the same population of high energy elections participate in both the synchrotron and IC processes, resulting in even higher correlation.   A positive optical lag (in the sense that variable features in the \gama-ray light curves precede the ones in the optical) that was found to be pronounced in the case of the blazar S5 0716+714, can be explained in terms of opacity distance \citep[e.g.][]{Fuhrmann2014,Max-Moerbeck2014} and strongly stratified radiation field  in the ambient magnetic field profile (e. g. $B\propto 1/r$) such that the high energy emission, e. g. \gama-ray emission, is produced closer to the central engine \citep[see][]{Marscher2016}. The observed lag also suggests co-spatial nature of the particles emitting the radiation, in which case the spatial separation can be associated with the light travel distances.

  In the hadronic blazar models, \gama-ray emission could arise due to proton-proton interactions and/or proton-photon interactions  within and/or outside the jets, In such cases, neutral and charged pi-mesons are produced, of which, neutral pions being quite unstable quickly decay to \gama-ray.  On the other hand, the charged pions decay producing secondary electrons/positrons which subsequently can emit synctroncron in the ambient magnetic field.  It is also possible that the observed \gama-ray emission could be the direct result of proton synchrotron emission. Therefore, in the hadronic scenario, the correlation between the \gama-ray and the lower energy emission e. g. optical emission, could be much more complex.   Nonetheless, due to the lack of a complete understanding of the dominant jet particles and the emission processes they participate in, contribution of hadronic processes to the observed correlation between the optical and the \gama-ray emission can not be ruled out.

 In the case of distinct flaring events, \gama-ray  lead over the optical emission can be linked to the difference in the  profiles of the magnetic energy density and the external radiation energy density  during the Gaussian-type of particle injection of in the jet. \citep[see][]{Hayashida2012}. However, assuming the  cooling of the particle by synchrotron and inverse-Compton of external radiation field,  depending upon the profile of those parameters along the jet, both soft or hard lag can occur \citep[see][]{Janiak2012}.  In any case,  study of the individual flares in multiple emission bands, especially \emph{orphan} flares \citep[e. g. see][]{Rajput2020},  could provide further details necessary to identify the dominant process operating in blazar jets \citep[see][]{Liodakis2019}.

It is interesting to note that while rest of the sources in the sample exhibit a significant correlation between the two emission, blazar 3C 273 do not show such a correlation within the lag of a few hundreds of days. The apparent lack of correlation between the emission could be owing to the presence of the prominent big blue bump in the source \citep{Shang2005}. Also,  the jet in the source has been found to contribute only a small fraction ($\sim$10\% and $\sim$ 40\% at the minimum and  the maximum, respectively) to the total optical emission\citep[see][]{Li2020}. This clearly results in a weak correlation between the optical emission, that is a mixture of the emission from the jet and the disk, and \gama-ray emission that  primarily originates in the jets.

\item Quasi-periodic oscillations: \\
 In this work, search for QPOs in the optical light curves of the sample blazars was mainly focused on the possible QPOs in the sources for which high significance QPOs in the decade-long Fermi/LAT observations were reported in \citetalias{Bhatta2020}.  Indeed, in some of the sources, e. g., S5 0616+714, Mkr 421, Mrk 501, PKS 2155-304 and PKS 1424, the optical LSP peaks at the timescale close to the \gama-ray QPO timescales were observed as shown in panel f) of Figure \ref{fig:6}. The significance of the possible QPO features  against spurious detection due to the red-noise behavior of blazars were estimated employing a large number of simulated light-curves with similar statistical properties, e. g., mean, standard deviation, observation length, and sampling rates. However, the significance levels presented here should be interpreted in the light of a few of the important caveats. The first, possibly the most important, one is that the assumed single-power law model used in the simulation could be too simplistic, such that for the rigorous PSD estimation more complex models, e. g., model with a break frequency and  auto-regressive models, should be considered. Similarly, as the current source sample is derived from the \gama-ray sample defined in \citetalias{Bhatta2020}, there is no direct way to estimate the trial factor to account for the ``look elsewhere effect".  Consequently, this could lead to the overestimation of the confidence levels. Lastly,  the simulations of the light curves do not directly consider the uncertainties in  the PSD indexes reported in \citet{Nilsson2018}. Nevertheless, it is emphasized that the optical and \gama-ray light curves have different the sampling rates, duration and the natures of the inherent red-noise, e. g. power-law spectral slope indexes in the \gama-ray and optical band are $\sim1.0$ and $\sim1.5$, respectively. In such context, if a peak in the LSP appears at the same temporal frequency in both the bands, this would serve as a strong qualitative indication of the presence of QPO at that temporal frequency, and therefore it is less likely to have arisen due to correlated noise.

 QPOs in blazars can naturally arise  in gravitationally bound supermassive binary black hole (SMBBH) systems, especially lying at the 
milliparsec separation which is well within in the  gravitational-wave driven regime \citep{Begelman1980,Liao2021}. The observed timescales can be interpreted as the Keplerian periods of the secondary black hole around the central black hole.   The periodic timescale in such a close SMBBH systems can change due to emission of low frequency (a few tens of nano-Hertz) gravitational waves (GW); however, the changes in the periods can not be detectable before a few thousands years \cite[see][]{Peters1964}.

  The observed QPOs might as well have been originated at the innermost regions of the central engine where the effects of gravitational field is strong. As a result,  the rapidly spinning  supermassive black hole can warp  space-time and give rise to the  precession  of the disk owing to  the \emph{ Lense-Thirring precession}.   In blazars, such disk precession can lead to jet precession, which in turn can appear as QPOs \cite[e.g][]{Liska2018}.  Similarly,  jet precession \citep{Graham2015}, and jet precession  induced in SMBBH can also result in periodic optical outbursts  \citep{Qian2018,Caproni2017}.  Moreover, in the case of strong gravitational fields, relativistic orbit models can be employed to explain the QPOs \citep[see][]{Rana2020}.

In the accretion disk based models, various hydrodynamic  instabilities can lead to the formation of  bright hotspots which go revolving around the central black hole with a Keplerian period comparable to the periods of observed QPOs.   For a typical black hole of mass of $10^9 M_\odot$, the radius of the Keplerian orbit corresponding can be located at a few tens of gravitational radius ($r_g$).    Similarly, hydrodynamical instability at the disc could be induced in a binary system \citep[see][]{Kelley2019}. In such a scenario, if the secondary companion is orbiting in a plane making an angle to the plane of the accretion disk, it can exert a torque resulting in the precession of the disk. This in turn can cause the disk to precess with a characteristic timescale\cite[see][]{Romero2000,Katz1997}. Also, thick accretion disks can be globally perturbed  and  consequently undergo p-mode oscillations with a fundamental frequency similar to the temporal frequencies associated with the QPO periods  \citep[see][and the reference therein]{An2013}.  

QPOs also arise when an emission region follows a helical path along the magnetized jets \cite[e.g.,][]{Mohan2015,Rieger2004,Camenzind92}.  In the case of non-ballistic  motion with typical inclination angle, $i \sim 1/\Gamma_b$ and bulk Lorentz factors $\Gamma_b\sim 10$, the observed period P can be significantly shortened by the relation $P\simeq \Gamma _{b}^{2}/(1+z)P_{obs}$. In such a scenario relativistic effects become dominant and the periodic changes  in the viewing angle translates into  the periodic flux modulation. 

The presence of similar timescale QPOs in more than one wavebands supports the argument for the MWL nature of these oscillations. Periodic modulations in the optical emission could well be dictated by the processes at the innermost regions of the accretion disk. But the observed strong correlation in the optical and \gama-ray emission, and the fact that the QPOs were observed at the similar characteristic  temporal frequency in  both the bands suggest that the processes generating optical QPOs should be co-spatial to the regions where \gama-ray emission in blazar is produced.   
\end{itemize}

\section{Conclusion \label{sec:5}}
Decade-long observations  from 4 AGN optical data archives, viz, AAVSO, SMARTS, Catalina and Steward Observatory were gathered to construct densely sampled light curves of 12 \gama-ray bright blazars. The light curves were performed adopting several methods of time series analysis with a goal to study the long-term variability properties of the sample blazars.  The results of the analyses were compared with those from the similar analyses in the \gama-ray band performed on the same sources in our previous work \citetalias{Bhatta2020}. It was found that,  similar to the \gama-ray emission, the optical emission from the sample blazars was found to display pronounced flux modulation characterized by  linear RMS-flux relation and log-normal PDF. Therefore, as in the case of \gama-ray variability, the  processes driving optical variability  could be multiplicative non-linear processes correlated over diverse timescales and flux states.  When comparing the results of the similar analysis performed in the same source in \gama-ray and the optical band, the observed variability properties in \gama-ray were found to be more enhanced,  in the sense that, compared to optical emission, \gama-ray emission showed stronger variability, steeper RMS-flux relation and larger PDF skewness.  Furthermore, the results of cross-correlation between the variability features in the two bands resulted in a strong optical–gamma-ray correlations. The observed strong correlation is naturally explained within the frame work of  leptonic
 blazar emission models including both SSC and EC.  Additionally, to examine the multi-frequency nature of the QPOs  in the sources including S5 0716+714, Mrk 421, Mrk 501, PKS 1424-418 and PKS 2155-304,  the LSP features in the optical and \gama-ray bands were compared. The analysis in these sources revealed hints of MWL QPOs at the similar characteristic  timescales.


\acknowledgments

I acknowledge the financial support by the Narodowe Centrum Nauki (NCN) grant UMO-2017/26/D/ST9/01178. I also thank the anonymous referee for a careful and thorough review of this paper, which helped improve the quality of the work.
This paper has made use of up-to-date SMARTS optical/near-infrared light curves that are available at \url{www.astro.yale.edu/smarts/glast/home.php}. We acknowledge with thanks the variable star observations from the AAVSO International Database contributed by observers worldwide and used in this research. The CSS survey is funded by the National Aeronautics and Space
Administration under Grant No. NNG05GF22G issued through the Science Mission Directorate Near-Earth Objects Observations Program.  The CRTS survey is supported by the U.S.~National Science Foundation under grants AST-0909182 and AST-1313422. Data from the Steward Observatory spectropolarimetric monitoring project were used. This program is supported by Fermi Guest Investigator grants NNX08AW56G, NNX09AU10G, NNX12AO93G, and NNX15AU81G.  I  thank Prof. Michal Ostrowski and Prof. Niraj Dhital for their useful discussions. I am also grateful to Prof. James Webb and Ms. Beatriz F. Fernandez for reading the manuscript carefully and suggesting some of the corrections.

\appendix
In this section,  contemporaneous optical and \gama-ray light curves of 12 \gama-ray bright blazars are presented to complement the discrete cross-correlation analysis presented in  Section \ref{sec:dcf}. The \gama-ray Fermi/LAT observations (0.1--300 GeV) were processed during the work of \citetalias{Bhatta2020}. For better comparison the source light curves in the two band are  normalized and presented in the same panel. As the figures show, we find that in general the source flux vary harmoniously in the both the bands.
\begin{figure*}[hb!]
\plotone{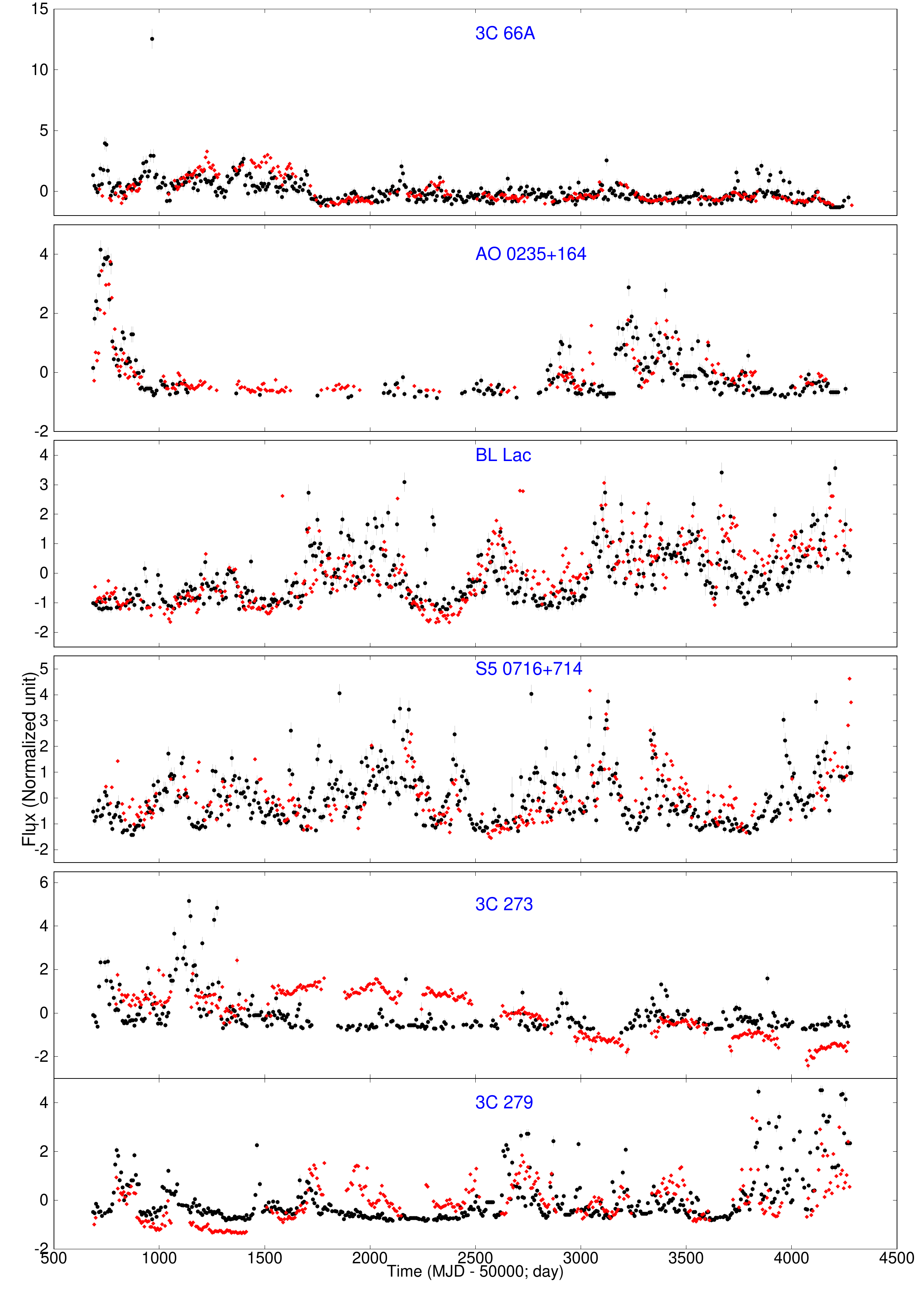}
\label{Fig:1}
\caption{Weekly binned optical (red)  and \gama-ray light curves from Fermi/LAT (black) of 12 \gama-ray bright blazars. 
} 
\end{figure*}

\renewcommand{\thefigure}{\arabic{figure} (Cont.)}
\addtocounter{figure}{-1}

\begin{figure*}[ht!]
\plotone{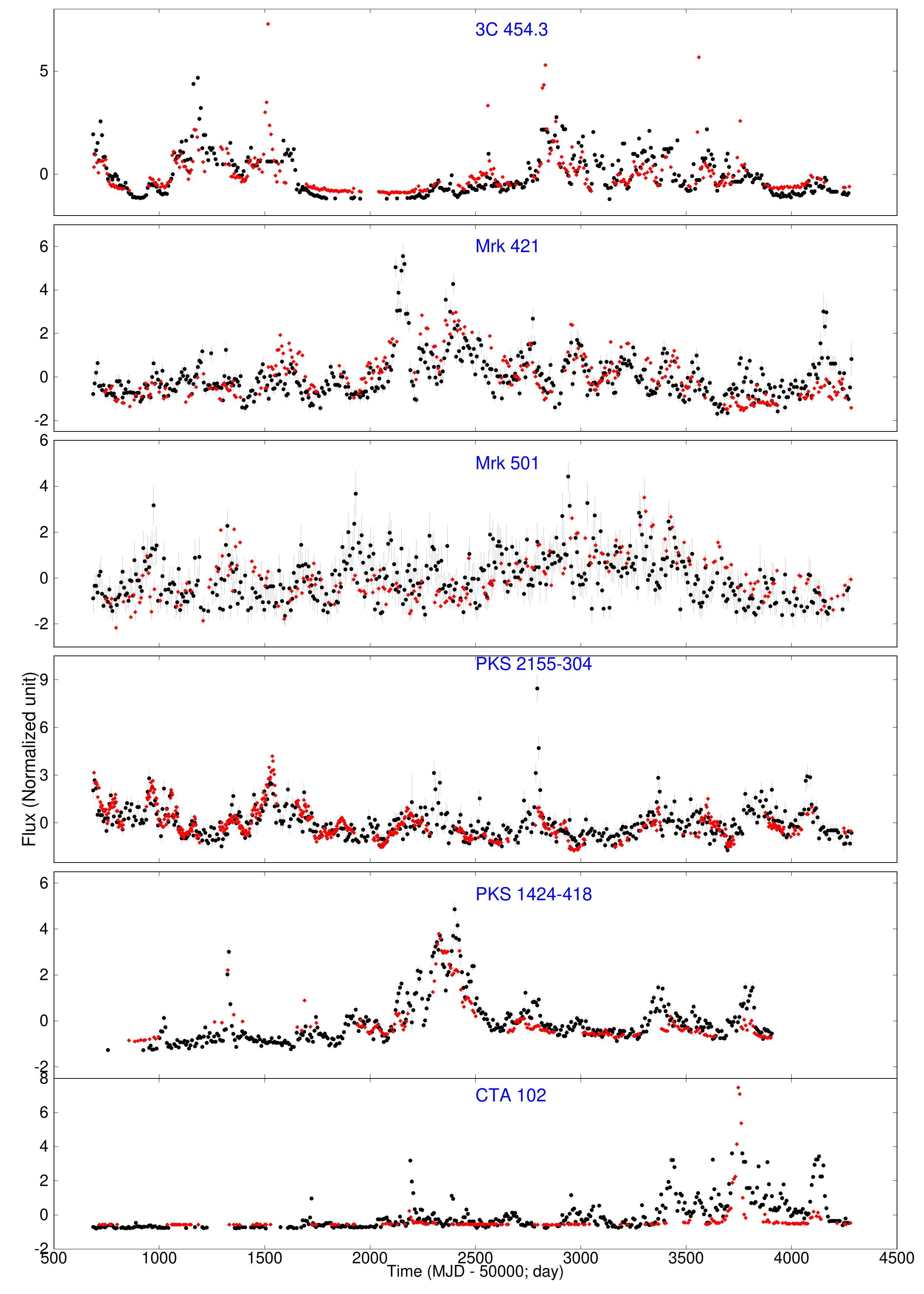}
\label{Fig:1}
\caption{
}
\end{figure*}

\end{document}